\documentclass[aps,prd,twocolumn,showkeys,amsmath,amssymb]{revtex4}
\usepackage{graphicx}
\usepackage{epstopdf}   
\usepackage{multirow}
\usepackage{subfigure}
\usepackage{extarrows}
\usepackage{feynmf}
\usepackage{enumitem}
\usepackage[colorlinks,citecolor=blue,anchorcolor=red,menucolor=red,linkcolor=red,filecolor=red,runcolor=red,urlcolor=blue,frenchlinks=red]{hyperref}
\usepackage{color}

\newcommand{\feynp}[1]{#1\kern-0.45em/}

\def\FF(s){\left[(\alpha+\beta)m_c^2-\alpha\beta s\right]}
\def\HH(s){\left[m_c^2-\alpha(1-\alpha) s\right]}
\def\KK(s){\left[\gamma m_c^2-\gamma(1-\gamma) s\right]}

\allowdisplaybreaks[3]

\begin{document}

\title{Hadronic molecules in B decays}

\author{Hua-Xing Chen}
\email{hxchen@seu.edu.cn}
\affiliation{
School of Physics, Southeast University, Nanjing 210094, China
}

\begin{abstract}
There are eighteen possibly existing $D^{(*)} \bar D^{(*)}$, $D^{(*)} \bar K^{(*)}$, and $D^{(*)} D_s^{(*)-}$ hadronic molecular states. We construct their corresponding interpolating currents and calculate their decay constants using QCD sum rules. Based on these results, we calculate their relative production rates in $B$ and $B^*$ decays through the current algebra, and calculate their relative branching ratios through the Fierz rearrangement, as summarized in Table~\ref{tab:width}. Our results support the interpretations of the $X(3872)$, $Z_c(3900)^0$, $Z_c(4020)^0$, and $X_0(2900)$ as the molecular states $D \bar D^*$ of $J^{PC} = 1^{++}$, $D \bar D^*$ of $J^{PC} = 1^{+-}$, $D^* \bar D^*$ of $J^{PC} = 1^{+-}$, and $D^* \bar K^*$ of $J^P = 0^{+}$, respectively. Our results also suggest that the $Z_{cs}(3985)^-$, $Z_{cs}(4000)^+$, and $Z_{cs}(4220)^+$ are strange partners of the $X(3872)$, $Z_c(3900)^0$, and $Z_c(4020)^0$, respectively. In the calculations we estimate the lifetime of a weakly-coupled composite particle $A = |BC\rangle$ to be $1/t_A \approx 1/t_B + 1/t_C + \Gamma_{A \to BC} + \cdots$, with $\cdots$ partial widths of other possible decay channels.
\end{abstract}

\maketitle

\section{Introduction}
\label{sec:intro}

Since the discovery of the $X(3872)$ in 2003 by Belle~\cite{Choi:2003ue}, a lot of charmonium-like $XYZ$ states were discovered in the past twenty years~\cite{pdg}. These structures are good candidates of tetraquark states, which are composed by two quarks and two antiquarks. Although there is still a long way to fully understand how the strong interaction binds these quarks and antiquarks together, this subject has become one of the most intriguing research topics in particle physics. Their theoretical and experimental studies are significantly improving our understanding of the strong interaction at the low energy region~\cite{Chen:2016qju,Liu:2019zoy,Lebed:2016hpi,Esposito:2016noz,Guo:2017jvc,Ali:2017jda,Olsen:2017bmm,Karliner:2017qhf,Guo:2019twa,Brambilla:2019esw}.

Among all the $XYZ$ states, the $X(3872)$, $X_0(2900)$, $Z_{cs}(4000)^+$, and $Z_{cs}(4220)^+$ have been observed in $B$ decays~\cite{Choi:2003ue,Aaij:2020hon,Aaij:2020ypa,Aaij:2021ivw}. Besides, the $Z_c(3900)^0$ and $Z_c(4020)^0$ (may) decay into the $J/\psi \pi^0$ final states~\cite{Ablikim:2013mio,Liu:2013dau,Voloshin:2013dpa,Chen:2019wjd}, so they are also possible to be observed in the $B^- \to K^- J/\psi \pi^0$ decay. In this paper we shall investigate these six exotic structures as a whole. We shall study their mass spectra, productions in $B$ and $B^*$ decays, and decay properties. Before doing this, let us briefly review some of their information, and we refer to the reviews~\cite{Chen:2016qju,Liu:2019zoy,Lebed:2016hpi,Esposito:2016noz,Guo:2017jvc,Ali:2017jda,Olsen:2017bmm,Karliner:2017qhf,Guo:2019twa,Brambilla:2019esw} for detailed discussions.

\begin{itemize}

\item The $X(3872)$ is the most puzzling state among all the charmonium-like $XYZ$ states. It is now denoted as the $\chi_{c1}(3872)$ in PDG2020~\cite{pdg}, but the mass of $\chi_{c1}(2P)$ was estimated to be $3.95$~GeV~\cite{Godfrey:1985xj}, which value is significantly larger than the mass of $X(3872)$. Consequently, various interpretations were proposed to explain it, such as a compact tetraquark state~\cite{Maiani:2004vq,Maiani:2014aja,Maiani:2020zhr,Hogaasen:2005jv,Ebert:2005nc,Barnea:2006sd,Chiu:2006hd,Padmanath:2015era}, a loosely-bound $D \bar D^*$ molecular state~\cite{Voloshin:1976ap,Voloshin:2003nt,Close:2003sg,Wong:2003xk,Braaten:2003he,Swanson:2003tb,Tornqvist:2004qy,Prelovsek:2013cra,Braaten:2019ags}, and a hybrid charmonium state~\cite{Close:2003mb,Li:2004sta}, etc. The $X(3872)$ was also studied as a conventional $c \bar c$ state in Refs.~\cite{Barnes:2003vb,Eichten:2004uh,Quigg:2004vf,Kong:2006ni}, and was suggested to be the mixture of a $c \bar c$ state with the $D \bar D^{*}$ component in Refs.~\cite{Meng:2005er,Meng:2014ota,Suzuki:2005ha}.

    The quantum number of $X(3872)$ was determined to be $I^GJ^{PC}=0^+1^{++}$~\cite{Aaij:2015eva}. It has been observed in the $J/\psi \pi \pi$, $J/\psi \pi \pi \pi$, $\gamma J/\psi$, $\gamma \psi(2S)$, $\chi_{c1} \pi$, and $D^{0} \bar D^{*0}$ channels~\cite{Gokhroo:2006bt,Aubert:2007rva,Adachi:2008sua,Aubert:2008ae,Bhardwaj:2011dj,Aaij:2014ala,Ablikim:2019soz,Li:2019kpj}. Especially, the $J/\psi \pi \pi$ and $J/\psi \pi \pi \pi$ channels have comparable branching ratios~\cite{Abe:2005ix,delAmoSanchez:2010jr,Choi:2011fc,Ablikim:2019zio}:
    \begin{equation}
    {\mathcal{B}(X(3872) \xrightarrow{~\,\omega\,~} J/\psi \pi \pi \pi) \over \mathcal{B}(X(3872) \xrightarrow{~~\rho~~} J/\psi \pi \pi)} = 1.6^{+0.4}_{-0.3} \pm 0.2 \, ,
    \end{equation}
    which implies a large isospin violation.

\item In 2013 BESIII discovered the charged $Z_c(3900)^\pm$ in the $Y(4260) \to J/\psi\pi^+\pi^-$ decay~\cite{Ablikim:2013mio}, which observation was confirmed by Belle~\cite{Liu:2013dau} and CLEO~\cite{Xiao:2013iha}. Later BESIII observed the charged $Z_c(4020)^\pm$ in the $e^+ e^- \to \pi^+ \pi^- h_c$ process~\cite{Ablikim:2013wzq}. Because the $Z_c(3900)^\pm$ and $Z_c(4020)^\pm$ both couple strongly to charmonia and yet they are charged, these two structures are definitely not conventional $c \bar c$ states. There have been various theoretical models developed to explain them, such as compact tetraquark states~\cite{Maiani:2004vq,Maiani:2014aja,Qiao:2013dda,Wang:2016mmg,Azizi:2020itk}, loosely-bound $D \bar D^*$ and $D^* \bar D^*$ molecular states~\cite{Voloshin:1976ap,Khemchandani:2013iwa,Wang:2013cya,Guo:2013sya,Wilbring:2013cha,He:2014nya,Prelovsek:2014swa,Chen:2015igx,Dong:2021juy,Yan:2021tcp}, hadro-quarkonia~\cite{Braaten:2013boa,Dubynskiy:2008mq}, or due to kinematical threshold effects~\cite{Chen:2011xk,Liu:2013vfa,Swanson:2014tra}, etc.

    The $Z_c(3900)$ has been observed in the $J/\psi \pi$, $h_c \pi$, and $D \bar D^*$ channels~\cite{Ablikim:2013mio,Liu:2013dau,Ablikim:2013xfr,Ablikim:2015swa}, and the quantum number of its neutral one was determined to be $I^GJ^{PC}=1^+1^{+-}$~\cite{Collaboration:2017njt}. The $Z_c(4020)$ has been observed in the $h_c \pi$ and $D^* \bar D^*$ channels~\cite{Ablikim:2013wzq,Ablikim:2013emm}, and the quantum number of its neutral one may also be $I^GJ^{PC}=1^+1^{+-}$. Evidence for the $Z_c(3900)^\pm \rightarrow \eta_c\rho^\pm$ decay was reported at $\sqrt{s} = 4.226$ GeV with the relative branching ratio~\cite{Ablikim:2019ipd}:
    \begin{equation}
    {\mathcal{B}(Z_c(3900)^\pm \rightarrow \eta_c\rho^\pm) \over \mathcal{B}(Z_c(3900)^\pm \rightarrow J/\psi\pi^\pm)} = 2.2 \pm 0.9 \, .
    \end{equation}
    This ratio was suggested in Ref.~\cite{Esposito:2014hsa} to be useful to discriminate the compact tetraquark and hadronic molecule scenarios, and it has been calculated by many theoretical methods in Refs.~\cite{Dias:2013xfa,Agaev:2016dev,Wang:2017lot,Goerke:2016hxf,Ke:2013gia,Patel:2014zja,Li:2014pfa,Dong:2013iqa,Faccini:2013lda,Xiao:2018kfx,Wang:2018pwi}.

\item In 2020 BESIII reported their observation of the $Z_{cs}(3985)^-$ in the $K^+$ recoil-mass spectra of the $e^+e^- \to K^+ (D^-_s D^{*0} + D^{*-}_s D^0)$ process~\cite{Ablikim:2020hsk}. Later LHCb reported their observation of the $Z_{cs}(4000)^+$ and $Z_{cs}(4220)^+$ in the $J/\psi K^+$ invariant mass spectrum of the $B^+ \to J/\psi \phi K^+$ decay~\cite{Aaij:2021ivw}, and one can naturally deduce the existence of the $Z_{cs}(4000)^-$ and $Z_{cs}(4220)^-$ as their antiparticles. These structures couple strongly to charmonia and yet they are charged, so they are definitely exotic hadrons/structures. There have been some theoretical methods developed to explain the $Z_{cs}(3985)^-$, $Z_{cs}(4000)^-$, and $Z_{cs}(4220)^-$, such as compact tetraquark states~\cite{Cui:2006mp,Wang:2009bd,Azizi:2020zyq,Wang:2020iqt,Jin:2020yjn,Sungu:2020zvk}, loosely-bound $D D_s^{*-}$, $D^* D_s^{-}$, and $D^* D_s^{*-}$ molecular states~\cite{Lee:2008uy,Meng:2020ihj,Yang:2020nrt,Liu:2020nge,Du:2020vwb,Chen:2020yvq,Cao:2020cfx,Sun:2020hjw,Ikeno:2021ptx,Wang:2020htx,Guo:2020vmu}, or due to kinematical threshold effects~\cite{Chen:2013wca,Wang:2020dmv,Wang:2020kej,Zhu:2021vtd,Ge:2021sdq}, etc.

\item In 2020 LHCb reported their observation of the $X_0(2900)$ in the $D^- K^+$ invariant mass spectrum of the $B^+ \to D^+ D^- K^+$ decay~\cite{Aaij:2020hon,Aaij:2020ypa}. This structure has the quark content $\bar c \bar s u d$, so it is an exotic hadron with valence quarks of four different flavors. Various theoretical methods were applied to explain it, such as a compact tetraquark state~\cite{Karliner:2020vsi,He:2020jna,Lu:2020qmp,Wang:2020xyc,Wang:2020prk,Yang:2021izl}, a loosely-bound $D^* \bar K^*$ molecular state~\cite{Molina:2010tx,Molina:2020hde,Liu:2020nil,Huang:2020ptc,Hu:2020mxp,Xiao:2020ltm,Tan:2020cpu,Abreu:2020ony,Qi:2021iyv}, or due to triangle singularities~\cite{Liu:2020orv}, etc.

\end{itemize}
Summarizing the above discussions, we quickly notice that the $X(3872)$, $Z_c(3900)^0$, $Z_c(4020)^0$, and $X_0(2900)$ can be interpreted in the hadronic molecular picture as the molecular states $D \bar D^*$ of $J^{PC} = 1^{++}$, $D \bar D^*$ of $J^{PC} = 1^{+-}$, $D^* \bar D^*$ of $J^{PC} = 1^{+-}$, and $D^* \bar K^*$ of $J^{P} = 0^{+}$, respectively; the $Z_{cs}(3985)^-/Z_{cs}(4000)^-$ and $Z_{cs}(4220)^-$ can be interpreted in the hadronic molecular picture as the molecular states $D D_s^{*-}/D^* D_s^-$ of $J^{P} = 1^{+}$ and $D^* D_s^{*-}$ of $J^{PC} = 1^{+}$. Besides, there may exist the molecular states $D \bar D$ of $J^{PC} = 0^{++}$, $D^* \bar D^*$ of $J^{PC} = 0^{++}/2^{++}$, $D \bar K$ of $J^P = 0^{+}$, $D \bar K^*/D^* \bar K$ of $J^P = 1^{+}$, $D^* \bar K^*$ of $J^P = 1^{+}/2^{+}$, $D D_s^-$ of $J^{P} = 0^{+}$, and $D^* D_s^{*-}$ of $J^{P} = 0^{+}/2^{+}$.

Altogether there are eighteen possibly existing $D^{(*)} \bar D^{(*)}$, $D^{(*)} \bar K^{(*)}$, and $D^{(*)} D^{(*)-}$ molecular states. We use the symbol
\begin{equation}
|D^{(*)} \bar D^{(*)}/ D^{(*)} \bar K^{(*)} / D^{(*)} \bar D_s^{(*)}; J^{P(C)} \rangle \, ,
\end{equation}
to denote them, where $\bar D_s^{(*)}$ is used to denote $D_s^{(*)-}$. To make this paper more understandable, we further simply our notations as follows:
\begin{itemize}

\item We shall not differentiate the charged $Z_c(3900)^\pm$ and the neutral $Z_c(3900)^0$. They are both denoted as the $Z_c(3900)$, which is interpreted as the $D \bar D^*$ molecular state of $J^{PC} = 1^{+-}$ in the present study. So does the $Z_c(4020)$, which is interpreted as the $D^* \bar D^*$ molecular state of $J^{PC} = 1^{+-}$. Note that their negative charge-conjugation parity $C=-$ is only due to the neutral ones, while the charged ones are not charge-conjugated.

\item We shall also denote the $Z_{cs}(3985)^-$, $Z_{cs}(4000)^\pm$, and $Z_{cs}(4220)^\pm$ as the $Z_{cs}(3985)$, $Z_{cs}(4000)$, and $Z_{cs}(4220)$, respectively. Both the $Z_{cs}(3985)$ and $Z_{cs}(4000)$ can be interpreted as the $D D_s^{*-}/D^* D_s^-$ molecular states of $J^{P} = 1^{+}$, and in the present study we shall further consider the mixing between the $D D_s^{*-}$ and $D^* D_s^{-}$ components, since their thresholds are quite close to each other. By doing this we can obtain strange partners of the $| D \bar D^*; 1^{++} \rangle$ and $| D \bar D^*; 1^{+-} \rangle$, which are denoted for convenience as $| D \bar D_s^{*}; 1^{++} \rangle$ and $| D \bar D_s^{*}; 1^{+-} \rangle$, respectively. We shall use them to separately explain the $Z_{cs}(3985)$ and $Z_{cs}(4000)$, but note that they are not charge-conjugated actually. See Sec.~\ref{sec:DDscurrent} for detailed discussions.

\end{itemize}
The above notations can be used to well differentiate the $X(3872)$, $Z_c(3900)$, and $Z_c(4020)$ as well as the $Z_{cs}(3985)$, $Z_{cs}(4000)$, and $Z_{cs}(4220)$. We refer to Refs.~\cite{Gamermann:2006nm,Dai:2015bcc,Mutuk:2018zxs,Wu:2019vbk,Cao:2020dlz,Dong:2020hxe,Prelovsek:2020eiw,Chen:2020eyu,Wang:2020dgr,Burns:2020xne,Wei:2021usz,Ozdem:2021yvo,Yang:2021sue} for some relevant theoretical studies, and to the reviews~\cite{Chen:2016qju,Liu:2019zoy,Lebed:2016hpi,Esposito:2016noz,Guo:2017jvc,Ali:2017jda,Olsen:2017bmm,Karliner:2017qhf,Guo:2019twa,Brambilla:2019esw} again for their detailed discussions.

In this paper we shall systematically study the eighteen possibly existing $D^{(*)} \bar D^{(*)}$, $D^{(*)} \bar K^{(*)}$, and $D^{(*)} \bar D_s^{(*)}$ hadronic molecular states. We shall systematically construct their corresponding interpolating currents, and apply the method of QCD sum rules to calculate their decay constants. The obtained results will be used to further study their production and decay properties. This method has been applied to systematically investigate $\bar D^{(*)} \Sigma_c^{(*)}$ hadronic molecular states in Ref.~\cite{Chen:2020opr}.

We use the isoscalar $D \bar D$ molecular state of $J^{PC} = 0^{++}$, $| D \bar D; 0^{++} \rangle$, as an example to briefly show our procedures. Firstly, we construct its corresponding hidden-charm tetraquark current:
\begin{equation}
\eta_1(x) = \bar q_a(x) \gamma_5 c_a(x) ~ \bar c_b(x) \gamma_5 q_b(x) \, ,
\end{equation}
which is a local meson-meson current coupling to $| D \bar D; 0^{++} \rangle$ through
\begin{equation}
\langle 0 | \eta_1 | D \bar D; 0^{++} \rangle = f_{| D \bar D; 0^{++} \rangle} \, .
\end{equation}
In principle, one needs to explicitly use the non-local current in order to exactly describe the $D \bar D$ molecular state, but this can not be done yet within the present QCD sum rule framework. In the above expressions, $a$ and $b$ are color indices; $q(x)$ is an $up$ or $down$ quark field, and $c(x)$ is a $charm$ quark field; the decay constant $f_{| D \bar D; 0^{++}\rangle}$ can be calculated using QCD sum rules, which is an important input parameter when studying production and decay properties of $| D \bar D; 0^{++}\rangle$.

Secondly, we investigate the three-body $B^- \to K^- D^0 \bar D^0$ decay, where $| D \bar D; 0^{++} \rangle$ can be produced. The total quark content of the final states is $\bar u c \bar c s \bar u u$. We apply the Fierz rearrangement to carefully examine the combination of these six quarks, from which we select the current $\eta_1$, so that the relative production rate of $| D \bar D; 0^{++} \rangle$ can be estimated.

Thirdly, we apply the Fierz rearrangement~\cite{fierz} of the Dirac and color indices to transform the current $\eta_1$ into
\begin{equation}
\eta_1 \rightarrow - {1\over12} ~ \bar q_a \gamma_5 q_a ~ \bar c_b \gamma_5 c_b
    + {1\over12} ~ \bar q_a \gamma_\mu q_a ~ \bar c_b \gamma^\mu c_b + \cdots \, .
\end{equation}
Accordingly, $\eta_1$ couples to the $\eta_c \eta$ and $J/\psi \omega$ channels simultaneously:
\begin{eqnarray}
\nonumber\langle 0 | \eta_1 | \eta_c \eta \rangle &\approx& - {1\over12}~\langle 0 | \bar q_a \gamma_5 q_a | \eta \rangle~\langle 0 | \bar c_b \gamma_5 c_b | \eta_c \rangle + \cdots \, ,
\\ \nonumber \langle 0 | \eta_1 | J/\psi \omega \rangle &\approx& {1\over12}~\langle 0 | \bar q_a \gamma_\mu q_a | \omega \rangle~ \langle 0 | \bar c_b \gamma^\mu c_b | J/\psi \rangle + \cdots \, .
\\
\end{eqnarray}
We can use these two equations to straightforwardly calculate the relative branching ratio of the $| D \bar D; 0^{++} \rangle$ decay into $\eta_c \eta$ to its decay into $J/\psi \omega$.

This paper is organized as follows. In Sec.~\ref{sec:current} we systematically construct hidden-charm tetraquark currents corresponding to the $D^{(*)} \bar D^{(*)}$, $D^{(*)} \bar K^{(*)}$, and $D^{(*)} \bar D_s^{(*)}$ hadronic molecular states. We use them to perform QCD sum rule analyses and calculate their decay constants. The obtained results are used in Sec.~\ref{sec:production} to study the productions of the $D^{(*)} \bar D^{(*)}$, $D^{(*)} \bar K^{(*)}$, and $D^{(*)} \bar D_s^{(*)}$ hadronic molecular states in $B$ and $B^*$ decays through the current algebra. In Sec.~\ref{sec:decay} we use the Fierz rearrangement of the Dirac and color indices to study their decay properties, and calculate some of their relative branching ratios. The obtained results are summarized and discussed in Sec.~\ref{sec:summary}. This paper has a supplementary file ``OPE.nb'' containing all the spectral densities.

\section{Interpolating currents}
\label{sec:current}

In this section we systematically construct (strange) hidden-charm and open-charm tetraquark interpolating currents corresponding to $D^{(*)} \bar D^{(*)}$, $D^{(*)} \bar K^{(*)}$, and $D^{(*)} \bar D_s^{(*)}$ hadronic molecular states. We separately construct them in the following subsections.

The isospin quantum number of these molecular states is important. In the present study,
\begin{itemize}

\item We assume that $|D \bar D; 0^{++} \rangle$, $|D^* \bar D^*; 0^{++} \rangle$, and $|D^* \bar D^*; 2^{++} \rangle$ have $I=0$.

\item The isospin of the $X(3872)$ as $|D \bar D^*; 1^{++} \rangle$ will be separately investigated in Sec.~\ref{sec:DD1pp}. Before that we simply assume it also has $I=0$.

\item We assume that $|D \bar D^*; 1^{+-} \rangle$ and $|D^* \bar D^*; 1^{+-} \rangle$ have $I=1$.

\item We assume that all the $D^{(*)} \bar K^{(*)}$ molecular states have $I=0$.

\item All the $D^{(*)} \bar D_s^{(*)}$ molecular states have $I=1/2$.

\end{itemize}
Keeping this in mind, we shall use $q$ to denote either the $up$ or $down$ quark, so that we do not need to explicitly write the isospin contents out.

\subsection{$D^{(*)} \bar D^{(*)}$ currents}
\label{sec:DDcurrent}

\begin{figure}[hbt]
\begin{center}
\subfigure[~\mbox{$\theta(x)$ currents}]{\includegraphics[width=0.2\textwidth]{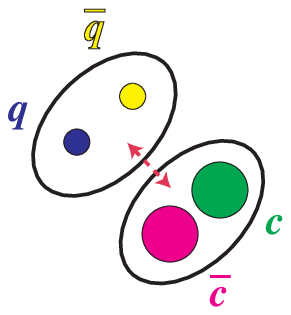}}
\subfigure[~\mbox{$\eta(x)$ currents}]{\includegraphics[width=0.2\textwidth]{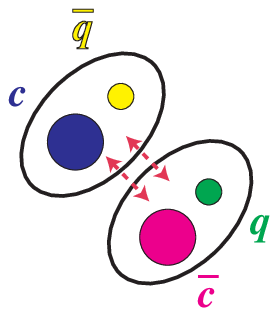}}
\caption{Two types of hidden-charm tetraquark currents, $\theta(x)$ and $\eta(x)$. Quarks are shown in red/green/blue color, and antiquarks are shown in cyan/magenta/yellow color.}
\label{fig:DDcurrent}
\end{center}
\end{figure}

In this subsection we use the $\bar c$, $c$, $\bar q$, and $q$ ($q=u/d$) quarks to construct hidden-charm tetraquark interpolating currents. We consider two types of currents, as illustrated in Fig.~\ref{fig:DDcurrent}:
\begin{eqnarray}
\theta(x) &=& [\bar q_a(x) \Gamma^\theta_1 q_b(x)] ~ [\bar c_c(x) \Gamma^\theta_2 c_d(x)] \, ,
\\[1mm] \eta(x)   &=& [\bar q_a(x) \Gamma^\eta_1 c_b(x)]   ~ [\bar c_c(x) \Gamma^\eta_2 q_d(x)] \, ,
\end{eqnarray}
where $a \cdots d$ are color indices and $\Gamma_{1/2}^{\theta/\eta}$ are Dirac matrices. We can relate these two configurations through the Fierz rearrangement in the Lorentz space and the color rearrangement in the color space:
\begin{equation}
\delta_{ab} \delta_{cd} = {1\over3} \delta_{ad} \delta_{cb} + {1\over2} \lambda^n_{ad} \lambda^n_{cb} \, .
\label{eq:color}
\end{equation}
We shall detailedly discuss this rearrangement in Sec.~\ref{sec:DDdecay} when investigating decay properties of $D^{(*)} \bar D^{(*)}$ molecular states.

%
%

There can exist altogether six $D^{(*)} \bar D^{(*)}$ hadronic molecular states:
\begin{eqnarray}
| D \bar D; 0^{++} \rangle &=& | D \bar D \rangle_{J=0} \, ,
\\[1mm]
\sqrt2| D \bar D^*; 1^{++} \rangle &=& | D \bar D^* \rangle_{J=1} + | D^* \bar D \rangle_{J=1} \, ,
\\[1mm]
\sqrt2| D \bar D^*; 1^{+-} \rangle &=& | D \bar D^* \rangle_{J=1} - | D^* \bar D \rangle_{J=1} \, ,
\\[1mm]
| D^* \bar D^*; 0^{++} \rangle &=& | D^* \bar D^* \rangle_{J=0} \, ,
\\[1mm]
| D^* \bar D^*; 1^{+-} \rangle &=& | D^* \bar D^* \rangle_{J=1} \, ,
\\[1mm]
| D^* \bar D^*; 2^{++} \rangle &=& | D^* \bar D^* \rangle_{J=2} \, .
\end{eqnarray}
Their corresponding currents are:
\begin{eqnarray}
\eta_1(x)              &=& \bar q_a(x) \gamma_5 c_a(x) ~ \bar c_b(x) \gamma_5 q_b(x) \, ,
\label{def:eta1}
\\[1mm] \nonumber
\eta_2^\alpha(x)       &=& \bar q_a(x) \gamma_5 c_a(x) ~ \bar c_b(x) \gamma^\alpha q_b(x)
\label{def:eta2}
\\               && ~~~~~~~~~~~~~~~~ - \{ \gamma_5 \leftrightarrow \gamma^\alpha \} \, ,
\\[1mm] \nonumber
\eta_3^\alpha(x)       &=& \bar q_a(x) \gamma_5 c_a(x) ~ \bar c_b(x) \gamma^\alpha q_b(x)
\\               && ~~~~~~~~~~~~~~~~ + \{ \gamma_5 \leftrightarrow \gamma^\alpha \} \, ,
\label{def:eta3}
\\[1mm]
\eta_4(x)              &=& \bar q_a(x) \gamma^\mu c_a(x) ~ \bar c_b(x) \gamma_\mu q_b(x) \, ,
\label{def:eta4}
\\[1mm] \nonumber
\eta_5^\alpha(x)       &=& \bar q_a(x) \gamma_\mu c_a(x) ~ \bar c_b(x)\sigma^{\alpha\mu}\gamma_5 q_b(x)
\\               && ~~~~~~~~~~~~~~~~ - \{ \gamma_\mu \leftrightarrow \sigma^{\alpha\mu}\gamma_5 \} \, ,
\label{def:eta5}
\\[1mm]
\eta_6^{\alpha\beta}(x) &=& P^{\alpha\beta,\mu\nu} \bar q_a(x) \gamma_\mu c_a(x) \bar c_b(x) \gamma_\nu q_b(x) \, ,
\label{def:eta6}
\end{eqnarray}
where $P^{\alpha\beta,\mu\nu}$ is the spin-2 projection operator
\begin{equation}
P^{\alpha\beta,\mu\nu} = {1\over2} g^{\alpha\mu} g^{\beta\nu} + {1\over2} g^{\alpha\nu} g^{\beta\mu} - {1 \over 4} g^{\alpha\beta} g^{\mu\nu} \, .
\end{equation}

In the above expressions we have used the tensor field $\bar c_b \sigma_{\mu\nu}\gamma_5 q_b$ ($\mu, \nu=0\cdots3$) to construct the current $\eta_5^\alpha(x)$ of $J^{PC} = 1^{+-}$. In principle, this tensor field contains both $J^P = 1^+$ and $1^-$ components, but its negative-parity component $\bar c_b \sigma_{ij}\gamma_5 q_b$ ($i, j=1, 2, 3$) gives the dominant contribution to $\eta_5^\alpha(x)$. Hence, the tetraquark current $\eta_5^\alpha(x)$ of $J^{PC} = 1^{+-}$ corresponds to $| D^* \bar D^*; 1^{+-} \rangle$. Beside it, there exists another current directly corresponding to $| D^* \bar D^*; 1^{+-} \rangle$:
\begin{equation}
\eta_7^{\alpha\beta}(x) = \bar q_a(x) \gamma^\alpha c_a(x)~\bar c_b(x) \gamma^\beta q_b(x) - \{ \alpha \leftrightarrow \beta \} \, ,
\end{equation}
but this current itself contains both positive- and negative-parity components, so we do not use it in the present study.

We use $X$ to generally denote the $D^{(*)} \bar D^{(*)}$ molecular states, and assume that the currents $\eta_{1\cdots6}^{\alpha_1\cdots\alpha_J}$ of spin-$J$ couple to them through
\begin{equation}
\langle 0 | \eta^{\alpha_1\cdots\alpha_J} | X \rangle = f_X \epsilon^{\alpha_1\cdots\alpha_J} \, ,
\end{equation}
where $f_X$ is the decay constant, and $\epsilon^{\alpha_1\cdots\alpha_J}$ is the traceless and symmetric polarization tensor.

We apply the method of QCD sum rules~\cite{Shifman:1978bx,Reinders:1984sr} to study the $D^{(*)} \bar D^{(*)}$ hadronic molecular states through the currents $\eta_{1\cdots6}$. We calculate their spectral densities using the method of operator product expansion (OPE) at the leading order of $\alpha_s$ and up to the $D({\rm imension}) = 8$ terms. In the calculations we calculate the perturbative term, the quark condensates $\langle \bar q q \rangle$/$\langle \bar s s \rangle$, the gluon condensate $\langle g_s^2 GG \rangle$, the quark-gluon mixed condensates $\langle g_s \bar q \sigma G q \rangle$/$\langle g_s \bar s \sigma G s \rangle$, and their combinations $\langle \bar q q \rangle^2$, $\langle \bar q q \rangle \langle \bar s s \rangle$, $\langle \bar q q \rangle\langle g_s \bar q \sigma G q \rangle$, $\langle \bar q q \rangle\langle g_s \bar s \sigma G s \rangle$, and $\langle \bar s s \rangle\langle g_s \bar q \sigma G q \rangle$. We take into account the $charm$ and $strange$ quark masses, but ignore the $up$ and $down$ quark masses. We adopt the factorization assumption of vacuum saturation for higher dimensional condensates. The $D=3$ quark condensates $\langle \bar q q \rangle$/$\langle \bar s s \rangle$ and the $D=5$ mixed condensates $\langle g_s \bar q \sigma G q \rangle$/$\langle g_s \bar s \sigma G s \rangle$ are both multiplied by the charm quark mass $m_c$, which are thus important power corrections.

The obtained spectral densities are too length, so we list them in the supplementary file ``OPE.nb''. In the calculations we use the following values for various QCD sum rule parameters~\cite{pdg,Yang:1993bp,Ellis:1996xc,Eidemuller:2000rc,Narison:2002pw,Gimenez:2005nt,Jamin:2002ev,Ioffe:2002be,Ovchinnikov:1988gk,colangelo}:
%
\begin{eqnarray}
\nonumber m_s &=& 96 ^{+8}_{-4} \mbox{ MeV} \, ,
\\ \nonumber m_c &=& 1.275 ^{+0.025}_{-0.035} \mbox{ GeV} \, ,
\\ \nonumber  \langle\bar qq \rangle &=& -(0.240 \pm 0.010)^3 \mbox{ GeV}^3 \, ,
\\ \langle\bar ss \rangle &=& (0.8\pm 0.1)\times \langle\bar qq \rangle \, ,
\label{eq:condensates}
\\ \nonumber  \langle g_s^2GG\rangle &=& 0.48\pm 0.14 \mbox{ GeV}^4 \, ,
\\
\nonumber \langle g_s\bar q\sigma G q\rangle &=& - M_0^2\times\langle\bar qq\rangle \, ,
\\
\nonumber \langle g_s\bar s\sigma G s\rangle &=& - M_0^2\times\langle\bar ss\rangle \, ,
\\
\nonumber M_0^2 &=& 0.8 \pm 0.2 \mbox{ GeV}^2 \, ,
\end{eqnarray}
where the running mass in the $\overline{MS}$ scheme is used for the $charm$ quark.

Based on these spectral densities, we calculate masses and decay constants of the $D^{(*)} \bar D^{(*)}$ hadronic molecular states. The results are summarized in Table~\ref{tab:mass}, supporting the interpretations of the $X(3872)$, $Z_c(3900)$, and $Z_c(4020)$ as the $|D \bar D^*; 1^{++} \rangle$, $|D \bar D^*; 1^{+-} \rangle$, and $|D^* \bar D^*; 1^{+-} \rangle$ molecular states, respectively. The accuracy of our QCD sum rule results is moderate but not enough to extract their binding energies. Therefore, our results can only suggest but not determine: a) whether these $D^{(*)} \bar D^{(*)}$ molecular states exist or not, and b) whether they are bound states or resonance states. Note that we have not taken into account the radiative corrections in our QCD sum rule calculations, which lead to even more theoretical uncertainties. However, in the present study we are more concerned about the ratios, {\it i.e.}, the relative production rates and the relative branching ratios, whose uncertainties are significantly reduced. We refer to Refs.~\cite{Navarra:2006nd,Matheus:2006xi,Zhang:2009em,Chen:2010ze,Wang:2013vex,Wang:2013daa,Cui:2013yva,Zhang:2013aoa,Chen:2015ata,Chen:2015moa,Chen:2016otp,Azizi:2017ubq,Ozdem:2017jqh,Chen:2019osl,Wang:2019hnw,Xu:2020qtg,Zhang:2020oze,Chen:2020aos,Agaev:2020nrc,Albuquerque:2020ugi,Albuquerque:2021tqd,Wan:2020oxt,Wang:2020rcx} for more QCD sum rule studies.

\begin{table*}[hpt]
\begin{center}
\renewcommand{\arraystretch}{1.5}
\caption{Masses and decay constants of $D^{(*)} \bar D^{(*)}$, $D^{(*)} \bar K^{(*)}$, and $D^{(*)} \bar D_s^{(*)}$ hadronic molecular states, extracted from the currents $\eta_{1\cdots6}$, $\xi_{1\cdots6}$, and $\zeta_{1\cdots6}$, respectively.}
\begin{tabular}{c | c | c | c | c | c | c | c | c}
\hline\hline
\multirow{2}{*}{\,Currents\,} & \multirow{2}{*}{\,Configuration\,} & \multirow{2}{*}{\,$s_0^{min}~[{\rm GeV}^2]$\,} & \multicolumn{2}{c|}{Working Regions} & \multirow{2}{*}{\,Pole~[\%]\,} & \multirow{2}{*}{\,Mass~[GeV]\,} & \multirow{2}{*}{~~$f_X$~[GeV$^5$]~~} & \multirow{2}{*}{\,Candidate\,}
\\ \cline{4-5} & & & ~$s_0~[{\rm GeV}^2]$~ & \,$M_B^2~[{\rm GeV}^2]$\, & & &
\\ \hline \hline
$\eta_1$ & $|D \bar D; 0^{++} \rangle$           & 16.6 & $18.0\pm1.0$ & $2.77$--$3.08$ & $40$--$50$ & $3.73^{+0.08}_{-0.08}$ & $\left(1.42^{+0.28}_{-0.25}\right) \times 10^{-2}$ &
\\
$\eta_2^\alpha$ & $|D \bar D^*; 1^{++} \rangle$  & 17.3 & $19.0\pm1.0$ & $2.77$--$3.15$ & $40$--$51$ & $3.87^{+0.08}_{-0.09}$ & $\left(2.16^{+0.40}_{-0.36}\right) \times 10^{-2}$ & $X(3872)$
\\
$\eta_3^\alpha$ & $|D \bar D^*; 1^{+-} \rangle$  & 17.3 & $19.0\pm1.0$ & $2.77$--$3.15$ & $40$--$51$ & $3.87^{+0.08}_{-0.09}$ & $\left(2.16^{+0.40}_{-0.36}\right) \times 10^{-2}$ & $Z_c(3900)$
\\
$\eta_4$ & $|D^* \bar D^*; 0^{++} \rangle$       & 18.7 & $20.0\pm1.0$ & $2.72$--$3.00$ & $40$--$49$ & $4.01^{+0.11}_{-0.11}$ & $\left(3.07^{+0.58}_{-0.52}\right) \times 10^{-2}$ &
\\
$\eta_5^\alpha$ & $|D^* \bar D^*; 1^{+-} \rangle$& 19.6 & $20.0\pm1.0$ & $3.05$--$3.14$ & $40$--$42$ & $4.06^{+0.14}_{-0.13}$ & $\left(4.96^{+0.89}_{-0.80}\right) \times 10^{-2}$ & $Z_c(4020)$
\\
$\eta_6^{\alpha\beta}$ & $|D^* \bar D^*; 2^{++} \rangle$
                                                 & 17.9 & $20.0\pm1.0$ & $2.68$--$3.20$ & $40$--$55$ & $4.04^{+0.13}_{-0.12}$ & $\left(1.71^{+0.30}_{-0.27}\right) \times 10^{-2}$ &
\\ \hline
$\xi_1$ & $|D \bar K; 0^{+} \rangle$             &  9.2 & $10.0\pm1.0$ & $2.08$--$2.26$ & $40$--$47$ & $2.72^{+0.10}_{-0.10}$ & $\left(0.77^{+0.18}_{-0.17}\right) \times 10^{-2}$ &
\\
$\xi_2^\alpha$ & $|D \bar K^*; 1^{+} \rangle$    & 10.4 & $11.0\pm1.0$ & $2.24$--$2.37$ & $40$--$45$ & $2.89^{+0.10}_{-0.11}$ & $\left(0.86^{+0.19}_{-0.17}\right) \times 10^{-2}$ &
\\
$\xi_3^\alpha$ & $|D^* \bar K; 1^{+} \rangle$    & 10.0 & $11.0\pm1.0$ & $2.13$--$2.35$ & $40$--$49$ & $2.85^{+0.10}_{-0.11}$ & $\left(0.82^{+0.18}_{-0.17}\right) \times 10^{-2}$ &
\\
$\xi_4$ & $|D^* \bar K^*; 0^{+} \rangle$         & 10.6 & $11.5\pm1.0$ & $2.05$--$2.23$ & $40$--$48$ & $2.91^{+0.10}_{-0.11}$ & $\left(1.29^{+0.31}_{-0.28}\right) \times 10^{-2}$ & $X_0(2900)$
\\
$\xi_5^\alpha$ & $|D^* \bar K^*; 1^{+} \rangle$  & 12.0 & $12.5\pm1.0$ & $2.46$--$2.58$ & $40$--$44$ & $3.13^{+0.15}_{-0.12}$ & $\left(2.96^{+0.62}_{-0.55}\right) \times 10^{-2}$ &
\\
$\xi_6^{\alpha\beta}$ & $|D^* \bar K^*; 2^{+} \rangle$
                                                 & 11.8 & $12.5\pm1.0$ & $2.40$--$2.56$ & $40$--$45$ & $3.09^{+0.12}_{-0.11}$ & $\left(1.10^{+0.23}_{-0.21}\right) \times 10^{-2}$ &
\\ \hline \hline
$\zeta_1$ & $|D \bar D_s; 0^{+} \rangle$            & 18.4 & $19.0\pm1.0$ & $3.08$--$3.22$ & $40$--$44$ & $3.86^{+0.07}_{-0.08}$ & $\left(1.74^{+0.32}_{-0.29}\right) \times 10^{-2}$ &
\\
$\zeta_2^\alpha$ & $|D \bar D_s^*; 1^{++} \rangle$  & 19.3 & $20.0\pm1.0$ & $3.14$--$3.31$ & $40$--$45$ & $3.99^{+0.08}_{-0.08}$ & $\left(2.65^{+0.46}_{-0.42}\right) \times 10^{-2}$ & $Z_{cs}(3985)$
\\
$\zeta_3^\alpha$ & $|D \bar D_s^*; 1^{+-} \rangle$  & 19.3 & $20.0\pm1.0$ & $3.14$--$3.31$ & $40$--$45$ & $3.99^{+0.08}_{-0.08}$ & $\left(2.65^{+0.46}_{-0.42}\right) \times 10^{-2}$ & $Z_{cs}(4000)$
\\
$\zeta_4$ & $|D^* \bar D_s^*; 0^{+} \rangle$        & 21.1 & $22.0\pm1.0$ & $3.20$--$3.38$ & $40$--$45$ & $4.20^{+0.08}_{-0.08}$ & $\left(4.44^{+0.77}_{-0.69}\right) \times 10^{-2}$ &
\\
$\zeta_5^\alpha$ & $|D^* \bar D_s^*; 1^{+} \rangle$ & 22.6 & $22.0\pm1.0$ & $\sim 3.69$    & $\sim 37$  & $4.22^{+0.09}_{-0.09}$ & $\left(6.96^{+1.15}_{-1.03}\right) \times 10^{-2}$ & $Z_{cs}(4220)$
\\
$\zeta_6^{\alpha\beta}$ & $|D^* \bar D_s^*; 2^{+} \rangle$
                                                    & 20.0 & $22.0\pm1.0$ & $3.16$--$3.62$ & $40$--$52$ & $4.20^{+0.09}_{-0.10}$ & $\left(2.39^{+0.36}_{-0.34}\right) \times 10^{-2}$ &
\\ \hline\hline
\end{tabular}
\label{tab:mass}
\end{center}
\end{table*}

\subsection{$D^{(*)} \bar K^{(*)}$ currents}
\label{sec:DKcurrent}

\begin{figure}[hbt]
\begin{center}
\includegraphics[width=0.2\textwidth]{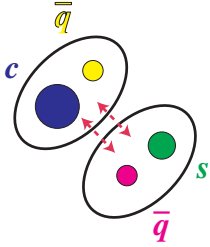}
\caption{Open-charm tetraquark currents $\xi(x)$.}
\label{fig:DKcurrent}
\end{center}
\end{figure}

In this subsection we use the $c$, $s$, $\bar q$, and $\bar q$ ($q=u/d$) quarks to construct open-charm tetraquark interpolating currents. We consider the following type of currents, as illustrated in Fig.~\ref{fig:DKcurrent}:
\begin{equation}
\xi(x)   = [\bar q_a(x) \Gamma^\xi_1 c_b(x)]   ~ [\bar q_c(x) \Gamma^\xi_2 s_d(x)] \, ,
\end{equation}
where $\Gamma_{1/2}^{\xi}$ are Dirac matrices. We shall use the Fierz rearrangement to study this configuration in Sec.~\ref{sec:DKdecay}.

%
%

There can exist altogether six $D^{(*)} \bar K^{(*)}$ hadronic molecular states:
\begin{eqnarray}
| D \bar K; 0^{+} \rangle &=& | D \bar K \rangle_{J=0} \, ,
\\[1mm]
| D \bar K^*; 1^{+} \rangle &=& | D \bar K^* \rangle_{J=1} \, ,
\\[1mm]
| D^* \bar K; 1^{+} \rangle &=& | D^* \bar K \rangle_{J=1} \, ,
\\[1mm]
| D^* \bar K^*; 0^{+} \rangle &=& | D^* \bar K^* \rangle_{J=0} \, ,
\\[1mm]
| D^* \bar K^*; 1^{+} \rangle &=& | D^* \bar K^* \rangle_{J=1} \, ,
\\[1mm]
| D^* \bar K^*; 2^{+} \rangle &=& | D^* \bar K^* \rangle_{J=2} \, .
\end{eqnarray}
Their corresponding currents are:
\begin{eqnarray}
\xi_1(x)              &=& \bar q_a(x) \gamma_5 c_a(x) ~ \bar q_b(x) \gamma_5 s_b(x) \, ,
\label{def:xi1}
\\[1mm]
\xi_2^\alpha(x)       &=& \bar q_a(x) \gamma_5 c_a(x) ~ \bar q_b(x) \gamma^\alpha s_b(x) \, ,
\label{def:xi2}
\\[1mm]
\xi_3^\alpha(x)       &=& \bar q_a(x) \gamma^\alpha c_a(x) ~ \bar q_b(x) \gamma_5 s_b(x) \, ,
\label{def:xi3}
\\[1mm]
\xi_4(x)              &=& \bar q_a(x) \gamma^\mu c_a(x) ~ \bar q_b(x) \gamma_\mu s_b(x) \, ,
\label{def:xi4}
\\[1mm] \nonumber
\xi_5^\alpha(x)       &=& \bar q_a(x) \gamma_\mu c_a(x) ~ \bar q_b(x)\sigma^{\alpha\mu}\gamma_5 s_b(x)
\\               && ~~~~~~~~~~~~~~~~ - \{ \gamma_\mu \leftrightarrow \sigma^{\alpha\mu}\gamma_5 \} \, ,
\label{def:xi5}
\\[1mm]
\xi_6^{\alpha\beta}(x) &=& P^{\alpha\beta,\mu\nu} \bar q_a(x) \gamma_\mu c_a(x) \bar q_b(x) \gamma_\nu s_b(x) \, .
\label{def:xi6}
\end{eqnarray}

We use the currents $\xi_{1\cdots6}$ to perform numerical analyses, and calculate masses and decay constants of $D^{(*)} \bar K^{(*)}$ molecular states. The obtained results are summarized in Table~\ref{tab:mass}. Our results support the interpretation of the $X_0(2900)$ as the $|D^* \bar K^*; 0^{+} \rangle$ molecular state. However, masses of $|D \bar K; 0^{+} \rangle$ and $|D^* \bar K; 1^{+} \rangle$ are significantly larger than the $D \bar K$ and $D^* \bar K$ thresholds at about 2360~MeV and 2500~MeV, respectively. This diversity may be due to the nature of $K$ mesons as Nambu-Goldstone bosons. Again, the accuracy of our QCD sum rule results is moderate but not enough to determine: a) whether these $D^{(*)} \bar K^{(*)}$ molecular states exist or not, and b) whether they are bound states or resonance states.

\subsection{$D^{(*)} \bar D_s^{(*)}$ currents}
\label{sec:DDscurrent}

In this subsection we use the $c$, $\bar c$, $s$, and $\bar q$ ($q=u/d$) quarks to construct strange hidden-charm tetraquark interpolating currents. We consider the following type of currents:
\begin{equation}
\zeta(x)   = [\bar q_a(x) \Gamma^\zeta_1 c_b(x)]   ~ [\bar c_c(x) \Gamma^\zeta_2 s_d(x)] \, ,
\end{equation}
where $\Gamma_{1/2}^{\zeta}$ are Dirac matrices. Their Fierz rearrangements are quite similar to those for $\eta(x)$, just with one light $up$/$down$ quark replaced by another $strange$ quark.

There can exist altogether six $D^{(*)} \bar D_s^{(*)}$ hadronic molecular states:
\begin{eqnarray}
| D \bar D_s; 0^{+} \rangle &=& | D D_s^- \rangle_{J=0} \, ,
\\[1mm]
\sqrt2| D \bar D_s^{*}; 1^{++} \rangle &=& | D D_s^{*-} \rangle_{J=1} + | D^* D_s^- \rangle_{J=1} \, ,
\label{def:DDs2}
\\[1mm]
\sqrt2| D \bar D_s^{*}; 1^{+-} \rangle &=& | D D_s^{*-} \rangle_{J=1} - | D^* D_s^- \rangle_{J=1} \, ,
\label{def:DDs3}
\\[1mm]
| D^* \bar D_s^{*}; 0^{+} \rangle &=& | D^* D_s^{*-} \rangle_{J=0} \, ,
\\[1mm]
| D^* \bar D_s^{*}; 1^{+} \rangle &=& | D^* D_s^{*-} \rangle_{J=1} \, ,
\\[1mm]
| D^* \bar D_s^{*}; 2^{+} \rangle &=& | D^* D_s^{*-} \rangle_{J=2} \, .
\end{eqnarray}
Their corresponding currents are:
\begin{eqnarray}
\zeta_1(x)              &=& \bar q_a(x) \gamma_5 c_a(x) ~ \bar c_b(x) \gamma_5 s_b(x) \, ,
\label{def:zeta1}
\\[1mm] \nonumber
\zeta_2^\alpha(x)       &=& \bar q_a(x) \gamma_5 c_a(x) ~ \bar c_b(x) \gamma^\alpha s_b(x)
\label{def:zeta2}
\\               && ~~~~~~~~~~~~~~~~ - \{ \gamma_5 \leftrightarrow \gamma^\alpha \} \, ,
\\[1mm] \nonumber
\zeta_3^\alpha(x)       &=& \bar q_a(x) \gamma_5 c_a(x) ~ \bar c_b(x) \gamma^\alpha s_b(x)
\\               && ~~~~~~~~~~~~~~~~ + \{ \gamma_5 \leftrightarrow \gamma^\alpha \} \, ,
\label{def:zeta3}
\\[1mm]
\zeta_4(x)              &=& \bar q_a(x) \gamma^\mu c_a(x) ~ \bar c_b(x) \gamma_\mu s_b(x) \, ,
\label{def:zeta4}
\\[1mm] \nonumber
\zeta_5^\alpha(x)       &=& \bar q_a(x) \gamma_\mu c_a(x) ~ \bar c_b(x)\sigma^{\alpha\mu}\gamma_5 s_b(x)
\\               && ~~~~~~~~~~~~~~~~ - \{ \gamma_\mu \leftrightarrow \sigma^{\alpha\mu}\gamma_5 \} \, ,
\label{def:zeta5}
\\[1mm]
\zeta_6^{\alpha\beta}(x) &=& P^{\alpha\beta,\mu\nu} \bar q_a(x) \gamma_\mu c_a(x) \bar c_b(x) \gamma_\nu s_b(x) \, .
\label{def:zeta6}
\end{eqnarray}
In the above expressions we have considered the mixing between the $D D_s^{*-}$ and $D^* D_s^{-}$ components, because their thresholds are very close to each other. After doing this, $| D \bar D_s^{*}; 1^{++} \rangle$ is the strange partner of $| D \bar D^*; 1^{++} \rangle$, so we denote its quantum number as $J^{PC} = 1^{++}$ for convenience; $| D \bar D_s^{*}; 1^{+-} \rangle$ is the strange partner of $| D \bar D^*; 1^{+-} \rangle$, so we also denote its quantum number as $J^{PC} = 1^{+-}$.

We use the currents $\zeta_{1\cdots6}$ to perform numerical analyses, and calculate masses and decay constants of $D^{(*)} \bar D_s^{(*)}$ molecular states. The obtained results are summarized in Table~\ref{tab:mass}. Our results support the interpretations of the $Z_{cs}(3985)$, $Z_{cs}(4000)$, and $Z_{cs}(4220)$ as the $|D \bar D_s^*; 1^{++} \rangle$, $|D \bar D_s^*; 1^{+-} \rangle$, and $|D^* \bar D_s^*; 1^{+} \rangle$ molecular states, respectively. However, the accuracy of our QCD sum rule results is not enough to differentiate the $Z_{cs}(3985)$ and $Z_{cs}(4000)$.

To better understand $D^{(*)} \bar D^{(*)}$, $D^{(*)} \bar K^{(*)}$, and $D^{(*)} \bar D_s^{(*)}$ molecular states, we shall further study their production and decay properties in the following sections. Especially, the decay constants $f_X$ calculated in this section are important input parameters.

\section{Productions through the current algebra}
\label{sec:production}

In this section we study productions of $D^{(*)} \bar D^{(*)}$, $D^{(*)} \bar K^{(*)}$, and $D^{(*)} \bar D_s^{(*)}$ hadronic molecular states in $B$ and $B^{*}$ decays through the current algebra. We shall calculate their relative production rates, such as $\mathcal{B}(B^- \to K^- X):\mathcal{B}(B^- \to K^- X^\prime)$, with $X$ and $X^\prime$ different molecular states. This method has been applied in Ref.~\cite{Chen:2020opr} to systematically study productions of $\bar D^{(*)} \Sigma_c^{(*)}$ hadronic molecular states in $\Lambda_b^0$ decays.

%
\begin{figure*}[hbt]
\begin{center}
\subfigure[~$B^- \to K^- D^{(*)0} {\bar D}^{(*)0}$]{\includegraphics[width=0.45\textwidth]{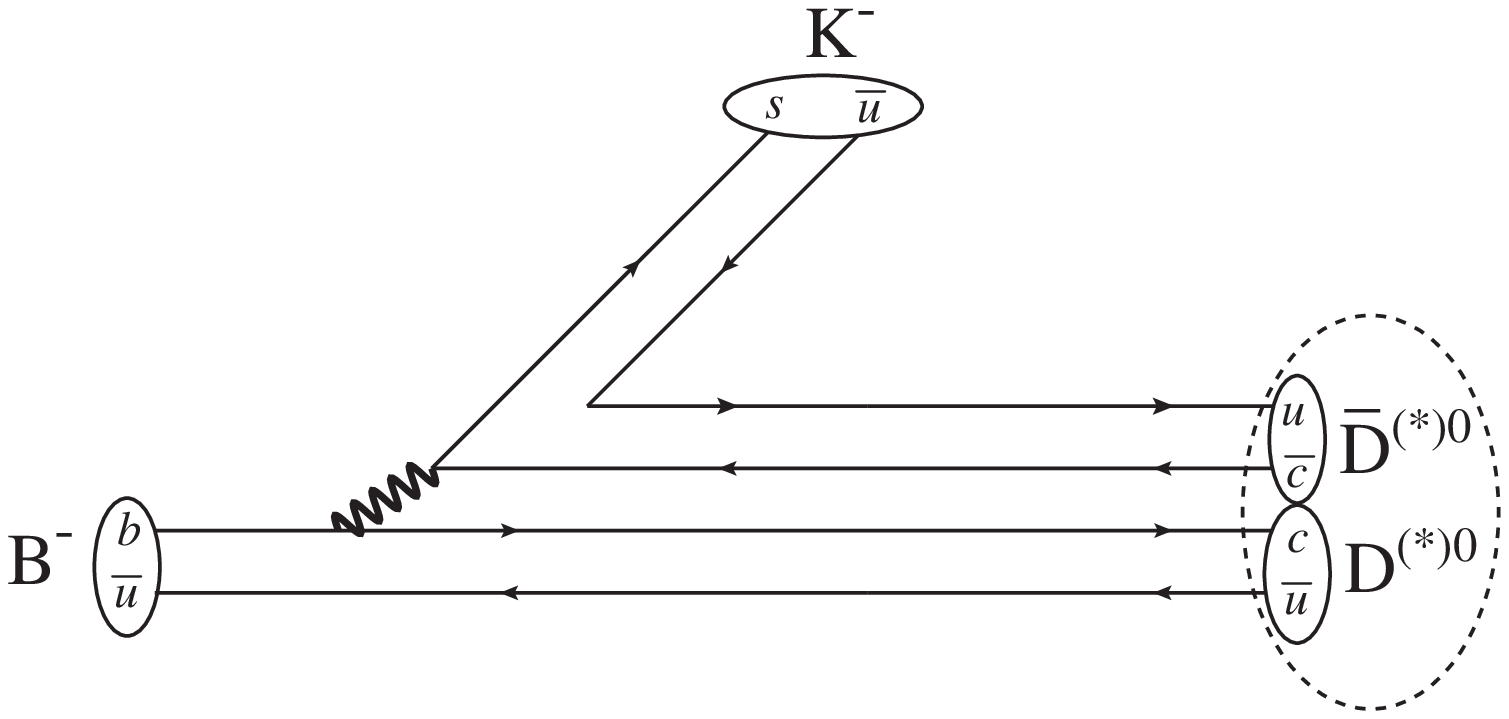}}
~~~~~~~~~~
\subfigure[~$B^- \to D^- D^{(*)0} {\bar K}^{(*)0}$]{\includegraphics[width=0.45\textwidth]{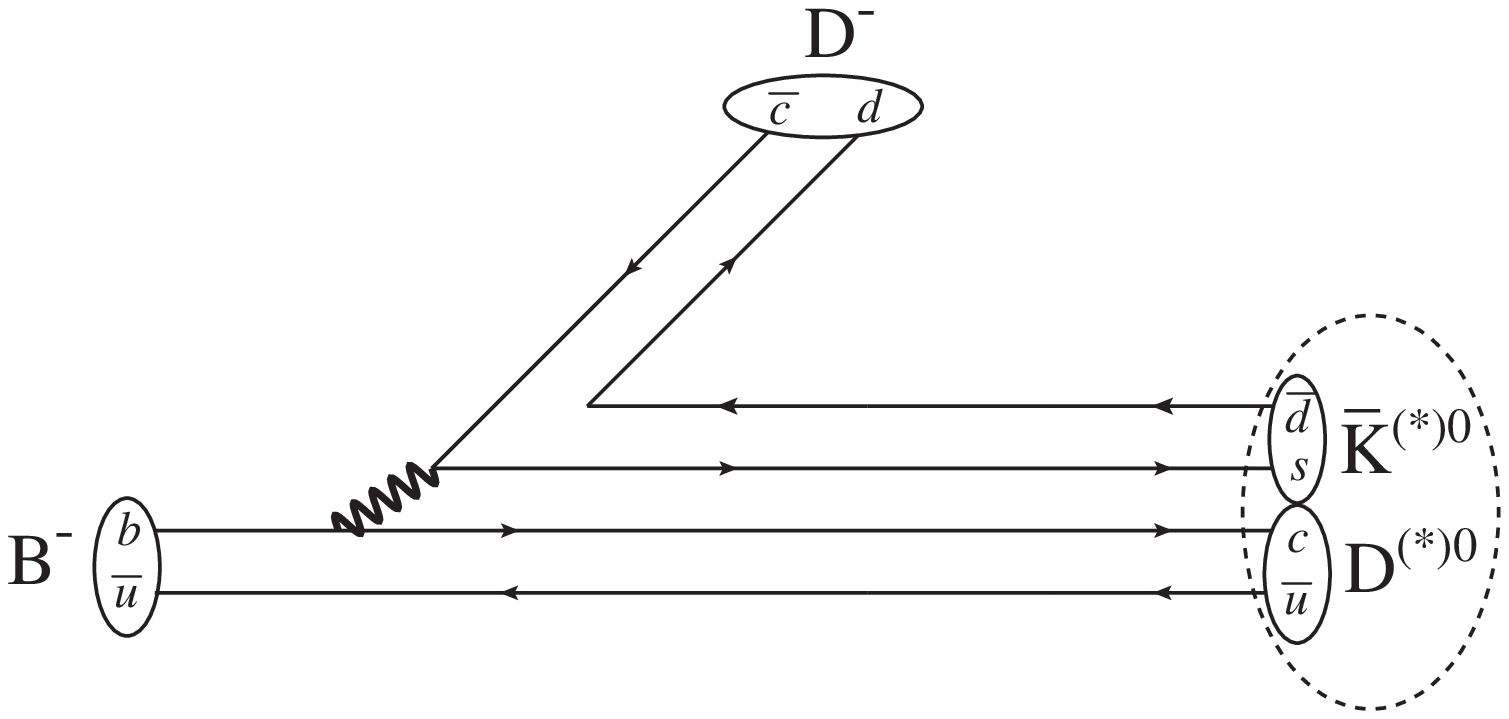}}
\\
\subfigure[~$B^- \to \phi D^{(*)0} D_s^{(*)-}$]{\includegraphics[width=0.45\textwidth]{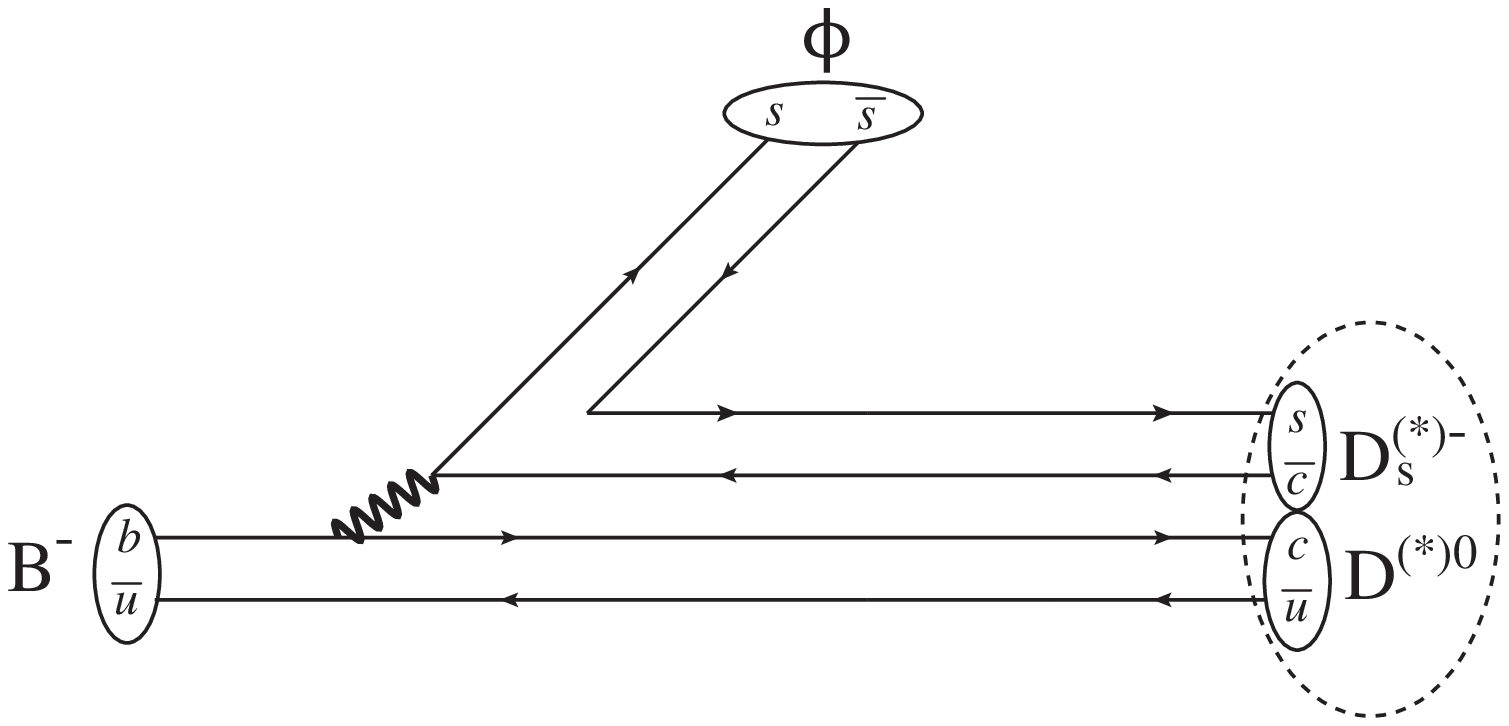}}
\\
\subfigure[~$B^- \to K^- D^{(*)+} {D}^{(*)-}$]{\includegraphics[width=0.45\textwidth]{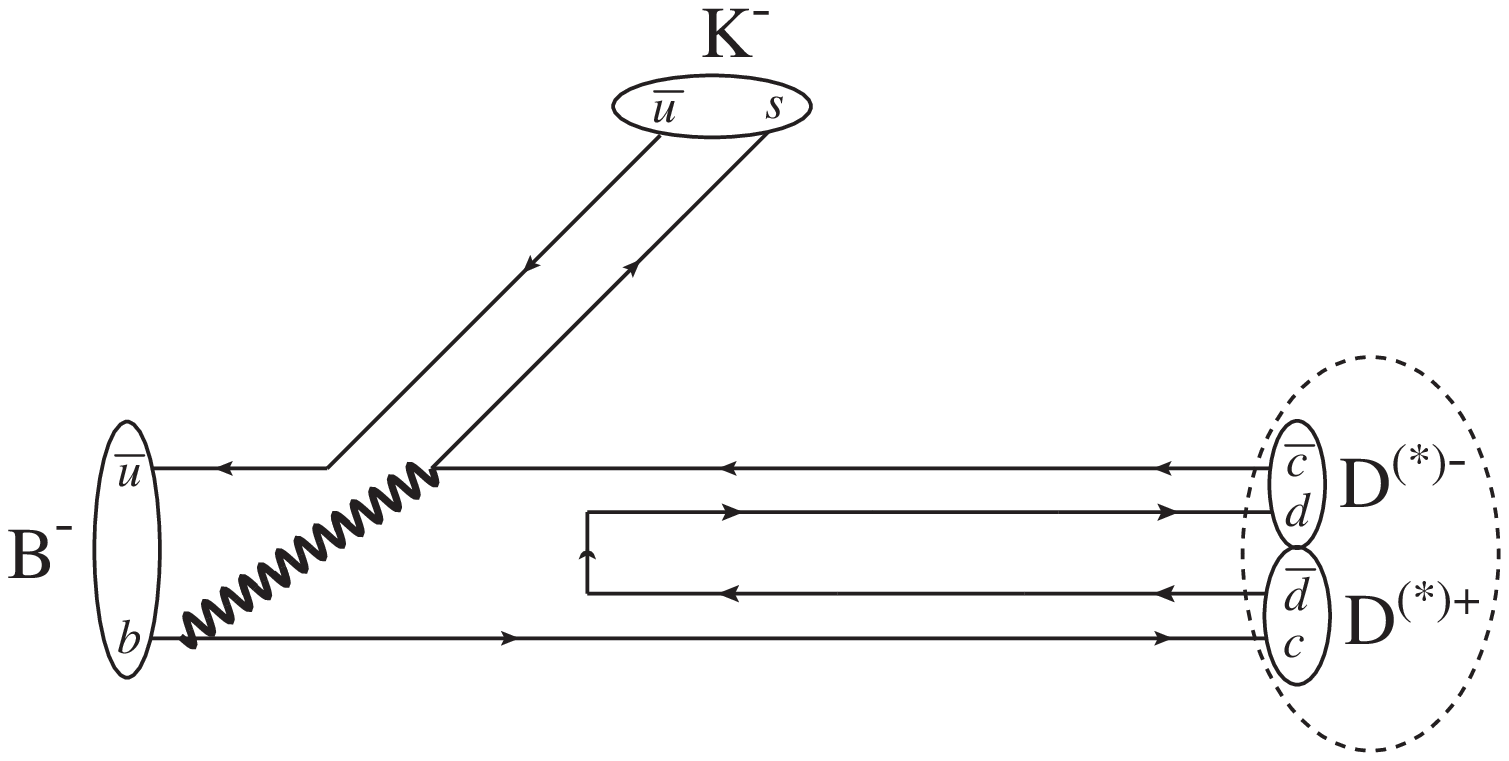}}
~~~~~~~~~~
\subfigure[~$B^- \to D^- D^{(*)+} {K}^{(*)-}$]{\includegraphics[width=0.45\textwidth]{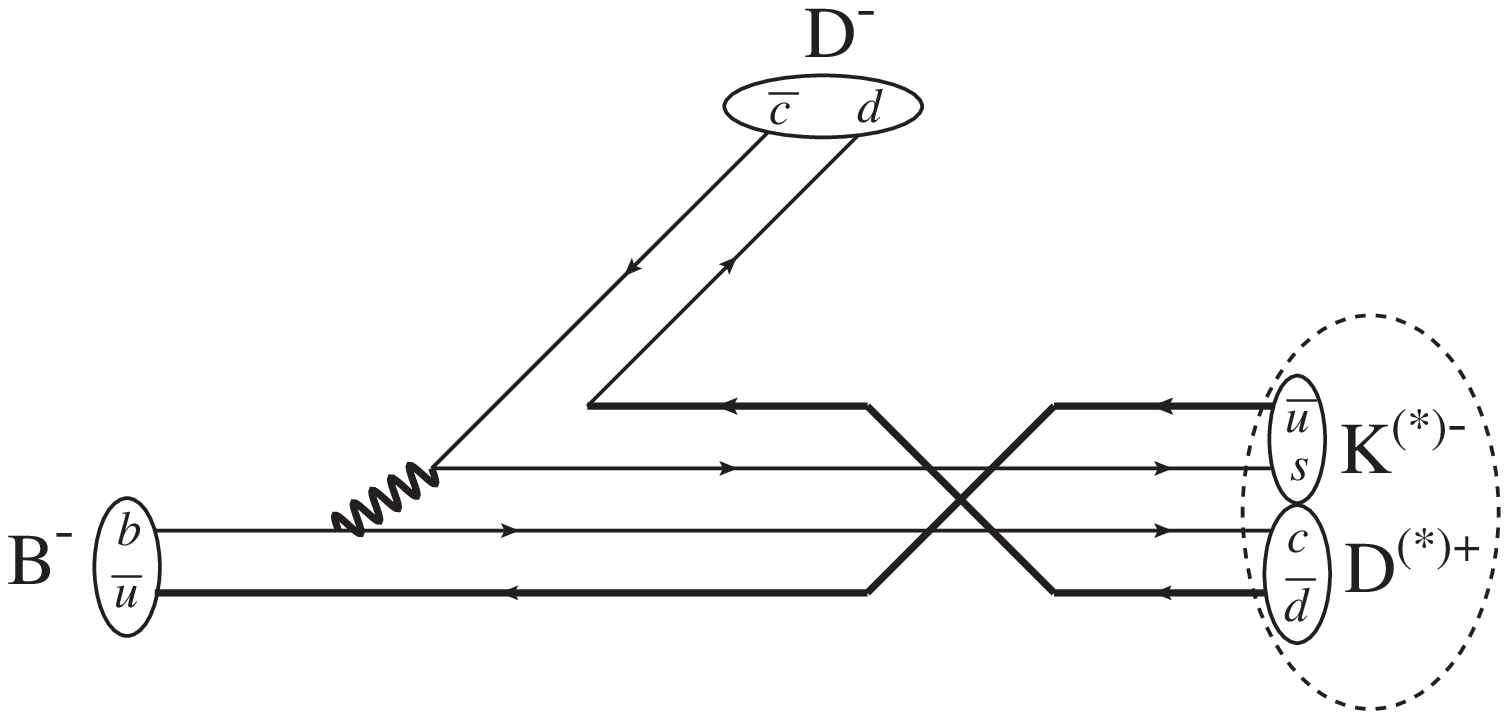}}
\caption{Possible production mechanisms of $D^{(*)} \bar D^{(*)}$, $D^{(*)} \bar K^{(*)}$, and $D^{(*)} \bar D_s^{(*)}$ hadronic molecular states in $B^-$ decays.}
\label{fig:production}
\end{center}
\end{figure*}
%

As depicted in Fig.~\ref{fig:production}, the quark content of the initial state $B^-$ is $\bar u b$. When it decays, first the $b$ quark decays into a $c$ quark by emitting a $W^-$ boson, and this $W^-$ boson translates into a pair of $\bar c$ and $s$ quarks, both of which are Cabibbo-favored; then they pick up an isoscalar quark-antiquark pair $\bar u u + \bar d d + \bar s s$ from the vacuum; finally they hadronize into the $D^{(*)} \bar D^{(*)} \bar K^{(*)}$ and $D^{(*)} \bar D_s^{(*)} \phi$ final states:
\begin{eqnarray}
\nonumber B^- &=& \bar u b \to \bar u c~\bar c s \to \bar u c~\bar c s~(\bar u u + \bar d d + \bar s s)
\\ &\to& D^{(*)} \bar D^{(*)} \bar K^{(*)} + D^{(*)} \bar D_s^{(*)} \phi + \cdots \, .
\end{eqnarray}
Among all the possible final states, $K^- D^{(*)0} \bar D^{(*)0}$, $K^- D^{(*)+} {D}^{(*)-}$, $D^- D^{(*)0} \bar K^{(*)0}$, and $D^- D^{(*)+} K^{(*)-}$ can further produce the neutral $D^{(*)} \bar D^{(*)}$ and $D^{(*)} \bar K^{(*)}$ molecular states, and $\phi D^{(*)0} D_s^{(*)-}$ can further produce the charged $D^{(*)} \bar D_s^{(*)}$ molecular states.


We shall develop three Fierz rearrangements in Sec.~\ref{sec:producionequation} to describe the production mechanisms depicted in Fig.~\ref{fig:production}(a,b,c), and use them to perform numerical analyses separately in Sec.~\ref{sec:productionDD}, Sec.~\ref{sec:productionDK}, and Sec.~\ref{sec:productionDDs}.  Productions of $D^{(*)} \bar D^{(*)}$, $D^{(*)} \bar K^{(*)}$, and $D^{(*)} \bar D_s^{(*)}$ molecular states in $B^{*}$ decays will be similarly investigated in Sec.~\ref{sec:Bsproducion}.

\subsection{Fierz rearrangement}
\label{sec:producionequation}

\begin{widetext}
To describe the production mechanism depicted in Fig.~\ref{fig:production}(a), we use the color rearrangement given in Eq.~(\ref{eq:color}), and apply the Fierz transformation to interchange the $s_d$ and $u_f$ quarks:
\begin{eqnarray}
B^- &\xrightarrow{~~~~~~~~~~~~~}& J_{B^-} = [\delta^{ab} \bar u_a \gamma_5 b_b]
\label{eq:stepDD}
\\[1mm] &\xrightarrow{~~~~\rm weak~~~~}&       [\delta^{ab} \bar u_a \gamma_\rho(1 - \gamma_5) c_b] ~\times~ [\delta^{cd}\bar c_c \gamma^\rho(1 - \gamma_5) s_d]
\label{eq:step1}
\\[1mm] &\xrightarrow{~~~~\rm QPC~~~~}&        [\delta^{ab} \bar u_a \gamma_\rho(1 - \gamma_5) c_b] ~\times~ [\delta^{cd}\bar c_c \gamma^\rho(1 - \gamma_5) s_d] ~\times~ [\delta^{ef}\bar u_e u_f]
\label{eq:step2}
\\[1mm] &\xlongequal{~~~~\rm color~~~~}&       {\delta^{ab} \delta^{cf} \delta^{ed} \over 3} \times \bar u_a \gamma_\rho(1 - \gamma_5) c_b \times \bar c_c \gamma^\rho(1 - \gamma_5) s_d \times \bar u_e u_f ~+~ \cdots
\label{eq:step3}
\\[1mm] &\xlongequal{{\rm Fierz:}s_d \leftrightarrow u_f}&
\label{eq:step5}
-{ 1 \over 12} \times [\delta^{ab}\bar u_a \gamma_\rho(1 - \gamma_5) c_b] \times [\delta^{cf}\bar c_c \gamma_5 u_f] \times [\delta^{ed}\bar u_e \gamma^\rho(1 - \gamma_5) s_d]
\\ \nonumber &&
+{ 1 \over 12} \times [\delta^{ab}\bar u_a \gamma_\rho(1 - \gamma_5) c_b] \times [\delta^{cf}\bar c_c \gamma^\rho(1 - \gamma_5) u_f] \times [\delta^{ed}\bar u_e \gamma_5 s_d]
\\ \nonumber &&
-{ i \over 12} \times [\delta^{ab}\bar u_a \gamma_\rho(1 - \gamma_5) c_b] \times [\delta^{cf}\bar c_c \sigma^{\mu\rho} \gamma_5 u_f] \times [\delta^{ed}\bar u_e \gamma_\mu (1-\gamma_5) s_d]
~+~ \cdots
\\[1mm] &\xlongequal{~~~~~~~~~~~~~}&
\label{eq:step6}
+{ 1 \over 24} \times \Big( \eta_3^\alpha - \eta_2^\alpha \Big) \times [\bar u_a \gamma_\alpha \gamma_5 s_a]
\\ \nonumber &&
+{ 1 \over 12} \times \eta_4 \times [\bar u_a \gamma_5 s_a]
\\ \nonumber &&
+{ i \over 24} \times \eta_5^\alpha \times [\bar u_a \gamma_\alpha \gamma_5 s_a]
~+~ \cdots \, .
\end{eqnarray}
Similarly, we describe the production mechanism depicted in Fig.~\ref{fig:production}(b) as:
\begin{eqnarray}
B^- &\xrightarrow{~~~~~~~~~~~~~}& J_{B^-} = [\delta^{ab} \bar u_a \gamma_5 b_b]
\label{eq:stepDK}
\\[1mm] &\xrightarrow{~~~~\rm weak~~~~}&       [\delta^{ab} \bar u_a \gamma_\rho(1 - \gamma_5) c_b] ~\times~ [\delta^{cd}\bar c_c \gamma^\rho(1 - \gamma_5) s_d]
\label{eq:step7}
\\[1mm] &\xrightarrow{~~~~\rm QPC~~~~}&        [\delta^{ab} \bar u_a \gamma_\rho(1 - \gamma_5) c_b] ~\times~ [\delta^{cd}\bar c_c \gamma^\rho(1 - \gamma_5) s_d] ~\times~ [\delta^{ef}\bar d_e d_f]
\label{eq:step8}
\\[1mm] &\xlongequal{~~~~\rm color~~~~}&       {\delta^{ab} \delta^{cf} \delta^{ed} \over 3} \times \bar u_a \gamma_\rho(1 - \gamma_5) c_b \times \bar c_c \gamma^\rho(1 - \gamma_5) s_d \times \bar d_e d_f ~+~ \cdots
\label{eq:step9}
\\[1mm] &\xlongequal{{\rm Fierz:}s_d \leftrightarrow d_f}&
\label{eq:step11}
-{ 1 \over 12} \times [\delta^{ab}\bar u_a \gamma_\rho(1 - \gamma_5) c_b] \times [\delta^{cf}\bar c_c \gamma_5 d_f] \times [\delta^{ed}\bar d_e \gamma^\rho(1 - \gamma_5) s_d]
\\ \nonumber &&
+{ 1 \over 12} \times [\delta^{ab}\bar u_a \gamma_\rho(1 - \gamma_5) c_b] \times [\delta^{cf}\bar c_c \gamma^\rho(1 - \gamma_5) d_f] \times [\delta^{ed}\bar d_e \gamma_5 s_d]
\\ \nonumber &&
-{ i \over 12} \times [\delta^{ab}\bar u_a \gamma_\rho(1 - \gamma_5) c_b] \times [\delta^{cf}\bar c_c \gamma_\mu (1-\gamma_5) d_f] \times [\delta^{ed}\bar d_e \sigma^{\mu\rho} \gamma_5 s_d]
~+~ \cdots
\\[1mm] &\xlongequal{~~~~~~~~~~~~~}&
\label{eq:step12}
-{ 1 \over 12} \times \xi_3^\alpha \times [\bar c_a \gamma_\alpha \gamma_5 d_a]
\\ \nonumber &&
-{ 1 \over 12} \times \xi_4 \times [\bar c_a \gamma_5 d_a]
\\ \nonumber &&
+{ i \over 24} \times \xi_5^\alpha \times [\bar c_a \gamma_\alpha \gamma_5 d_a]
~+~ \cdots \, ,
\end{eqnarray}
and the production mechanism depicted in Fig.~\ref{fig:production}(c) as:
\begin{eqnarray}
B^- &\xrightarrow{~~~~~~~~~~~~~}& J_{B^-} = [\delta^{ab} \bar u_a \gamma_5 b_b]
\label{eq:stepDDs}
\\[1mm] &\xrightarrow{~~~~\rm weak~~~~}&       [\delta^{ab} \bar u_a \gamma_\rho(1 - \gamma_5) c_b] ~\times~ [\delta^{cd}\bar c_c \gamma^\rho(1 - \gamma_5) s_d]
\label{eq:step13}
\\[1mm] &\xrightarrow{~~~~\rm QPC~~~~}&        [\delta^{ab} \bar u_a \gamma_\rho(1 - \gamma_5) c_b] ~\times~ [\delta^{cd}\bar c_c \gamma^\rho(1 - \gamma_5) s_d] ~\times~ [\delta^{ef}\bar s_e s_f]
\label{eq:step14}
\\[1mm] &\xlongequal{~~~~\rm color~~~~}&       {\delta^{ab} \delta^{cf} \delta^{ed} \over 3} \times \bar u_a \gamma_\rho(1 - \gamma_5) c_b \times \bar c_c \gamma^\rho(1 - \gamma_5) s_d \times \bar s_e s_f ~+~ \cdots
\label{eq:step15}
\\[1mm] &\xlongequal{{\rm Fierz:}s_d \leftrightarrow s_f}&
\label{eq:step17}
-{ 1 \over 12} \times [\delta^{ab}\bar u_a \gamma_\rho(1 - \gamma_5) c_b] \times [\delta^{cf}\bar c_c \gamma_5 s_f] \times [\delta^{ed}\bar s_e \gamma^\rho(1 - \gamma_5) s_d]
\\ \nonumber &&
-{ i \over 12} \times [\delta^{ab}\bar u_a \gamma_\rho(1 - \gamma_5) c_b] \times [\delta^{cf}\bar c_c \sigma^{\mu\rho} \gamma_5 s_f] \times [\delta^{ed}\bar s_e \gamma_\mu (1-\gamma_5) s_d]
~+~ \cdots
\\[1mm] &\xlongequal{~~~~~~~~~~~~~}&
\label{eq:step18}
-{ 1 \over 24} \times \Big( \zeta_3^\alpha - \zeta_2^\alpha \Big) \times [\bar s_a \gamma_\alpha s_a]
\\ \nonumber &&
-{ i \over 24} \times \zeta_5^\alpha \times [\bar s_a \gamma_\alpha s_a]
~+~ \cdots \, .
\end{eqnarray}
\end{widetext}

The brief explanations to the Fierz rearrangement given in Eq.~(\ref{eq:stepDD}) are as follows:
\begin{itemize}

\item Eq.~(\ref{eq:step1}) describes the Cabibbo-favored weak decay process, $b\to c + \bar{c}s$, via the $V$-$A$ current.

\item Eq.~(\ref{eq:step2}) describes the production of the $\bar u u$ pair from the vacuum via the $^3P_0$ quark pair creation mechanism.

\item In Eq.~(\ref{eq:step3}) we apply the color rearrangement given in Eq.~(\ref{eq:color}).

\item In Eq.~(\ref{eq:step5}) we apply the Fierz transformation to interchange the $s_d$ and $u_f$ quarks.

\item In Eq.~(\ref{eq:step6}) we combine the four quarks $\bar u_a c_b \bar c_c u_f$ together so that $D^{(*)0} \bar D^{(*)0}$ molecular states can be produced.

\end{itemize}
The other two Fierz rearrangements given in Eqs.~(\ref{eq:stepDK}) and (\ref{eq:stepDDs}) can be similarly investigated.

In the above expression, we only consider $\eta_{1\cdots6}$, $\xi_{1\cdots6}$, and $\zeta_{1\cdots6}$ defined in Eqs.~(\ref{def:eta1}--\ref{def:eta6}), Eqs.~(\ref{def:xi1}--\ref{def:xi6}), and Eqs.~(\ref{def:zeta1}--\ref{def:zeta6}). These currents couple to $D^{(*)} \bar D^{(*)}$, $D^{(*)} \bar K^{(*)}$, and $D^{(*)} \bar D_s^{(*)}$ molecular states through $S$-wave. Besides, there may exist some other currents coupling to these states through $P$-wave, but we do not take them into account in the present study. Hence, some other molecular states such as $|D \bar D; 0^{++} \rangle$ may still be produced in $B^-$ decays, and omissions of these currents cause some theoretical uncertainties.

Beside the three production mechanisms depicted in Fig.~\ref{fig:production}(a,b,c), there exist some other possible production mechanisms, such as those depicted in Fig.~\ref{fig:production}(d,e). However, the mechanism depicted in Fig.~\ref{fig:production}(d) contains the color factor ${\delta^{ad} \delta^{cf} \delta^{eb} \over 9} \times \bar u_a c_b \bar c_c s_d \bar d_e d_f$, and the one depicted in Fig.~\ref{fig:production}(e) contains the color factor ${\delta^{ad} \delta^{cf} \delta^{eb} \over 9} \times \bar u_a c_b \bar c_c s_d \bar d_e d_f$, so both of them are suppressed.

Very recently, the LHCb Collaboration reported their observations of $X(4630)$ and $X(4685)$ in the $B^+ \to K^+ X \to K^+ J/\psi \phi$ decay process~\cite{Aaij:2021ivw}. Although these two structures were also observed in $B$ decays, their production mechanisms may not be (completely) the same as those depicted in Fig.~\ref{fig:production}(a,b,c), so we do not investigate them in the present study.

\subsection{Productions of $D^{(*)} \bar D^{(*)}$ molecules}
\label{sec:productionDD}

In this subsection we use the Fierz rearrangement given in Eq.~(\ref{eq:stepDD}) to perform numerical analyses, and calculate relative production rates of $D^{(*)} \bar D^{(*)}$ molecular states in $B^- \to K^- X$ decays. To do this we need the following couplings to the $K^-$ meson:
\begin{eqnarray}
\langle 0 | \bar u_a i \gamma_5 s_a | K^-(q) \rangle &=& \lambda_{K} \, ,
\\ \nonumber \langle 0 | \bar u_a \gamma_\mu \gamma_5 s_a | K^-(q) \rangle &=& i q_\mu f_{K} \, ,
\end{eqnarray}
where $f_{K}= 155.6$~MeV~\cite{pdg} and $\lambda_{K} = {f_{K}^2 m_K \over m_q + m_s}$ with the isospin-averaged current quark mass $m_q = {m_u + m_d \over 2}$.

We extract from Eq.~(\ref{eq:stepDD}) the following decay channels:
\begin{enumerate}

\item The decay of $B^-$ into $|D \bar D^*; 1^{++} \rangle K^-$ is contributed by $\eta_2^\alpha \times [\bar u_a \gamma_\alpha \gamma_5 s_a]$:
\begin{eqnarray}
&& \langle B^-(q) ~|~D \bar D^*; 1^{++}(\epsilon_1, q_1)~K^-(q_2) \rangle
\\ \nonumber &=& - {i ~ d_1 \over 24}  f_K f_{|D \bar D^*; 1^{++} \rangle} ~ \epsilon_1 \cdot q_2  \, .
\end{eqnarray}
The decay constant $f_{|D \bar D^*; 1^{++} \rangle}$ has been calculated in the previous section and given in Table~\ref{tab:mass}. The overall factor $d_1$ is related to: a) the coupling of $J_{B^-}$ to $B^-$, b) the weak and $^3P_0$ decay processes described by Eqs.~(\ref{eq:step1}) and (\ref{eq:step2}), c) the isospin factor between $|D^0 \bar D^{*0}; 1^{++} \rangle$ and $|D \bar D^*; 1^{++} \rangle$, and d) the dynamical process depicted in Fig.~\ref{fig:production}(a).

\item The decay of $B^-$ into $|D \bar D^*; 1^{+-} \rangle K^-$ is contributed by $\eta_3^\alpha \times [\bar u_a \gamma_\alpha \gamma_5 s_a]$:
\begin{eqnarray}
&& \langle B^-(q) ~|~D \bar D^*; 1^{+-}(\epsilon_1, q_1)~K^-(q_2) \rangle
\\ \nonumber &=& {i ~ d_1 \over 24}  f_K f_{|D \bar D^*; 1^{+-} \rangle} ~ \epsilon_1 \cdot q_2  \, .
\end{eqnarray}

\item The decay of $B^-$ into $|D^* \bar D^*; 0^{++} \rangle K^-$ is contributed by $\eta_4 \times [\bar u_a \gamma_5 s_a]$:
\begin{eqnarray}
&& \langle B^-(q) ~|~D^* \bar D^*; 0^{++}(q_1)~K^-(q_2) \rangle
\\ \nonumber &=& {d_1 \over 12}  \lambda_K f_{|D^* \bar D^*; 0^{++} \rangle}  \, .
\end{eqnarray}

\item The decay of $B^-$ into $|D^* \bar D^*; 1^{+-} \rangle K^-$ is contributed by $\eta_5^\alpha \times [\bar u_a \gamma_\alpha \gamma_5 s_a]$:
\begin{eqnarray}
&& \langle B^-(q) ~|~D^* \bar D^*; 1^{+-}(\epsilon_1, q_1)~K^-(q_2) \rangle
\\ \nonumber &=& -{d_1 \over 24}  f_K f_{|D^* \bar D^*; 1^{+-} \rangle} ~ \epsilon_1 \cdot q_2  \, .
\end{eqnarray}

\end{enumerate}

In the present study we adopt the possible interpretations of the $X(3872)$, $Z_c(3900)$, and $Z_c(4020)$ as the $|D \bar D^*; 1^{++} \rangle$, $|D \bar D^*; 1^{+-} \rangle$, and $|D^* \bar D^*; 1^{+-} \rangle$ molecular states, respectively.
As summarized in Table~\ref{tab:mass}, the masses of the $D^{(*)} \bar D^{(*)}$ molecular states calculated through the QCD sum rule method have considerable uncertainties, based on which we are not able to study their production and decay properties, {\it e.g.}, we are not able to calculate the partial decay width of the $| D \bar D^*; 1^{+-} \rangle$ molecular state into the $D \bar D^*$ channel due to the ambiguous phase space, as will be studied in Sec.~\ref{sec:DD1pm}. Hence, we assume the masses of the $D^{(*)} \bar D^{(*)}$ molecular states to be either the realistic masses or at the lowest $D^{(*)} \bar D^{(*)}$ thresholds~\cite{pdg}:
\begin{eqnarray}
     \nonumber    M_{|D \bar D;     0^{++} \rangle} &\approx& M_{D^0} + M_{\bar D^0}       = 3730   ~{\rm MeV} \, ,
\\[1mm] \nonumber M_{|D \bar D^*;   1^{++} \rangle} &=&       M_{X(3872)}            = 3871.69~{\rm MeV} \, ,
\\[1mm] M_{|D \bar D^*;   1^{+-} \rangle} &=&       M_{Z_c(3900)}          = 3888.4 ~{\rm MeV} \, ,
\label{eq:DDmass}
\\[1mm] \nonumber M_{|D^* \bar D^*; 0^{++} \rangle} &\approx& M_{D^{*0}} + M_{\bar D^{*0}} = 4014   ~{\rm MeV} \, ,
\\[1mm] \nonumber M_{|D^* \bar D^*; 1^{+-} \rangle} &\approx& M_{Z_c(4020)}          = 4024.1 ~{\rm MeV} \, ,
\\[1mm] \nonumber M_{|D^* \bar D^*; 2^{++} \rangle} &\approx& M_{D^{*0}} + M_{\bar D^{*0}} = 4014   ~{\rm MeV} \, .
\end{eqnarray}
Then we can evaluate the above production amplitudes to obtain the following partial decay widths:
\begin{eqnarray}
\nonumber         \Gamma(B^- \to K^- |D \bar D;     0^{++} \rangle) &=& 0 \, ,
\\[1mm] \nonumber \Gamma(B^- \to K^- |D \bar D^*;   1^{++} \rangle) &=& d_1^2~0.40 \times 10^{-10}~{\rm GeV}^{13} \, ,
\\[1mm] \nonumber \Gamma(B^- \to K^- |D \bar D^*;   1^{+-} \rangle) &=& d_1^2~0.38 \times 10^{-10}~{\rm GeV}^{13} \, ,
\\[1mm] \nonumber \Gamma(B^- \to K^- |D^* \bar D^*; 0^{++} \rangle) &=& d_1^2~4.21 \times 10^{-10}~{\rm GeV}^{13} \, ,
\\[1mm] \nonumber \Gamma(B^- \to K^- |D^* \bar D^*; 1^{+-} \rangle) &=& d_1^2~1.35 \times 10^{-10}~{\rm GeV}^{13} \, ,
\\[1mm]           \Gamma(B^- \to K^- |D^* \bar D^*; 2^{++} \rangle) &=& 0 \, .
\label{result:productionDD}
\end{eqnarray}

\subsection{Productions of $D^{(*)} \bar K^{(*)}$ molecules}
\label{sec:productionDK}

Similarly, we use the Fierz rearrangement given in Eq.~(\ref{eq:stepDK}) to perform numerical analyses, and calculate relative production rates of $D^{(*)} \bar K^{(*)}$ molecular states in $B^- \to D^- X$ decays:
\begin{enumerate}
\setcounter{enumi}{4}

\item The decay of $B^-$ into $|D^* \bar K; 1^{+} \rangle D^-$ is contributed by $\xi_3^\alpha \times [\bar c_a \gamma_\alpha \gamma_5 d_a]$:
\begin{eqnarray}
&& \langle B^-(q) ~|~D^* \bar K; 1^{+}(\epsilon_1, q_1)~D^-(q_2) \rangle
\\ \nonumber &=& - {i ~ d_2 \over 12}  f_D f_{|D^* \bar K; 1^{+} \rangle} ~ \epsilon_1 \cdot q_2  \, .
\end{eqnarray}
The overall factor $d_2$ is related to: a) the coupling of $J_{B^-}$ to $B^-$, b) the weak and $^3P_0$ decay processes described by Eqs.~(\ref{eq:step7}) and (\ref{eq:step8}), c) the isospin factor between $|D^{*0} \bar K^{0}; 1^{+} \rangle$ and $|D^* \bar K; 1^{+} \rangle$, and d) the dynamical process depicted in Fig.~\ref{fig:production}(b).

\item The decay of $B^-$ into $|D^* \bar K^*; 0^{+} \rangle D^-$ is contributed by $\xi_4 \times [\bar c_a \gamma_5 d_a]$:
\begin{eqnarray}
&& \langle B^-(q) ~|~D^* \bar K^*; 0^{+}(q_1)~D^-(q_2) \rangle
\\ \nonumber &=& - {d_2 \over 12}  \lambda_D f_{|D^* \bar K^*; 0^{+} \rangle}  \, .
\end{eqnarray}

\item The decay of $B^-$ into $|D^* \bar K^*; 1^{+} \rangle D^-$ is contributed by $\xi_5^\alpha \times [\bar c_a \gamma_\alpha \gamma_5 d_a]$:
\begin{eqnarray}
&& \langle B^-(q) ~|~D^* \bar K^*; 1^{+}(\epsilon_1, q_1)~D^-(q_2) \rangle
\\ \nonumber &=& - {d_2 \over 24}  f_D f_{|D^* \bar K^*; 1^{+} \rangle} ~ \epsilon_1 \cdot q_2  \, .
\end{eqnarray}

\end{enumerate}
In the above expressions $f_D$ and $\lambda_D$ are decay constants of the $D^-$ meson, whose definitions and values can be found in Table~\ref{tab:coupling}.

In the present study we adopt the possible interpretation of the $X_0(2900)$ as the $|D^* \bar K^*; 0^{+} \rangle$ molecular state. Accordingly, we assume masses of $D^{(*)} \bar K^{(*)}$ molecular states to be either the realistic mass or at the lowest $D^{(*)} \bar K^{(*)}$ thresholds~\cite{pdg}:
\begin{eqnarray}
\nonumber    M_{|D \bar K;     0^{+} \rangle} &\approx& M_{D^0} + M_{K^+}       = 2359   ~{\rm MeV} \, ,
\\[1mm] \nonumber M_{|D \bar K^*;   1^{+} \rangle} &\approx& M_{D^0} + M_{K^{*+}}     = 2756   ~{\rm MeV} \, ,
\\[1mm] M_{|D^* \bar K;   1^{+} \rangle} &\approx& M_{D^{*0}} + M_{K^+}   = 2501   ~{\rm MeV} \, ,
\label{eq:DKmass}
\\[1mm] \nonumber M_{|D^* \bar K^*; 0^{+} \rangle} &=&       M_{X_0(2900)}          = 2866   ~{\rm MeV} \, ,
\\[1mm] \nonumber M_{|D^* \bar K^*; 1^{+} \rangle} &\approx& M_{D^{*0}} + M_{K^{*+}} = 2899   ~{\rm MeV} \, ,
\\[1mm] \nonumber M_{|D^* \bar K^*; 2^{+} \rangle} &\approx& M_{D^{*0}} + M_{K^{*+}} = 2899   ~{\rm MeV} \, .
\end{eqnarray}
Then we can evaluate the above production amplitudes to obtain the following partial decay widths:
\begin{eqnarray}
\nonumber         \Gamma(B^- \to D^- |D \bar K;     0^{+} \rangle) &=& 0 \, ,
\\[1mm] \nonumber \Gamma(B^- \to D^- |D \bar K^*;   1^{+} \rangle) &=& 0 \, ,
\\[1mm] \nonumber \Gamma(B^- \to D^- |D^* \bar K;   1^{+} \rangle) &=& d_2^2~2.13 \times 10^{-10}~{\rm GeV}^{13} \, ,
\\[1mm] \nonumber \Gamma(B^- \to D^- |D^* \bar K^*; 0^{+} \rangle) &=& d_2^2~3.18 \times 10^{-10}~{\rm GeV}^{13} \, ,
\\[1mm] \nonumber \Gamma(B^- \to D^- |D^* \bar K^*; 1^{+} \rangle) &=& d_2^2~2.23 \times 10^{-10}~{\rm GeV}^{13} \, ,
\\[1mm]           \Gamma(B^- \to D^- |D^* \bar K^*; 2^{+} \rangle) &=& 0 \, .
\label{result:productionDK}
\end{eqnarray}

\subsection{Productions of $D^{(*)} \bar D_s^{(*)}$ molecules}
\label{sec:productionDDs}

Similarly, we use the Fierz rearrangement given in Eq.~(\ref{eq:stepDDs}) to perform numerical analyses, and calculate relative production rates of $D^{(*)} \bar D_s^{(*)}$ molecular states in $B^- \to \phi X$ decays:
\begin{enumerate}
\setcounter{enumi}{7}

\item The decay of $B^-$ into $|D \bar D_s^{*}; 1^{++} \rangle \phi$ is contributed by $\zeta_2^\alpha \times [\bar s_a \gamma_\alpha s_a]$:
\begin{eqnarray}
&& \langle B^-(q) ~|~D \bar D_s^{*}; 1^{++}(\epsilon_1, q_1)~\phi(\epsilon_2, q_2) \rangle
\\ \nonumber &=& {d_3 \over 24}  m_\phi f_\phi f_{|D \bar D_s^{*}; 1^{++} \rangle} ~ \epsilon_1 \cdot \epsilon_2  \, .
\end{eqnarray}
The overall factor $d_3$ is related to: a) the coupling of $J_{B^-}$ to $B^-$, b) the weak and $^3P_0$ decay processes described by Eqs.~(\ref{eq:step13}) and (\ref{eq:step14}), and c) the dynamical process depicted in Fig.~\ref{fig:production}(c). In this case there does not exist the isospin factor.

\item The decay of $B^-$ into $|D \bar D_s^{*}; 1^{+-} \rangle \phi$ is contributed by $\zeta_3^\alpha \times [\bar s_a \gamma_\alpha s_a]$:
\begin{eqnarray}
&& \langle B^-(q) ~|~D \bar D_s^{*}; 1^{+-}(\epsilon_1, q_1)~\phi(\epsilon_2, q_2) \rangle
\\ \nonumber &=& -{d_3 \over 24}  m_\phi f_\phi f_{|D \bar D_s^{*}; 1^{+-} \rangle} ~ \epsilon_1 \cdot \epsilon_2  \, .
\end{eqnarray}

\item The decay of $B^-$ into $|D^* \bar D_s^{*}; 1^{+} \rangle \phi$ is contributed by $\zeta_5^\alpha \times [\bar s_a \gamma_\alpha s_a]$:
\begin{eqnarray}
&& \langle B^-(q) ~|~D^* \bar D_s^{*}; 1^{+}(\epsilon_1, q_1)~\phi(\epsilon_2, q_2) \rangle
\\ \nonumber &=& -{i~d_3 \over 24}  m_\phi f_\phi f_{|D^* \bar D_s^{*}; 1^{+} \rangle} ~ \epsilon_1 \cdot \epsilon_2  \, .
\end{eqnarray}

\end{enumerate}
In the above expressions $f_\phi$ is decay constant of the $\phi$ meson. Besides, we also need $f_\phi^T$, and their definitions are:
\begin{eqnarray}
\langle 0 | \bar s_a \gamma_\mu s_a | \phi(\epsilon, q) \rangle &=& m_\phi f_\phi \epsilon_\mu \, ,
\\ \nonumber \langle 0 | \bar s_a \sigma_{\mu\nu} s_a | \phi(\epsilon, q) \rangle &=& i f^T_{\phi} (q_\mu\epsilon_\nu - q_\nu\epsilon_\mu) \, ,
\end{eqnarray}
where $f_{\phi}= 233$~MeV~\cite{pdg} and $f_{\phi}^T= 177$~MeV~\cite{Becirevic:2004qv}.

In the present study we adopt the possible interpretations of the $Z_{cs}(3985)$, $Z_{cs}(4000)$, and $Z_{cs}(4220)$ as the $|D \bar D_s^*; 1^{++} \rangle$, $|D \bar D_s^*; 1^{+-} \rangle$, and $|D^* \bar D_s^*; 1^{+} \rangle$ molecular states, respectively. Accordingly, we assume masses of $D^{(*)} \bar D_s^{(*)}$ molecular states to be either the realistic masses or at the lowest $D^{(*)} \bar D_s^{(*)}$ thresholds~\cite{pdg}:
\begin{eqnarray}
\nonumber    M_{|D \bar D_s; 0^{+} \rangle}          &\approx& M_{D^0} + M_{D_s^-}       = 3833   ~{\rm MeV} \, ,
\\[1mm] \nonumber M_{|D \bar D_s^*; 1^{++} \rangle}  &\approx& M_{Z_{cs}(3985)}          = 3982.5 ~{\rm MeV} \, ,
\\[1mm] M_{|D \bar D_s^*; 1^{+-} \rangle}  &\approx& M_{Z_{cs}(4000)}          = 4003   ~{\rm MeV} \, ,
\label{eq:DDsmass}
\\[1mm] \nonumber M_{|D^* \bar D_s^*; 0^{+} \rangle} &=&       M_{D^{*0}} + M_{D_s^{*-}} = 4119   ~{\rm MeV} \, ,
\\[1mm] \nonumber M_{|D^* \bar D_s^*; 1^{+} \rangle} &\approx& M_{Z_{cs}(4220)}          = 4216   ~{\rm MeV} \, ,
\\[1mm] \nonumber M_{|D^* \bar D_s^*; 2^{+} \rangle} &\approx& M_{D^{*0}} + M_{D_s^{*-}} = 4119   ~{\rm MeV} \, .
\end{eqnarray}
Then we can evaluate the above production amplitudes to obtain the following partial decay widths:
\begin{eqnarray}
\nonumber         \Gamma(B^- \to \phi |D \bar D_s; 0^{+} \rangle    ) &=& 0 \, ,
\\[1mm] \nonumber \Gamma(B^- \to \phi |D \bar D_s^*; 1^{++} \rangle ) &=& d_3^2~2.53 \times 10^{-10}~{\rm GeV}^{13} \, ,
\\[1mm] \nonumber \Gamma(B^- \to \phi |D \bar D_s^*; 1^{+-} \rangle ) &=& d_3^2~2.38 \times 10^{-10}~{\rm GeV}^{13} \, ,
\\[1mm] \nonumber \Gamma(B^- \to \phi |D^* \bar D_s^*; 0^{+} \rangle) &=& 0 \, ,
\\[1mm] \nonumber \Gamma(B^- \to \phi |D^* \bar D_s^*; 1^{+} \rangle) &=& d_3^2~5.77 \times 10^{-10}~{\rm GeV}^{13} \, ,
\\[1mm]           \Gamma(B^- \to \phi |D^* \bar D_s^*; 2^{+} \rangle) &=& 0 \, .
\label{result:productionDDs}
\end{eqnarray}

From Eqs.~(\ref{result:productionDD}), (\ref{result:productionDK}), and (\ref{result:productionDDs}), we can derive the relative production rate $\mathcal{R}_1 \equiv {\mathcal{B}\left(B^- \rightarrow K^- X / D^- X / \phi X \right) \over \mathcal{B}\left(B^- \rightarrow D^-|D^* \bar K^*; 0^{+} \rangle \right)/d_2}$. We summarize the obtained results in Table~\ref{tab:width}, which will be further discussed in Sec.~\ref{sec:summary}. The difference among the three overall factors $d_{1,2,3}$ is partly caused by the different dynamical processes described by Eqs.~(\ref{eq:step6}), (\ref{eq:step12}), and (\ref{eq:step18}), {\it i.e.}, the attraction between $D^{(*)}$ and $\bar D^{(*)}$, the attraction between $D^{(*)}$ and $\bar K^{(*)}$, and the attraction between $D^{(*)}$ and $\bar D_s^{(*)}$ can be different. However, we are not able to estimate this in the present study.

\subsection{Productions in $B^{*}$ decays}
\label{sec:Bsproducion}

In this subsection we follow the same procedures to study productions of $D^{(*)} \bar D^{(*)}$, $D^{(*)} \bar K^{(*)}$, and $D^{(*)} \bar D_s^{(*)}$ molecular states in $B^{*}$ decays.

\begin{widetext}
Similar to Eq.~(\ref{eq:stepDD}), we apply the Fierz transformation to obtain:
\begin{eqnarray}
B^{*-} &\xrightarrow{~~~~~~~~~~~~~}& J^\alpha_{B^{*-}} = [\delta^{ab} \bar u_a \gamma^\alpha b_b]
\label{eq:BsDD}
\\[1mm] \nonumber &\xrightarrow{~~~~\rm weak~~~~}&       [\delta^{ab} \bar u_a \gamma^\alpha \gamma_\rho(1 - \gamma_5) c_b] ~\times~ [\delta^{cd}\bar c_c \gamma^\rho(1 - \gamma_5) s_d]
\\[1mm] \nonumber &\xrightarrow{~~~~\rm QPC~~~~}&        [\delta^{ab} \bar u_a \gamma^\alpha \gamma_\rho(1 - \gamma_5) c_b] ~\times~ [\delta^{cd}\bar c_c \gamma^\rho(1 - \gamma_5) s_d] ~\times~ [\delta^{ef}\bar u_e u_f]
\\[1mm] \nonumber &\xlongequal{~~~~\rm color~~~~}&       {\delta^{ab} \delta^{cf} \delta^{ed} \over 3} \times \bar u_a \gamma^\alpha \gamma_\rho(1 - \gamma_5) c_b \times \bar c_c \gamma^\rho(1 - \gamma_5) s_d \times \bar u_e u_f ~+~ \cdots
\\[1mm] \nonumber &\xlongequal{{\rm Fierz:}s_d \leftrightarrow u_f}&
+{ 1 \over 12} \times [\delta^{ab}\bar u_a \gamma_5 c_b] \times [\delta^{cf}\bar c_c \gamma_5 u_f] \times [\delta^{ed}\bar u_e \gamma^\alpha(1 - \gamma_5) s_d]
\\ \nonumber &&
-{ 1 \over 12} \times [\delta^{ab}\bar u_a \gamma_5 c_b] \times [\delta^{cf}\bar c_c \gamma^\alpha(1 - \gamma_5) u_f] \times [\delta^{ed}\bar u_e \gamma_5 s_d]
\\ \nonumber &&
+{ 1 \over 12} \times [\delta^{ab}\bar u_a i \sigma^{\alpha\rho} \gamma_5 c_b] \times [\delta^{cf}\bar c_c \gamma^\rho(1 - \gamma_5) u_f] \times [\delta^{ed}\bar u_e \gamma_5 s_d]
~+~ \cdots
\\[1mm] \nonumber &\xlongequal{~~~~~~~~~~~~~}&
-{ 1 \over 12} \times \eta_1 \times [\bar u_a \gamma^\alpha \gamma_5 s_a]
\\ \nonumber &&
-{ 1 \over 24} \times \Big( \eta_2^\alpha + \eta_3^\alpha \Big) \times [\bar u_a \gamma_5 s_a]
\\ \nonumber &&
-{ i \over 24} \times \eta_5^\alpha \times [\bar u_a \gamma_5 s_a]
~+~ \cdots \, .
\end{eqnarray}
Similar to Eq.~(\ref{eq:stepDK}), we apply the Fierz transformation to obtain:
\begin{eqnarray}
B^{*-} &\xrightarrow{~~~~~~~~~~~~~}& J^\alpha_{B^{*-}} = [\delta^{ab} \bar u_a \gamma^\alpha b_b]
\label{eq:BsDK}
\\[1mm] \nonumber &\xrightarrow{~~~~\rm weak~~~~}&       [\delta^{ab} \bar u_a \gamma^\alpha \gamma_\rho(1 - \gamma_5) c_b] ~\times~ [\delta^{cd}\bar c_c \gamma^\rho(1 - \gamma_5) s_d]
\\[1mm] \nonumber &\xrightarrow{~~~~\rm QPC~~~~}&        [\delta^{ab} \bar u_a \gamma^\alpha \gamma_\rho(1 - \gamma_5) c_b] ~\times~ [\delta^{cd}\bar c_c \gamma^\rho(1 - \gamma_5) s_d] ~\times~ [\delta^{ef}\bar d_e d_f]
\\[1mm] \nonumber &\xlongequal{~~~~\rm color~~~~}&       {\delta^{ab} \delta^{cf} \delta^{ed} \over 3} \times \bar u_a \gamma^\alpha \gamma_\rho(1 - \gamma_5) c_b \times \bar c_c \gamma^\rho(1 - \gamma_5) s_d \times \bar d_e d_f ~+~ \cdots
\\[1mm] \nonumber &\xlongequal{{\rm Fierz:}s_d \leftrightarrow u_f}&
+{ 1 \over 12} \times [\delta^{ab}\bar u_a \gamma_5 c_b] \times [\delta^{cf}\bar c_c \gamma_5 d_f] \times [\delta^{ed}\bar d_e \gamma^\alpha(1 - \gamma_5) s_d]
\\ \nonumber &&
-{ 1 \over 12} \times [\delta^{ab}\bar u_a \gamma_5 c_b] \times [\delta^{cf}\bar c_c \gamma^\alpha(1 - \gamma_5) d_f] \times [\delta^{ed}\bar d_e \gamma_5 s_d]
\\ \nonumber &&
-{ 1 \over 12} \times [\delta^{ab}\bar u_a i \sigma^{\alpha\rho} \gamma_5 c_b] \times [\delta^{cf}\bar c_c \gamma_5 d_f] \times [\delta^{ed}\bar d_e \gamma^\rho(1 - \gamma_5) s_d]
~+~ \cdots
\\[1mm] \nonumber &\xlongequal{~~~~~~~~~~~~~}&
+{ 1 \over 12} \times \xi_1 \times [\bar c_a \gamma^\alpha \gamma_5 d_a]
\\ \nonumber &&
+{ 1 \over 12} \times \xi_2^\alpha \times [\bar c_a \gamma_5 d_a]
\\ \nonumber &&
+{ i \over 24} \times \xi_5^\alpha \times [\bar c_a \gamma_5 d_a]
~+~ \cdots \, .
\end{eqnarray}
Similar to Eq.~(\ref{eq:stepDDs}), we apply the Fierz transformation to obtain:
\begin{eqnarray}
B^{*-} &\xrightarrow{~~~~~~~~~~~~~}& J^\alpha_{B^{*-}} = [\delta^{ab} \bar u_a \gamma^\alpha b_b]
\label{eq:BsDDs}
\\[1mm] \nonumber &\xrightarrow{~~~~\rm weak~~~~}&       [\delta^{ab} \bar u_a \gamma^\alpha \gamma_\rho(1 - \gamma_5) c_b] ~\times~ [\delta^{cd}\bar c_c \gamma^\rho(1 - \gamma_5) s_d]
\\[1mm] \nonumber &\xrightarrow{~~~~\rm QPC~~~~}&        [\delta^{ab} \bar u_a \gamma^\alpha \gamma_\rho(1 - \gamma_5) c_b] ~\times~ [\delta^{cd}\bar c_c \gamma^\rho(1 - \gamma_5) s_d] ~\times~ [\delta^{ef}\bar s_e s_f]
\\[1mm] \nonumber &\xlongequal{~~~~\rm color~~~~}&       {\delta^{ab} \delta^{cf} \delta^{ed} \over 3} \times \bar u_a \gamma^\alpha \gamma_\rho(1 - \gamma_5) c_b \times \bar c_c \gamma^\rho(1 - \gamma_5) s_d \times \bar s_e s_f ~+~ \cdots
\\[1mm] \nonumber &\xlongequal{{\rm Fierz:}s_d \leftrightarrow s_f}&
+{ 1 \over 12} \times [\delta^{ab}\bar u_a \gamma_5 c_b] \times [\delta^{cf}\bar c_c \gamma_5 s_f] \times [\delta^{ed}\bar s_e \gamma^\alpha(1 - \gamma_5) s_d]
\\ \nonumber &&
-{ i \over 12} \times [\delta^{ab}\bar u_a \gamma_5 c_b] \times [\delta^{cf}\bar c_c \gamma_\mu (1-\gamma_5) s_f] \times [\delta^{ed}\bar s_e \sigma^{\alpha\mu} \gamma_5 s_d]
~+~ \cdots
\\[1mm] \nonumber &\xlongequal{~~~~~~~~~~~~~}&
+{ 1 \over 12} \times \zeta_1 \times [\bar s_a \gamma^\alpha s_a]
\\ \nonumber &&
-{ i \over 24}\times\Big( \zeta_{2\mu} + \zeta_{3\mu} \Big)\times[\bar s_a \sigma^{\alpha\mu} \gamma_5 s_a]
~+~ \cdots \, .
\end{eqnarray}
\end{widetext}
From Eq.~(\ref{eq:BsDD}), we obtain the following partial decay widths
\begin{eqnarray}
\nonumber         \Gamma(B^{*-} \to K^- |D \bar D;     0^{++} \rangle) &=& d_4^2~0.16 \times 10^{-10}~{\rm GeV}^{13} \, ,
\\[1mm] \nonumber \Gamma(B^{*-} \to K^- |D \bar D^*;   1^{++} \rangle) &=& d_4^2~0.62 \times 10^{-10}~{\rm GeV}^{13} \, ,
\\[1mm] \nonumber \Gamma(B^{*-} \to K^- |D \bar D^*;   1^{+-} \rangle) &=& d_4^2~0.61 \times 10^{-10}~{\rm GeV}^{13} \, ,
\\[1mm] \nonumber \Gamma(B^{*-} \to K^- |D^* \bar D^*; 0^{++} \rangle) &=& 0 \, ,
\\[1mm] \nonumber \Gamma(B^{*-} \to K^- |D^* \bar D^*; 1^{+-} \rangle) &=& d_4^2~2.84 \times 10^{-10}~{\rm GeV}^{13} \, ,
\\[1mm]           \Gamma(B^{*-} \to K^- |D^* \bar D^*; 2^{++} \rangle) &=& 0 \, .
\label{result:productionBsDD}
\end{eqnarray}
From Eq.~(\ref{eq:BsDK}), we obtain the following partial decay widths
\begin{eqnarray}
\nonumber         \Gamma(B^{*-} \to D^- |D \bar K;     0^{+} \rangle) &=& d_5^2~0.18 \times 10^{-10}~{\rm GeV}^{13} \, ,
\\[1mm] \nonumber \Gamma(B^{*-} \to D^- |D \bar K^*;   1^{+} \rangle) &=& d_5^2~1.69 \times 10^{-10}~{\rm GeV}^{13} \, ,
\\[1mm] \nonumber \Gamma(B^{*-} \to D^- |D^* \bar K;   1^{+} \rangle) &=& 0 \, ,
\\[1mm] \nonumber \Gamma(B^{*-} \to D^- |D^* \bar K^*; 0^{+} \rangle) &=& 0 \, ,
\\[1mm] \nonumber \Gamma(B^{*-} \to D^- |D^* \bar K^*; 1^{+} \rangle) &=& d_5^2~4.40 \times 10^{-10}~{\rm GeV}^{13} \, ,
\\[1mm]           \Gamma(B^{*-} \to D^- |D^* \bar K^*; 2^{+} \rangle) &=& 0 \, .
\label{result:productionBsDK}
\end{eqnarray}
From Eq.~(\ref{eq:BsDDs}), we obtain the following partial decay widths
\begin{eqnarray}
\nonumber         \Gamma(B^{*-} \to \phi |D \bar D_s;     0^{+} \rangle )  &=& d_6^2~1.78 \times 10^{-10}~{\rm GeV}^{13} \, ,
\\[1mm] \nonumber \Gamma(B^{*-} \to \phi |D \bar D_s^*;   1^{++} \rangle) &=& d_6^2~1.09 \times 10^{-10}~{\rm GeV}^{13} \, ,
\\[1mm] \nonumber \Gamma(B^{*-} \to \phi |D \bar D_s^*;   1^{+-} \rangle) &=& d_6^2~1.02 \times 10^{-10}~{\rm GeV}^{13} \, ,
\\[1mm] \nonumber \Gamma(B^{*-} \to \phi |D^* \bar D_s^*; 0^{+} \rangle )  &=& 0 \, ,
\\[1mm] \nonumber \Gamma(B^{*-} \to \phi |D^* \bar D_s^*; 1^{+} \rangle )  &=& 0 \, ,
\\[1mm]           \Gamma(B^{*-} \to \phi |D^* \bar D_s^*; 2^{+} \rangle )  &=& 0 \, .
\label{result:productionBsDDs}
\end{eqnarray}
In the above expressions $d_{4,5,6}$ are three overall factors.

From Eqs.~(\ref{result:productionBsDD}--\ref{result:productionBsDDs}), we can derive the relative production rate $\mathcal{R}_2 \equiv {\mathcal{B}\left(B^{*-} \rightarrow K^- X / D^- X/ \phi X \right) \over \mathcal{B}\left(B^{*-} \rightarrow K^-|D \bar D; 0^{++} \rangle \right)/d_4}$. We summarize the obtained results in Table~\ref{tab:width}, which will be further discussed in Sec.~\ref{sec:summary}. Again, the difference among the three overall factors $d_{4,5,6}$ is partly due to that the attraction between $D^{(*)}$ and $\bar D^{(*)}$, the attraction between $D^{(*)}$ and $\bar K^{(*)}$, and the attraction between $D^{(*)}$ and $\bar D_s^{(*)}$ can be different.

\section{Decay properties through the Fierz rearrangement}
\label{sec:decay}

The Fierz rearrangement~\cite{fierz} of the Dirac and color indices has been applied in Refs.~\cite{Chen:2019wjd,Chen:2019eeq} to study strong decay properties of the $X(3872)$ and $Z_c(3900)$ as $D \bar D^{*}$ molecular states. In this section we follow the same procedures to study decay properties of other $D^{(*)} \bar D^{(*)}$, $D^{(*)} \bar K^{(*)}$, and $D^{(*)} \bar D_s^{(*)}$ hadronic molecular states, separately in Sec.~\ref{sec:DDdecay}, Sec.~\ref{sec:DKdecay}, and Sec.~\ref{sec:decaysummary}. We shall study their two-body and three-body hadronic decays as well as the four-body decay process $X \to J/\psi \omega \to J/\psi \pi \pi \pi$.

This method has been systematically applied to study light baryon and tetraquark currents in Refs.~\cite{Chen:2006hy,Chen:2006zh,Chen:2007xr,Chen:2008ej,Chen:2008ne,Chen:2012ut,Chen:2008qv,Chen:2009sf,Chen:2012ex}, and applied to study strong decay properties of the $P_c$ and $X(6900)$ states in Refs.~\cite{Chen:2020pac,Chen:2020xwe}. A similar arrangement of the spin and color indices in the nonrelativistic case can be found in Refs.~\cite{Voloshin:2013dpa,Maiani:2017kyi,Voloshin:2018pqn,Voloshin:2019aut,Wang:2019spc,Xiao:2019spy,Cheng:2020nho}, which was applied to study decay properties of the $P_c$ and $XYZ$ states.

Before doing this, we note that the Fierz rearrangement in the Lorentz space is a matrix identity, valid if each quark/antiquark field in the initial and final currents is at the same location. For example, we can apply the Fierz rearrangement to transform a non-local current $\eta = [\bar q(x) c(x)] ~ [\bar c(y) q(y)]$ into the combination of many non-local currents $\theta = [\bar q(x) q(y)] ~ [\bar c(y) c(x)]$, with all the quark fields remaining at the same locations. We shall keep this in mind, and omit the coordinates in this section.

\begin{table*}[hbt]
\begin{center}
\renewcommand{\arraystretch}{1.36}
\caption{Couplings of meson operators to meson states, where color indices are omitted for simplicity. All the isovector meson operators have the quark content $\bar q \Gamma q = \left(\bar u \Gamma u - \bar d \Gamma d\right)/\sqrt2$, and all the isoscalar light meson operators have the quark content $\bar q \Gamma q = \left(\bar u \Gamma u + \bar d \Gamma d\right)/\sqrt2$.}
\begin{tabular}{ c | c | c | c | c | c}
\hline\hline
~Operators--$J_{(\mu\nu)}$~ & ~~$I^GJ^{PC}$~~ & ~~~~~~~~Mesons~~~~~~~~ & ~~$I^GJ^{PC}$~~ & ~~~Couplings~~~ & ~~~~~~~~Decay Constants~~~~~~~~
\\ \hline\hline
$\bar c c$                & $0^+0^{++}$                  & $\chi_{c0}(1P)$      & $0^+0^{++}$ & $\langle 0 | J | \chi_{c0} \rangle = m_{\chi_{c0}} f_{\chi_{c0}}$          & $f_{\chi_{c0}} = 343$~MeV~\mbox{\cite{Veliev:2010gb}}
\\ \hline
$\bar c i\gamma_5 c$      & $0^+0^{-+}$                  & $\eta_c$             & $0^+0^{-+}$ & $\langle 0 | J | \eta_c \rangle = \lambda_{\eta_c}$                      & $\lambda_{\eta_c} = {f_{\eta_c} m_{\eta_c}^2 \over 2 m_c}$
\\ \hline
$\bar c \gamma_\mu c$ & $0^-1^{--}$                  & $J/\psi$             & $0^-1^{--}$ & $\langle0| J_\mu | J/\psi \rangle = m_{J/\psi} f_{J/\psi} \epsilon_\mu$  & $f_{J/\psi} = 418$~MeV~\mbox{\cite{Becirevic:2013bsa}}
\\ \hline
\multirow{2}{*}{$\bar c \gamma_\mu \gamma_5 c$}
                               & \multirow{2}{*}{$0^+1^{++}$}    & $\eta_c$             & $0^+0^{-+}$ & $\langle 0 | J_\mu | \eta_c \rangle = i p_\mu f_{\eta_c}$                & $f_{\eta_c} = 387$~MeV~\mbox{\cite{Becirevic:2013bsa}}
\\ \cline{3-6}
                               &                              & $\chi_{c1}(1P)$      & $0^+1^{++}$ & $\langle 0 | J_\mu | \chi_{c1} \rangle = m_{\chi_{c1}} f_{\chi_{c1}} \epsilon_\mu $
                               &  $f_{\chi_{c1}} = 335$~MeV~\mbox{\cite{Novikov:1977dq}}
\\ \hline
\multirow{2}{*}{$\bar c \sigma_{\mu\nu} c$}
                               & \multirow{2}{*}{$0^-1^{\pm-}$}  & $J/\psi$             & $0^-1^{--}$ & $\langle 0 | J_{\mu\nu} | J/\psi \rangle = i f^T_{J/\psi} (p_\mu\epsilon_\nu - p_\nu\epsilon_\mu) $
                               &  $f_{J/\psi}^T = 410$~MeV~\mbox{\cite{Becirevic:2013bsa}}
\\ \cline{3-6}
                               &                              & $h_c(1P)$            & $0^-1^{+-}$ & $\langle 0 | J_{\mu\nu} | h_c \rangle = i f^T_{h_c} \epsilon_{\mu\nu\alpha\beta} \epsilon^\alpha p^\beta $
                               &  $f_{h_c}^T = 235$~MeV~\mbox{\cite{Becirevic:2013bsa}}
\\ \hline\hline
$\bar c q$                & $0^{+}$                   & $\bar D_0^{*}$           & $0^{+}$  & $\langle 0 | J | \bar D_0^{*} \rangle = m_{D_0^{*}} f_{D_0^{*}}$             &  $f_{D_0^{*}} = 410$~MeV~\mbox{\cite{Narison:2015nxh}}
\\ \hline
$\bar c i\gamma_5 q$      & $0^{-}$                   & $\bar D$                & $0^{-}$  & $\langle 0 | J | \bar D \rangle = \lambda_D$                                &  $\lambda_D = {f_D m_D^2 \over {m_c + m_d}}$
\\ \hline
$\bar c \gamma_\mu q$ & $1^{-}$                   & $\bar D^{*}$        & $1^{-}$  & $\langle0| J_\mu | \bar D^{*} \rangle = m_{D^*} f_{D^*} \epsilon_\mu$   &  $f_{D^*} = 253$~MeV~\mbox{\cite{Chang:2018aut}}
\\ \hline
\multirow{2}{*}{$\bar c \gamma_\mu \gamma_5 q$}
                               & \multirow{2}{*}{$1^{+}$}      & $\bar D$         & $0^{-}$  & $\langle 0 | J_\mu | \bar D \rangle = i p_\mu f_{D}$                 &  $f_{D} = 211.9$~MeV~\mbox{\cite{pdg}}
\\ \cline{3-6}
                                  &                            & $\bar D_1$                & $1^{+}$  & $\langle 0 | J_\mu | \bar D_1 \rangle = m_{D_1} f_{D_1} \epsilon_\mu $        &  $f_{D_1} = 356$~MeV~\mbox{\cite{Narison:2015nxh}}
\\ \hline
\multirow{2}{*}{$\bar c \sigma_{\mu\nu} q$}
                               & \multirow{2}{*}{$1^{\pm}$}    & $\bar D^{*}$        & $1^{-}$  &  $\langle 0 | J_{\mu\nu} | \bar D^{*} \rangle = i f_{D^*}^T (p_\mu\epsilon_\nu - p_\nu\epsilon_\mu) $
                               &  $f_{D^*}^T \approx 220$~MeV
\\ \cline{3-6}
                               &                               &  --                  & $1^{+}$  &  --  &  --
\\ \hline\hline
$\bar c s$                & $0^{+}$                   & --           & $0^{+}$  & --        &  --
\\ \hline
$\bar c i\gamma_5 s$      & $0^{-}$                   & $\bar D_s$                & $0^{-}$  & $\langle 0 | J | \bar D_s \rangle = \lambda_{D_s}$                                &  $\lambda_{D_s} = {f_{D_s} m_{D_s}^2 \over {m_c + m_s}}$
\\ \hline
$\bar c \gamma_\mu s$ & $1^{-}$                   & $\bar D_s^{*}$        & $1^{-}$  & $\langle0| J_\mu | \bar D^{*}_s \rangle = m_{D_s^*} f_{D_s^*} \epsilon_\mu$   &  $f_{D_s^*} = 301$~MeV~\mbox{\cite{pdg}}
\\ \hline
\multirow{2}{*}{$\bar c \gamma_\mu \gamma_5 s$}
                               & \multirow{2}{*}{$1^{+}$}      & $\bar D_s$         & $0^{-}$  & $\langle 0 | J_\mu | \bar D \rangle = i p_\mu f_{D_s}$                 &  $f_{D_s} = 257$~MeV~\mbox{\cite{pdg}}
\\ \cline{3-6}
                                  &                            & --                & $1^{+}$      &  --
\\ \hline
\multirow{2}{*}{$\bar c \sigma_{\mu\nu} s$}
                               & \multirow{2}{*}{$1^{\pm}$}    & $\bar D^{*}_s$        & $1^{-}$  &  $\langle 0 | J_{\mu\nu} | \bar D_s^{*} \rangle = i f_{D_s^*}^T (p_\mu\epsilon_\nu - p_\nu\epsilon_\mu) $
                               &  $f_{D_s^*}^T \approx f_{D_s^*}\times{f_{D^*}^T \over f_{D^*}}$
\\ \cline{3-6}
                               &                               &  --                  & $1^{+}$  &  --  &  --
\\ \hline\hline
$\bar q q$                & $0^+0^{++}$                  & $f_0(500)$ & $0^+0^{++}$ & $\langle 0 | J | f_0 \rangle = m_{f_0} f_{f_0}$          & $f_{f_0} \sim 380$~MeV~\mbox{\cite{Chen:2019eeq}}
\\ \hline
$\bar q i\gamma_5 q$      & $0^+0^{-+}$                  & $\eta$               & $0^+0^{-+}$ & -- & -- 
\\ \hline
$\bar q \gamma_\mu q$ & $0^-1^{--}$                  & $\omega$             & $0^-1^{--}$ & $\langle 0 | J_\mu | \omega \rangle = m_\omega f_\omega \epsilon_\mu$
                               & $f_\omega \approx f_{\rho} = 216$~MeV~\mbox{\cite{Jansen:2009hr}}
\\ \hline
\multirow{2}{*}{$\bar q \gamma_\mu \gamma_5 q$}
                               & \multirow{2}{*}{$0^+1^{++}$} & $\eta$               & $0^+0^{-+}$ & $\langle 0 | J_\mu | \eta \rangle = i p_\mu f_{\eta}$                    & $f_{\eta} = 97$~MeV~\mbox{\cite{Ottnad:2017bjt,Guo:2015xva}}
\\ \cline{3-6}
                               &                              & $f_1(1285)$          & $0^+1^{++}$ & -- & -- 
\\ \hline
\multirow{2}{*}{$\bar q \sigma_{\mu\nu} q$}
                               & \multirow{2}{*}{$0^-1^{\pm-}$} & $\omega$             & $0^-1^{--}$ & $\langle 0 | J_{\mu\nu} | \omega \rangle = i f^T_{\omega} (p_\mu\epsilon_\nu - p_\nu\epsilon_\mu) $
                               &  $f^T_{\omega} \approx f_{\rho}^T = 159$~MeV~\mbox{\cite{Jansen:2009hr}}
\\ \cline{3-6}
                               &                              & $h_1(1170)$          & $0^-1^{+-}$ & $\langle 0 | J_{\mu\nu} | h_1 \rangle = i f^T_{h_1} \epsilon_{\mu\nu\alpha\beta} \epsilon^\alpha p^\beta $
                               &  $f_{h_1}^T \approx f_{b_1}^T = 180$~MeV~\mbox{\cite{Ball:1996tb}}
\\ \hline\hline
$\bar q q$ & $1^-0^{++}$ & -- & $1^-0^{++}$ & -- & --
\\ \hline
$\bar q i\gamma_5 q$ & $1^-0^{-+}$ &  $\pi^0$  & $1^-0^{-+}$ &  $\langle 0 | J | \pi^0 \rangle = \lambda_\pi$  &  $\lambda_\pi = {f_\pi m_\pi^2 \over m_u + m_d}$
\\ \hline
$\bar q \gamma_\mu q$ & $1^+1^{--}$ &  $\rho^0$  & $1^+1^{--}$ &  $\langle 0 | J_\mu | \rho^0 \rangle = m_\rho f_{\rho} \epsilon_\mu$  &  $f_{\rho} = 216$~MeV~\mbox{\cite{Jansen:2009hr}}
\\ \hline
\multirow{2}{*}{$\bar q \gamma_\mu \gamma_5 q$} & \multirow{2}{*}{$1^-1^{++}$} & $\pi^0$  & $1^-0^{-+}$ & $\langle 0 | J_\mu | \pi^0 \rangle = i p_\mu f_{\pi}$  &  $f_{\pi} = 130.2$~MeV~\mbox{\cite{pdg}}
\\ \cline{3-6}
                                                         & &  $a_1(1260)$  & $1^-1^{++}$ & $\langle 0 | J_\mu | a_1 \rangle = m_{a_1} f_{a_1} \epsilon_\mu $  &  $f_{a_1} = 254$~MeV~\mbox{\cite{Wingate:1995hy}}
\\ \hline
\multirow{2}{*}{$\bar q \sigma_{\mu\nu} q$} & \multirow{2}{*}{$1^+1^{\pm-}$} &  $\rho^0$ & $1^+1^{--}$   &  $\langle 0 | J_{\mu\nu} | \rho^0 \rangle = i f^T_{\rho} (p_\mu\epsilon_\nu - p_\nu\epsilon_\mu) $  &  $f_{\rho}^T = 159$~MeV~\mbox{\cite{Jansen:2009hr}}
\\ \cline{3-6}
                                                             & &  $b_1(1235)$  & $1^+1^{+-}$ &  $\langle 0 | J_{\mu\nu} | b_1 \rangle = i f^T_{b_1} \epsilon_{\mu\nu\alpha\beta} \epsilon^\alpha p^\beta $  &  $f_{b_1}^T = 180$~MeV~\mbox{\cite{Ball:1996tb}}
\\ \hline\hline
\end{tabular}
\label{tab:coupling}
\end{center}
\end{table*}

In the calculations we need couplings of charmonium operators to charmonium states, which are listed in Table~\ref{tab:coupling}. We also need couplings of charmed/light meson operators to charmed/light meson states, which are also listed in Table~\ref{tab:coupling}. We refer to Refs.~\cite{Chen:2019wjd,Chen:2019eeq} for detailed discussions of these couplings.

\subsection{Decay properties of $D^{(*)} \bar D^{(*)}$ molecules}
\label{sec:DDdecay}

In this subsection we perform the Fierz rearrangement for the currents $\eta_{1,4,5,6}$, and use the obtained results to study strong decay properties of $D^{(*)} \bar D^{(*)}$ molecular states. The results obtained in Refs.~\cite{Chen:2019wjd,Chen:2019eeq} using $\eta_{2,3}^\alpha$ are also summarized here for completeness.

We adopt the possible interpretations of the $X(3872)$, $Z_c(3900)$, and $Z_c(4020)$ as the $|D \bar D^*; 1^{++} \rangle$, $|D \bar D^*; 1^{+-} \rangle$, and $|D^* \bar D^*; 1^{+-} \rangle$ molecular states, respectively. Masses of $D^{(*)} \bar D^{(*)}$ molecular states are accordingly chosen, as given in Eqs.~(\ref{eq:DDmass}). Besides, we assume that $|D \bar D^*; 1^{+-} \rangle$ and $|D^* \bar D^*; 1^{+-} \rangle$ have $I=1$; $|D \bar D; 0^{++} \rangle$, $|D^* \bar D^*; 0^{++} \rangle$, and $|D^* \bar D^*; 2^{++} \rangle$ have $I=0$; the isospin of the $X(3872)$ as $|D \bar D^*; 1^{++} \rangle$ will be separately investigated in Sec.~\ref{sec:DD1pp}.

%
\begin{figure*}[hbt]
\begin{center}
\subfigure[~\mbox{Fierz rearrangement: $\eta \rightarrow \theta$}]{\includegraphics[width=0.32\textwidth]{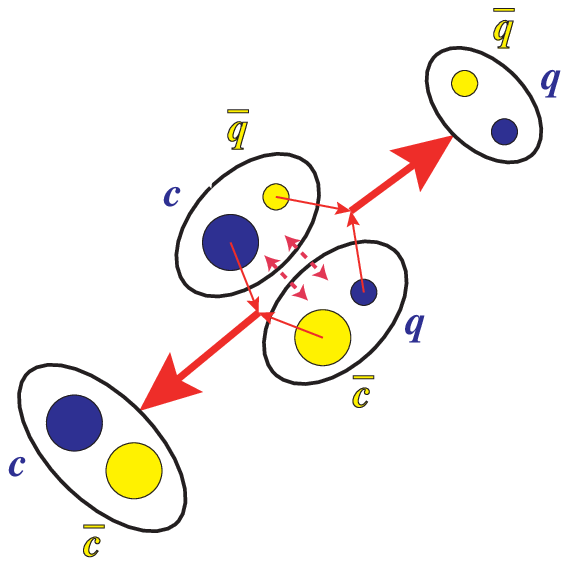}}
~~~~~~~~~~~~~~~~~~~~
\subfigure[~\mbox{direct fall-apart process}]{\includegraphics[width=0.32\textwidth]{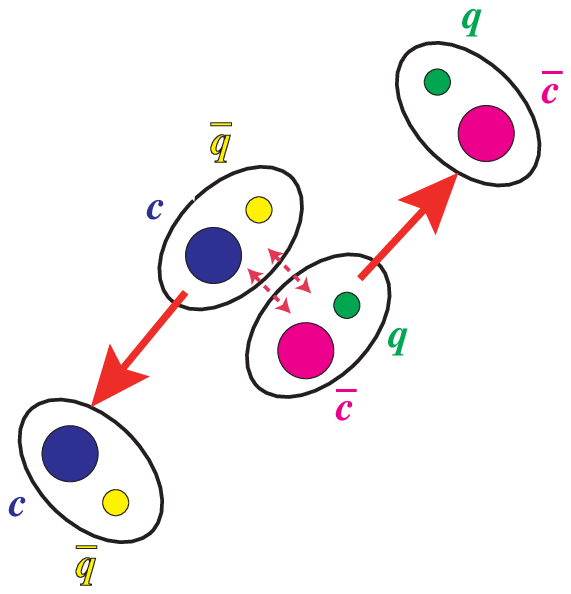}}
\caption{Fall-apart decay processes of $D^{(*)} \bar D^{(*)}$ molecular states, investigated through the $\eta(x)$ currents.}
\label{fig:decay}
\end{center}
\end{figure*}
%

\subsubsection{$| D \bar D; 0^{++} \rangle$}
\label{sec:DD0pp}

The current $\eta_1$ corresponds to the $| D \bar D; 0^{++} \rangle$ molecular state. We can apply the Fierz rearrangement to interchange the $c_a$ and $q_b$ quarks, and transform it into:
\begin{eqnarray}
\eta_1              &=& \bar q_a \gamma_5 c_a ~ \bar c_b \gamma_5 q_b
\label{eq:eta1}
\\ \nonumber   &\rightarrow& - {1\over12} ~ \bar q_a q_a ~ \bar c_b c_b                       - {1\over12} ~ \bar q_a \gamma_5 q_a ~ \bar c_b \gamma_5 c_b
\\ \nonumber         &&      + {1\over12} ~ \bar q_a \gamma_\mu q_a ~ \bar c_b \gamma^\mu c_b - {1\over12} ~ \bar q_a \gamma_\mu \gamma_5 q_a ~ \bar c_b \gamma^\mu \gamma_5 c_b
\\ \nonumber         &&      - {1\over24} ~ \bar q_a \sigma_{\mu\nu} q_a ~ \bar c_b \sigma^{\mu\nu} c_b + \cdots \, ,
\end{eqnarray}
where we have only kept the leading-order fall-apart decays described by color-singlet-color-singlet meson-meson currents, but neglect the $\mathcal{O}(\alpha_s)$ corrections described by color-octet-color-octet meson-meson currents.

As an example, we investigate $| D \bar D; 0^{++} \rangle$ through $\eta_1(x)$ and the above Fierz rearrangement. As depicted in Fig.~\ref{fig:decay}(a), when the $\bar q_a$ and $q_d$ quarks meet each other and the $c_b$ and $\bar c_c$ quarks meet each other at the same time, $| D \bar D; 0^{++} \rangle$ can decay into one charmonium meson and one light meson:
\begin{eqnarray}
&& \left[\delta^{ab} \bar q_a c_b\right] ~ \left[\delta^{cd} \bar c_c q_d\right]
\\[1mm] \nonumber &\xlongequal{~\rm color~}& {1\over3} \delta^{ad} \delta^{cb} ~ \bar q_a c_b ~ \bar c_c q_d + \cdots
\\[1mm] \nonumber &\xrightarrow{~\rm Fierz~}& {1\over3} ~ \left[\delta^{ad}\bar q_a q_d\right] ~ \left[\delta^{cb} c_b c_b\right] + \cdots .
\end{eqnarray}
This decay process can be described by the Fierz rearrangement given in Eq.~(\ref{eq:eta1}).

Assuming the isospin of $| D \bar D; 0^{++} \rangle$ to be $I=0$, we extract the following three decay channels:
\begin{enumerate}

\item The decay of $| D \bar D; 0^{++} \rangle$ into $\eta_c \eta$ is contributed by both $\bar q_a \gamma_5 q_a \times \bar c_b \gamma_5 c_b$ and $\bar q_a \gamma_\mu \gamma_5 q_a \times \bar c_b \gamma^\mu \gamma_5 c_b$:
\begin{eqnarray}
&& \langle D \bar D; 0^{++}(p) | \eta_c(p_1)~\eta(p_2) \rangle
\\ \nonumber &=&  {a_1 \over 12}~\lambda_{\eta_c} \lambda_{\eta}
                      + {a_1 \over 12}~f_{\eta_c} f_{\eta}~p_1 \cdot p_2 \, ,
\end{eqnarray}
where $a_1$ is an overall factor, related to the coupling of $\eta_1$ to $| D \bar D; 0^{++} \rangle$ as well as the dynamical process depicted in Fig.~\ref{fig:decay}(a).

\item The decay of $| D \bar D; 0^{++} \rangle$ into $J/\psi \omega$ is contributed by both $\bar q_a \gamma_\mu q_a \times \bar c_b \gamma^\mu c_b$ and $\bar q_a \sigma_{\mu\nu} q_a \times \bar c_b \sigma^{\mu\nu} c_b$:
\begin{eqnarray}
&& \langle D \bar D; 0^{++}(p) | J/\psi(\epsilon_1, p_1)~\omega(\epsilon_2, p_2) \rangle
\\ \nonumber &=&  {a_1 \over 12}~m_{J/\psi} f_{J/\psi} m_{\omega} f_{\omega} ~ \epsilon_1 \cdot \epsilon_2
\\ \nonumber &+& {a_1 \over 12}~f_{J/\psi}^T f_{\omega}^T \left( \epsilon_1 \cdot \epsilon_2 ~ p_1 \cdot p_2 - \epsilon_1 \cdot p_2 ~ \epsilon_2 \cdot p_1 \right) \, .
\end{eqnarray}

\item The decay of $| D \bar D; 0^{++} \rangle$ into $\chi_{c0} f_0(500)$ is contributed by $\bar q_a q_a \times \bar c_b c_b$:
\begin{eqnarray}
&& \langle D \bar D; 0^{++}(p) | \chi_{c0}(p_1)~f_0(p_2) \rangle
\\ \nonumber &=&  -{a_1 \over 12}~m_{\chi_{c0}} f_{\chi_{c0}} m_{f_{0}} f_{f_{0}} \, .
\end{eqnarray}

\end{enumerate}

Assuming the mass of $| D \bar D; 0^{++} \rangle$ to be about $M_{D^0} + M_{\bar D^0} = 3730$~MeV, we summarize the above amplitudes to obtain the following partial decay widths:
\begin{eqnarray}
\nonumber && \Gamma(| D \bar D; 0^{++} \rangle \to \eta_c \eta ) = a_1^2 ~ 1.7 \times 10^{-4} ~{\rm GeV}^7 \, ,
\\[1mm] && \Gamma(| D \bar D; 0^{++} \rangle \to J/\psi \omega \to J/\psi \pi\pi\pi )
\label{result:eta1}
\\ \nonumber && ~~~~~~~~~~~~~~~~~~~~~~~~~~~ = a_1^2 ~ 1.1 \times 10^{-10} ~{\rm GeV}^7 \, ,
\\[1mm] \nonumber && \Gamma(| D \bar D; 0^{++} \rangle \to \chi_{c0} f_0(500) \to \chi_{c0} \pi\pi )
\\ \nonumber && ~~~~~~~~~~~~~~~~~~~~~~~~~~~ = a_1^2 ~ 9.7 \times 10^{-10} ~{\rm GeV}^7 \, .
\end{eqnarray}

There are two different terms, $A \equiv \bar q_a \gamma_5 q_a \times \bar c_b \gamma_5 c_b$ and $B \equiv \bar q_a \gamma_\mu \gamma_5 q_a \times \bar c_b \gamma^\mu \gamma_5 c_b$, both of which can contribute to the decay of $| D \bar D; 0^{++} \rangle$ into the $\eta_c \eta$ final states. There are also two different terms, $C \equiv \bar q_a \gamma_\mu q_a \times \bar c_b \gamma^\mu c_b$ and $D \equiv \bar q_a \sigma_{\mu\nu} q_a \times \bar c_b \sigma^{\mu\nu} c_b$, both of which can contribute to the decay of $| D \bar D; 0^{++} \rangle$ into the $J/\psi \omega$ final states. Phase angles among them, such as the phase angle between the two coupling constants $f_{J/\psi}$ and $f_{J/\psi}^T$, can not be well determined in the present study. This causes some uncertainties, which will be investigated in Appendix~\ref{app:phase}.

\subsubsection{$| D^* \bar D^*; 0^{++} \rangle$}
\label{sec:DsDs0pp}

The current $\eta_4$ corresponds to the $| D^* \bar D^*; 0^{++} \rangle$ molecular state. We can apply the Fierz rearrangement and transform it into:
\begin{eqnarray}
\eta_4              &=& \bar q_a \gamma^\mu c_a ~ \bar c_b \gamma_\mu q_b
\label{eq:eta4}
\\ \nonumber   &\rightarrow& - {1\over3} ~ \bar q_a q_a ~ \bar c_b c_b                        + {1\over3} ~ \bar q_a \gamma_5 q_a ~ \bar c_b \gamma_5 c_b
\\ \nonumber         &&      + {1\over6} ~ \bar q_a \gamma_\mu q_a ~ \bar c_b \gamma^\mu c_b  + {1\over6} ~ \bar q_a \gamma_\mu \gamma_5 q_a ~ \bar c_b \gamma^\mu \gamma_5 c_b + \cdots \ .
\end{eqnarray}

Assuming the isospin of $| D^* \bar D^*; 0^{++} \rangle$ to be $I=0$ and its mass to be about $M_{D^{*0}} + M_{\bar D^{*0}} = 4014$~MeV, we obtain the following partial decay widths:
\begin{eqnarray}
\nonumber \Gamma(| D^* \bar D^*; 0^{++} \rangle \to \eta_c \eta ) &=& a_4^2 ~ 4.0 \times 10^{-3} ~{\rm GeV}^7 \, ,
\\[1mm] \nonumber \Gamma(| D^* \bar D^*; 0^{++} \rangle \to J/\psi \omega ) &=& a_4^2 ~ 4.8 \times 10^{-6} ~{\rm GeV}^7 \, ,
\\[1mm] \nonumber \Gamma(| D^* \bar D^*; 0^{++} \rangle \to \chi_{c0} f_0(500) ) &=& a_4^2 ~ 4.1 \times 10^{-6} ~{\rm GeV}^7 \, ,
\\ \label{result:eta4}
\end{eqnarray}
where $a_4$ is an overall factor.

Besides, $| D^* \bar D^*; 0^{++} \rangle$ can directly fall apart into the $D^*$ and $\bar D^*$ mesons, and further decay into the $D^* \bar D \pi$ and $D \bar D^* \pi$ final states. This process is depicted in Fig.~\ref{fig:decay}(b). We estimate its partial width to be
\begin{eqnarray}
&& \Gamma(| D^* \bar D^*; 0^{++} \rangle \to D^* \bar D^* \to D^* \bar D \pi + D \bar D^* \pi )
\label{result:eta4p}
\\ \nonumber && ~~~~~~~~~~~~~~~~~~~~~~~~~~~~~ =  a_4^{\prime2} ~ 1.7 \times 10^{-6} ~{\rm GeV}^7 \, ,
\end{eqnarray}
where $a_4^\prime$ is another overall factor. It is probably larger than $a_4$, and we define their ratio to be $t \equiv a_4^\prime / a_4$. This parameter measures the dynamical difference between the process depicted in Fig.~\ref{fig:decay}(a) and the one depicted Fig.~\ref{fig:decay}(b).

The uncertainty of Eq.~(\ref{result:eta4p}) is quite large, because we choose the mass of $| D^* \bar D^*; 0^{++} \rangle$ to be just at the $D^{*0} \bar D^{*0}$ threshold, so that only the $D^{*0} \bar D^0 \pi^0$ and $D^0 \bar D^{*0} \pi^0$ final states are kinematically allowed. However, if its realistic mass turns out to be slightly larger, some other final states such as $D^{*0} \bar D^{*0}$ and $D^{*0} D^+ \pi^-$ may also become possible. Moreover, phase spaces of the $D^{*0} \bar D^0 \pi^0$ and $D^0 \bar D^{*0} \pi^0$ final states depend significantly on the mass of $| D^* \bar D^*; 0^{++} \rangle$.

\subsubsection{$| D^* \bar D^*; 2^{++} \rangle$}
\label{sec:DD2pp}

The current $\eta_6^{\alpha\beta}$ corresponds to the $| D^* \bar D^*; 2^{++} \rangle$ molecular state. We can apply the Fierz rearrangement and transform it into:
\begin{eqnarray}
\eta_6^{\alpha\beta} &=& P^{\alpha\beta,\mu\nu} ~ \bar q_a \gamma_\mu c_a ~ \bar c_b \gamma_\nu q_b
\label{eq:eta6}
\\ \nonumber   &\rightarrow& - {1\over6} ~ P^{\alpha\beta,\mu\nu} \bar q_a \gamma_\mu q_a ~ \bar c_b \gamma_\nu c_b
\\ \nonumber         &&      - {1\over6} ~ P^{\alpha\beta,\mu\nu} \bar q_a \gamma_\mu \gamma_5 q_a ~ \bar c_b \gamma_\nu \gamma_5 c_b
\\ \nonumber         &&      + {1\over6} ~ P^{\alpha\beta,\mu\nu} \bar q_a \sigma_{\mu\rho} q_a ~ \bar c_b \sigma_{\nu\rho} c_b + \cdots \, .
\end{eqnarray}

Assuming the isospin of $| D^* \bar D^*; 2^{++} \rangle$ to be $I=0$ and its mass to be about $M_{D^{*0}} + M_{\bar D^{*0}} = 4014$~MeV, we obtain the following partial decay widths:
\begin{eqnarray}
\nonumber \Gamma(| D^* \bar D^*; 2^{++} \rangle \to \eta_c \eta ) &=& a_6^2 ~ 5.1 \times 10^{-7} ~{\rm GeV}^7 \, ,
\\[1mm] \nonumber \Gamma(| D^* \bar D^*; 2^{++} \rangle \to J/\psi \omega ) &=& a_6^2 ~ 3.7 \times 10^{-5} ~{\rm GeV}^7 \, ,
\\ \label{result:eta6}
\end{eqnarray}
where $a_6$ is an overall factor.

Besides, $| D^* \bar D^*; 2^{++} \rangle$ can directly fall apart into the $D^*$ and $\bar D^*$ mesons, and further decay into the $D^* \bar D \pi$ and $D \bar D^* \pi$ final states:
\begin{eqnarray}
&& \Gamma(| D^* \bar D^*; 2^{++} \rangle \to D^* \bar D^* \to D^* \bar D \pi + D \bar D^* \pi )
\label{result:eta6p}
\\ \nonumber && ~~~~~~~~~~~~~~~~~~~~~~~~~~~~~ = a_6^{\prime2} ~ 4.9 \times 10^{-6} ~{\rm GeV}^7 \, ,
\end{eqnarray}
where $a_6^\prime \approx t \times a_6$ is another overall factor.

\subsubsection{$| D \bar D^*; 1^{++} \rangle$}
\label{sec:DD1pp}

The current $\eta_2^{\alpha}$ corresponds to the $| D \bar D^*; 1^{++} \rangle$ molecular state. Based on this current, we have systematically studied decay properties of the $X(3872)$ as $| D \bar D^*; 1^{++} \rangle$ in Ref.~\cite{Chen:2019eeq}. We summarize some of those results here.

The isospin breaking effect of the $X(3872)$ is significant and important to understand its nature. In Ref.~\cite{Chen:2019eeq} we assumed it to be the combination of both $I=0$ and $I=1$ $| D \bar D^*; 1^{++} \rangle$ states:
\begin{equation}
X(3872) = \cos\theta~| D \bar D^*; 0^+1^{++} \rangle + \sin\theta~| D \bar D^*; 1^-1^{++} \rangle \, .
\end{equation}
After fine-tuning the isospin-breaking angle to be $\theta = \pm 15^\circ$, the BESIII measurement~\cite{Ablikim:2019zio},
\begin{equation}
{\mathcal{B}(X(3872) \to J/\psi \omega \to J/\psi \pi \pi \pi) \over \mathcal{B}(X(3872) \rightarrow J/\psi \rho \rightarrow J/\psi \pi \pi)} = 1.6^{+0.4}_{-0.3} \pm 0.2 \, ,
\end{equation}
can be explained.

In the present study we further select the positive angle $\theta = + 15^\circ$, so that the $D^0 \bar D^{*0}$ component is more than the $D^+ D^{*-}$ one:
\begin{eqnarray}
&& X(3872)
\\ \nonumber &=& \cos15^\circ~| D \bar D^*; 0^+1^{++} \rangle + \sin15^\circ~| D \bar D^*; 1^-1^{++} \rangle
\\ \nonumber &=& 0.8684~| D^0 \bar D^{*0}; 1^{++} \rangle + 0.4959~| D^+ D^{*-}; 1^{++} \rangle \, .
\end{eqnarray}
Using this angle, decay properties of the $X(3872)$ as $| D \bar D^*; 1^{++} \rangle$ were systematically studied in Ref.~\cite{Chen:2019eeq}, and the results are summarized in Eq.~(\ref{result:DDs1pp}) of Sec.~\ref{sec:decaysummary}.

Besides, productions of the $X(3872)$ as $| D \bar D^*; 1^{++} \rangle$ in $B$ and $B^{*}$ decays are increased from Eqs.~(\ref{result:productionDD}) and (\ref{result:productionBsDD}) to be
\begin{eqnarray}
\nonumber \Gamma(B^{-} \to K^- |D \bar D^*;   1^{++} \rangle) &=& d_1^2~0.60 \times 10^{-10}~{\rm GeV}^{13} \, ,
\\[1mm] \nonumber \Gamma(B^{*-} \to K^- |D \bar D^*;   1^{++} \rangle) &=& d_3^2~0.93 \times 10^{-10}~{\rm GeV}^{13} \, .
\\
\end{eqnarray}
The two ratios $\mathcal{R}_1$ and $\mathcal{R}_2$ given in Table~\ref{tab:width} need to be accordingly modified.

\subsubsection{$| D \bar D^*; 1^{+-} \rangle$ and $| D^* \bar D^*; 1^{+-} \rangle$}
\label{sec:DD1pm}

The currents $\eta_3^{\alpha}$ and $\eta_5^{\alpha}$ correspond to the $| D \bar D^*; 1^{+-} \rangle$ and $| D^* \bar D^*; 1^{+-} \rangle$ molecular states, respectively. Based on these two currents, we have systematically studied decay properties of the $Z_c(3900)$ as $| D \bar D^*; 1^{+-} \rangle$ in Ref.~\cite{Chen:2019wjd}. We summarize some of those results here. Especially, in Ref.~\cite{Chen:2019wjd} we have considered the mixing between $| D \bar D^*; 1^{+-} \rangle$ and $| D^* \bar D^*; 1^{+-} \rangle$:
\begin{eqnarray}
\nonumber | D \bar D^*; 1^{+-} \rangle^\prime &=& + \cos \phi~| D \bar D^* \rangle_{1^{+-}} + \sin\phi~| D^* \bar D^* \rangle_{1^{+-}} \, ,
\\[1mm] \nonumber | D^* \bar D^*; 1^{+-} \rangle^\prime &=& -\sin \phi~| D \bar D^* \rangle_{1^{+-}} + \cos\phi~| D^* \bar D^* \rangle_{1^{+-}} \, ,
\\
\end{eqnarray}
in order to explain the BESIII measurement~\cite{Ablikim:2019ipd}:
\begin{equation}
{\mathcal{B}(Z_c(3900)^\pm \rightarrow \eta_c\rho^\pm) \over \mathcal{B}(Z_c(3900)^\pm \rightarrow J/\psi\pi^\pm)} = 2.2 \pm 0.9 \, .
\label{eq:BESIII2}
\end{equation}

In the present study we do not consider this mixing any more. We just use the single current $\eta_3^{\alpha}$ to study decay properties of the $Z_c(3900)$ as $| D \bar D^*; 1^{+-} \rangle$, and use the single current $\eta_5^{\alpha}$ to study the $Z_c(4020)$ as $| D^* \bar D^*; 1^{+-} \rangle$. The results are summarized in Eqs.~(\ref{result:DDs1pm}) and (\ref{result:DsDs1pm}) of Sec.~\ref{sec:decaysummary}. Actually, as shown in Table~\ref{tab:uncertainty} of Appendix~\ref{app:phase}, the above BESIII measurement given in Eq.~(\ref{eq:BESIII2}) can also be explained if we take into account those unknown phase angles among different coupling constants.

\subsection{Decay properties of $D^{(*)} \bar K^{(*)}$ molecules}
\label{sec:DKdecay}

In this subsection we perform the Fierz rearrangement for $\xi_{1\cdots6}$, and use the obtained results to study strong decay properties of $D^{(*)} \bar K^{(*)}$ molecular states. Again, we adopt the possible interpretation of the $X_0(2900)$ as the $|D^* \bar K^*; 0^{+} \rangle$ molecular state. Masses of $D^{(*)} \bar K^{(*)}$ molecular states are accordingly chosen, as given in Eqs.~(\ref{eq:DKmass}).

\subsubsection{$| D \bar K; 0^{+} \rangle$ and $| D^* \bar K^*; 0^{+} \rangle$}
\label{sec:DK0p}

The currents $\xi_1$ and $\xi_4$ correspond to the $| D \bar K; 0^{+} \rangle$ and $| D^* \bar K^*; 0^{+} \rangle$ molecular states, respectively. We can apply the Fierz rearrangement and transform them into:
\begin{eqnarray}
\xi_1              &=& \bar q_a \gamma_5 c_a ~ \bar q_b \gamma_5 s_b
\label{eq:xi1}
\\ \nonumber   &\rightarrow& - {1\over12} ~ \bar q_a s_a ~ \bar q_b c_b                       - {1\over12} ~ \bar q_a \gamma_5 s_a ~ \bar q_b \gamma_5 c_b
\\ \nonumber         &&      + {1\over12} ~ \bar q_a \gamma_\mu s_a ~ \bar q_b \gamma^\mu c_b - {1\over12} ~ \bar q_a \gamma_\mu \gamma_5 s_a ~ \bar q_b \gamma^\mu \gamma_5 c_b
\\ \nonumber         &&      - {1\over24} ~ \bar q_a \sigma_{\mu\nu} s_a ~ \bar q_b \sigma^{\mu\nu} c_b + \cdots \, ,
\\[1mm]
\xi_4              &=& \bar q_a \gamma^\mu c_a ~ \bar q_b \gamma_\mu s_b
\label{eq:xi4}
\\ \nonumber   &\rightarrow& - {1\over3} ~ \bar q_a s_a ~ \bar q_b c_b                        + {1\over3} ~ \bar q_a \gamma_5 s_a ~ \bar q_b \gamma_5 c_b
\\ \nonumber         &&      + {1\over6} ~ \bar q_a \gamma_\mu s_a ~ \bar q_b \gamma^\mu c_b  + {1\over6} ~ \bar q_a \gamma_\mu \gamma_5 s_a ~ \bar q_b \gamma^\mu \gamma_5 c_b + \cdots \, .
\end{eqnarray}

Assuming the mass of $| D \bar K; 0^{+} \rangle$ to be about $M_{D^0} + M_{K^+} = 2359$~MeV, all its strong decay channels turn out to be kinematically forbidden.

Assuming the mass of $| D^* \bar K^*; 0^{+} \rangle$ to be $M_{X_0(2900)} = 2866$~MeV, we obtain the following partial decay widths:
\begin{eqnarray}
\nonumber && \Gamma(| D^* \bar K^*; 0^{+} \rangle \to D \bar K ) = b_4^2 ~ 1.6 \times 10^{-5} ~{\rm GeV}^7 \, ,
\\[1mm] && \Gamma(| D^* \bar K^*; 0^{+} \rangle \to D^* \bar K^* \to D^* \bar K \pi )
\label{result:xi4}
\\ \nonumber && ~~~~~~~~~~~~~~~~~~~~~~~~~~~~~ = b_4^2 ~ 5.4 \times 10^{-8} ~{\rm GeV}^7 \, ,
\end{eqnarray}
where $b_4$ is an overall factor.

Besides, $| D^* \bar K^*; 0^{+} \rangle$ can directly fall apart into the $D^*$ and $\bar K^*$ mesons, and further decay into the $D^* \bar K \pi$ final states:
\begin{eqnarray}
&& \Gamma(| D^* \bar K^*; 0^{+} \rangle \to D^* \bar K^* \to D^* \bar K \pi )
\label{result:xi4p}
\\ \nonumber && ~~~~~~~~~~~~~~~~~~~~~~~~~~~~~ = b_4^{\prime2} ~ 1.9 \times 10^{-6} ~{\rm GeV}^7 \, ,
\end{eqnarray}
where $b_4^\prime \approx t \times b_4$ is another overall factor.

\subsubsection{$| D \bar K^*; 1^{+} \rangle$, $| D^* \bar K; 1^{+} \rangle$, and $| D^* \bar K^*; 1^{+} \rangle$}
\label{sec:DK1p}

The currents $\xi_2^\alpha$, $\xi_3^\alpha$, and $\xi_5^\alpha$ correspond to the $| D \bar K^*; 1^{+} \rangle$, $| D^* \bar K; 1^{+} \rangle$, and $| D^* \bar K^*; 1^{+} \rangle$ molecular states, respectively. We can apply the Fierz rearrangement and transform them into:
\begin{eqnarray}
\xi_2^\alpha       &=& \bar q_a \gamma_5 c_a ~ \bar q_b \gamma^\alpha s_b
\label{eq:xi2}
\\ \nonumber   &\rightarrow& - {1\over12} ~ \bar q_a \gamma_5 s_a ~ \bar q_b \gamma^\alpha c_b   - {1\over12} ~ \bar q_a \gamma^\alpha s_a ~ \bar q_b \gamma_5 c_b
\\ \nonumber         &&      + {1\over12} ~ \bar q_a \gamma^\alpha \gamma_5 s_a ~ \bar q_b c_b   - {1\over12} ~ \bar q_a s_a ~ \bar q_b \gamma^\alpha \gamma_5 c_b
\\ \nonumber         &&      - {i\over12} ~ \bar q_a \gamma_\mu \gamma_5 s_a ~ \bar q_b \sigma^{\alpha\mu} c_b   - {i\over12} ~ \bar q_a \sigma^{\alpha\mu} s_a ~ \bar q_b \gamma_\mu \gamma_5 c_b
\\ \nonumber         &&      + {i\over12} ~ \bar q_a \gamma_\mu s_a ~ \bar q_b \sigma^{\alpha\mu} \gamma_5 c_b   - {i\over12} ~ \bar q_a \sigma^{\alpha\mu} \gamma_5 s_a ~ \bar q_b \gamma_\mu c_b
\\ \nonumber && + \cdots \, ,
\\[1mm]
\xi_3^\alpha       &=& \bar q_a \gamma^\alpha c_a ~ \bar q_b \gamma_5 s_b
\label{eq:xi3}
\\ \nonumber   &\rightarrow& - {1\over12} ~ \bar q_a \gamma_5 s_a ~ \bar q_b \gamma^\alpha c_b   - {1\over12} ~ \bar q_a \gamma^\alpha s_a ~ \bar q_b \gamma_5 c_b
\\ \nonumber         &&      - {1\over12} ~ \bar q_a \gamma^\alpha \gamma_5 s_a ~ \bar q_b c_b   + {1\over12} ~ \bar q_a s_a ~ \bar q_b \gamma^\alpha \gamma_5 c_b
\\ \nonumber         &&      - {i\over12} ~ \bar q_a \gamma_\mu \gamma_5 s_a ~ \bar q_b \sigma^{\alpha\mu} c_b   - {i\over12} ~ \bar q_a \sigma^{\alpha\mu} s_a ~ \bar q_b \gamma_\mu \gamma_5 c_b
\\ \nonumber         &&      - {i\over12} ~ \bar q_a \gamma_\mu s_a ~ \bar q_b \sigma^{\alpha\mu} \gamma_5 c_b   + {i\over12} ~ \bar q_a \sigma^{\alpha\mu} \gamma_5 s_a ~ \bar q_b \gamma_\mu c_b
\\ \nonumber && + \cdots \, ,
\\[1mm]
\xi_5^\alpha       &=& \bar q_a \gamma_\mu c_a ~ \bar q_b\sigma^{\alpha\mu}\gamma_5 s_b - \{ \gamma_\mu \leftrightarrow \sigma^{\alpha\mu}\gamma_5 \}
\label{eq:xi5}
\\ \nonumber   &\rightarrow& - {i\over2} ~ \bar q_a \gamma_5 s_a ~ \bar q_b \gamma^\alpha c_b   + {i\over2} ~ \bar q_a \gamma^\alpha s_a ~ \bar q_b \gamma_5 c_b
\\ \nonumber         &&      + {1\over6} ~ \bar q_a \gamma_\mu \gamma_5 s_a ~ \bar q_b \sigma^{\alpha\mu} c_b   - {1\over6} ~ \bar q_a \sigma^{\alpha\mu} s_a ~ \bar q_b \gamma_\mu \gamma_5 c_b
\\ \nonumber && + \cdots \, .
\end{eqnarray}

Assuming the mass of $| D \bar K^*; 1^{+} \rangle$ to be about $M_{D^0} + M_{K^{*+}} = 2756$~MeV, we obtain the following partial decay widths:
\begin{eqnarray}
\nonumber && \Gamma(| D \bar K^*; 1^{+} \rangle \to D^* \bar K ) = b_2^2 ~ 1.7 \times 10^{-7} ~{\rm GeV}^7 \, ,
\\[1mm] && \Gamma(| D \bar K^*; 1^{+} \rangle \to D \bar K^* \to D \bar K \pi )
\label{result:xi2}
\\ \nonumber && ~~~~~~~~~~~~~~~~~~~~~~~~~~~~~   =  b_2^2 ~ 4.7 \times 10^{-9} ~{\rm GeV}^7 \, ,
\\[1mm] \nonumber && \Gamma(| D \bar K^*; 1^{+} \rangle \to D^* \bar K^* \to D^* \bar K \pi )
\\ \nonumber && ~~~~~~~~~~~~~~~~~~~~~~~~~~~~~   =  b_2^2 ~ 7.0 \times 10^{-10} ~{\rm GeV}^7 \, ,
\end{eqnarray}
where $b_2$ is an overall factor.

Besides, $| D \bar K^*; 1^{+} \rangle$ can directly fall apart into the $D$ and $\bar K^*$ mesons, and further decay into the $D \bar K \pi$ final states:
\begin{eqnarray}
&& \Gamma(| D \bar K^*; 1^{+} \rangle \to D \bar K^* \to D \bar K \pi )
\label{result:xi2p}
\\ \nonumber && ~~~~~~~~~~~~~~~~~~~~~~~~~~~~~   = b_2^{\prime2} ~ 2.9 \times 10^{-6} ~{\rm GeV}^7 \, ,
\end{eqnarray}
where $b_2^\prime \approx t \times b_2$ is another overall factor.

Assuming the mass of $| D^* \bar K; 1^{+} \rangle$ to be about $M_{D^{*0}} + M_{K^+} = 2501$~MeV, we obtain the following partial decay widths:
\begin{eqnarray}
&& \Gamma(| D^* \bar K; 1^{+} \rangle \to D \bar K^* \to D \bar K \pi )
\label{result:xi3}
\\ \nonumber && ~~~~~~~~~~~~~~~~~~~~~~~~~~~~~   =  b_3^2 ~ 5.4 \times 10^{-16} ~{\rm GeV}^7 \, ,
\\[1mm] \nonumber && \Gamma(| D^* \bar K; 1^{+} \rangle \to D^* \bar K \to D \pi \bar K )
\\ \nonumber && ~~~~~~~~~~~~~~~~~~~~~~~~~~~~~   =  b_3^2 ~ 1.3 \times 10^{-12} ~{\rm GeV}^7 \, ,
\end{eqnarray}
where $b_3$ is an overall factor.

Besides, $| D^* \bar K; 1^{+} \rangle$ can directly fall apart into the $D^*$ and $\bar K$ mesons, and further decay into the $D \pi \bar K$ final states:
\begin{eqnarray}
&& \Gamma(| D^* \bar K; 1^{+} \rangle \to D^* \bar K \to D \pi \bar K )
\label{result:xi3p}
\\ \nonumber && ~~~~~~~~~~~~~~~~~~~~~~~~~~~~~   = b_3^{\prime2} ~ 3.2 \times 10^{-10} ~{\rm GeV}^7 \, ,
\end{eqnarray}
where $b_3^\prime \approx t \times b_3$ is another overall factor.

Assuming the mass of $| D^* \bar K^*; 1^{+} \rangle$ to be about $M_{D^{*0}} + M_{K^{*+}} = 2899$~MeV, we obtain the following partial decay widths:
\begin{eqnarray}
\nonumber \Gamma(| D^* \bar K^*; 1^{+} \rangle \to D \bar K^*) &=&  b_5^2 ~ 3.8 \times 10^{-6} ~{\rm GeV}^7 \, ,
\\[1mm] \nonumber \Gamma(| D^* \bar K^*; 1^{+} \rangle \to D^* \bar K ) &=&  b_5^2 ~ 1.2 \times 10^{-5} ~{\rm GeV}^7 \, ,
\\ \label{result:xi5}
\end{eqnarray}
where $b_5$ is an overall factor.

Besides, $| D^* \bar K^*; 1^{+} \rangle$ can directly fall apart into the $D^*$ and $\bar K^*$ mesons, and further decay into the $D^* \bar K \pi$ and $D \pi \bar K^*$ final states:
\begin{eqnarray}
&& \Gamma(| D^* \bar K^*; 1^{+} \rangle \to D^* \bar K^* \to D^* \bar K \pi )
\label{result:xi5p}
\\ \nonumber && ~~~~~~~~~~~~~~~~~~~~~~~~~~~~~   =  b_5^{\prime2} ~ 1.1 \times 10^{-5} ~{\rm GeV}^7 \, ,
\\[1mm] \nonumber && \Gamma(| D^* \bar K^*; 1^{+} \rangle \to D^* \bar K^* \to D \pi \bar K^* )
\\ \nonumber && ~~~~~~~~~~~~~~~~~~~~~~~~~~~~~   =  b_5^{\prime2} ~ 1.1 \times 10^{-9} ~{\rm GeV}^7 \, ,
\end{eqnarray}
where $b_5^\prime \approx t \times b_5$ is another overall factor.

\subsubsection{$| D^* \bar K^*; 2^{+} \rangle$}
\label{sec:DK2p}

The current $\xi_6^{\alpha\beta}$ corresponds to the $| D^* \bar K^*; 2^{+} \rangle$ molecular state. We can apply the Fierz rearrangement and transform it into:
\begin{eqnarray}
\xi_6^{\alpha\beta} &=& P^{\alpha\beta,\mu\nu} ~ \bar q_a \gamma_\mu c_a ~ \bar q_b \gamma_\nu s_b
\label{eq:xi6}
\\ \nonumber   &\rightarrow& - {1\over6} ~ P^{\alpha\beta,\mu\nu} \bar q_a \gamma_\mu s_a ~ \bar q_b \gamma_\nu c_b
\\ \nonumber         &&      - {1\over6} ~ P^{\alpha\beta,\mu\nu} \bar q_a \gamma_\mu \gamma_5 s_a ~ \bar q_b \gamma_\nu \gamma_5 c_b
\\ \nonumber         &&      + {1\over6} ~ P^{\alpha\beta,\mu\nu} \bar q_a \sigma_{\mu\rho} s_a ~ \bar q_b \sigma_{\nu\rho} c_b + \cdots \, .
\end{eqnarray}

Assuming the mass of $| D^* \bar K^*; 2^{+} \rangle$ to be about $M_{D^{*0}} + M_{K^{*+}} = 2899$~MeV, we obtain the following partial decay widths:
\begin{eqnarray}
\nonumber && \Gamma(| D^* \bar K^*; 2^{+} \rangle \to D \bar K ) = b_6^2 ~ 3.0 \times 10^{-7} ~{\rm GeV}^7 \, ,
\\[1mm] && \Gamma(| D^* \bar K^*; 2^{+} \rangle \to D^* \bar K^* \to D^* \bar K \pi )
\label{result:xi6}
\\ \nonumber && ~~~~~~~~~~~~~~~~~~~~~~~~~~~~~   =  b_6^2 ~ 1.3 \times 10^{-6} ~{\rm GeV}^7 \, ,
\\[1mm] \nonumber && \Gamma(| D^* \bar K^*; 2^{+} \rangle \to D^* \bar K^* \to D \pi \bar K^* )
\\ \nonumber && ~~~~~~~~~~~~~~~~~~~~~~~~~~~~~   =  b_6^2 ~ 7.4 \times 10^{-11} ~{\rm GeV}^7 \, ,
\end{eqnarray}
where $b_6$ is an overall factor.

Besides, $| D^* \bar K^*; 2^{+} \rangle$ can directly fall apart into the $D^*$ and $\bar K^*$ mesons, and further decay into the $D^* \bar K \pi$ and $D \pi \bar K^*$ final states:
\begin{eqnarray}
&& \Gamma(| D^* \bar K^*; 2^{+} \rangle \to D^* \bar K^* \to D^* \bar K \pi )
\label{result:xi6p}
\\ \nonumber && ~~~~~~~~~~~~~~~~~~~~~~~~~~~~~   =  b_6^{\prime2} ~ 1.7 \times 10^{-5} ~{\rm GeV}^7 \, ,
\\[1mm] \nonumber && \Gamma(| D^* \bar K^*; 2^{+} \rangle \to D^* \bar K^* \to D \pi \bar K^* )
\\ \nonumber && ~~~~~~~~~~~~~~~~~~~~~~~~~~~~~   = b_6^{\prime2} ~ 9.9 \times 10^{-10} ~{\rm GeV}^7 \, ,
\end{eqnarray}
where $b_6^\prime \approx t \times b_6$ is another overall factor.

\begin{widetext}
\subsection{Summary of decay properties}
\label{sec:decaysummary}

We summarize the relative branching ratios obtained in Sec.~\ref{sec:DDdecay} and Sec.~\ref{sec:DKdecay} here. The Fierz rearrangements for $\zeta_{1\cdots6}$ are quite similar to those for $\eta_{1\cdots6}$, just with one $up$/$down$ quark replaced by another $strange$ quark. Hence, decay properties of $D^{(*)} \bar D_s^{(*)}$ molecular states can be similarly investigated, and we also summarize their results here. We use the parameter $t \approx {a_i^\prime / a_i} \approx {b_i^\prime / b_i}$~($i=1\cdots6$) to measure the dynamical difference between the process depicted in Fig.~\ref{fig:decay}(a) and the one depicted in Fig.~\ref{fig:decay}(b).

We obtain the following relative branching ratios for $| D \bar D; 0^{++} \rangle$, $| D^* \bar D^*; 0^{++} \rangle$, and $| D^* \bar D^*; 2^{++} \rangle$, all of which are assumed to have $I=0$:
\begin{eqnarray}
\nonumber && \mathcal{B}\Big({| D \bar D; 0^{++} \rangle} \to ~\eta_c \eta~ : ~J/\psi \omega (\to \pi\pi\pi)~ : ~\chi_{c0} f_0(500) (\to \pi\pi)~ \Big)
\\        &\approx&   ~~~~~~~~~~~~~~~~~~~~~~~~~  1 ~~ :   ~~~~~~~\,10^{-6}\,~~~~~~~ : ~~~~~~~10^{-5}~ \, ,
\label{result:DD0pp}
\\[1mm] \nonumber && \mathcal{B}\Big({| D^* \bar D^*; 0^{++} \rangle} \to ~\eta_c \eta~ : ~~J/\psi \omega~~ : ~\chi_{c0} f_0(500)~ : ~D^* \bar D^*(\to \bar D \pi)~ \Big)
\\        &\approx&   ~~~~~~~~~~~~~~~~~~~~~~~~~~~~ 1 ~~ :   ~~\,0.001~~ : ~~~~\,0.001\,~~~~ : ~~~~~~~10^{-4}t~ \, ,
\label{result:DsDs0pp}
\\[1mm] \nonumber && \mathcal{B}\Big({| D^* \bar D^*; 2^{++} \rangle} \to ~\eta_c \eta~ : ~~J/\psi \omega~~ : ~D^* \bar D^*(\to \bar D \pi)~ \Big)
\\        &\approx&   ~~~~~~~~~~~~~~~~~~~~~~~~~~~~ 1 ~~ :   ~~~~73~~~~ : ~~~~~~~9.5t~ \, .
\label{result:DsDs2pp}
\end{eqnarray}

Decay properties of the $X(3872)$ as $| D \bar D^*; 1^{++} \rangle$ have been systematically studied in Ref.~\cite{Chen:2019eeq}, where we obtain
\begin{eqnarray}
\nonumber && {\mathcal{B}\Big(|D \bar D^*; 1^{++} \rangle \rightarrow
 J/\psi\omega (\rightarrow \pi \pi \pi)
: J/\psi\rho (\rightarrow \pi \pi)
: ~\chi_{c0}\pi~
: \eta_c f_0 (\rightarrow \pi \pi)
: \chi_{c1} f_0 (\rightarrow \pi \pi)
: D \bar D^{*} (\rightarrow \bar D \pi)~
\Big) }
\\ &\approx&
~~~~~~~~~~~~~~~~~~~~~~~~~~~~~~~~~~~1~~~~~~ : \,0.63~({\rm input})\, : ~0.015\, : ~~~~\,0.091~~~~ : ~~~~~0.086~~~~~ : ~~~~~~~7.4t~ \, .
\label{result:DDs1pp}
\end{eqnarray}
In the calculations we have assumed the $X(3872)$ to be the combination of both $I=0$ and $I=1$ $| D \bar D^*; 1^{++} \rangle$ molecular states.

Decay properties of the $Z_c(3900)$ as $| D \bar D^*; 1^{+-} \rangle$ have been systematically studied in Ref.~\cite{Chen:2019wjd}. In the present study we further study the $Z_c(4020)$ as $| D^* \bar D^*; 1^{+-} \rangle$. Altogether, we obtain
\begin{eqnarray}
\nonumber && {\mathcal{B}\Big(|D \bar D^*; 1^{+-} \rangle \rightarrow ~~J/\psi\pi~~
:~~~ \eta_c\rho ~~~
:~~~ h_c\pi ~~~
:~ \chi_{c1}\rho (\rightarrow \pi \pi)~
:~~ D \bar D^{*}(\to \bar D \pi)~
\Big) }
\\ &\approx&
~~~~~~~~~~~~~~~~~~~~~~~~~~~~~~1\,~~~ : ~~0.092\,~ : ~~0.011~~ : ~~~~~~10^{-6}~~~~~\, : ~~~~~~~74t~\, ,
\label{result:DDs1pm}
\\[1mm] \nonumber && {\mathcal{B}\Big(|D^* \bar D^*; 1^{+-} \rangle \rightarrow ~~J/\psi\pi~~
:~~ \eta_c\rho ~~
:~~ h_c\pi ~~
:~ \chi_{c1}\rho (\rightarrow \pi \pi)~
:~ D^* \bar D^{*}(\to \bar D \pi)~
\Big)}
\\ &\approx&
~~~~~~~~~~~~~~~~~~~~~~~~~~~~~~~1\,~~~ : ~\,0.33\,~ : ~0.002~ : ~~~~~~10^{-5}\,~~~~~ : ~~~~~~~11t~\, .
\label{result:DsDs1pm}
\end{eqnarray}
In the calculations we have assumed that the $Z_c(3900)$ and $Z_c(4020)$ both have $I=1$.

Assuming the mass of $| D \bar K; 0^{+} \rangle$ to be about $M_{D^0} + M_{K^+} = 2359$~MeV, all its strong decay channels turn out to be kinematically forbidden.

We obtain the following relative branching ratios for $| D^* \bar K; 1^{+} \rangle$:
\begin{eqnarray}
\nonumber && {\mathcal{B}\Big({| D^* \bar K; 1^{+} \rangle} \to  ~D \bar K^* (\to \bar K \pi)~ : ~\bar K D^* (\to D \pi)~ \Big)}
\\        &\approx&   ~~~~~~~~~~~~~~~~~~~~~~~~~~~~~~~  10^{-6} \,~~~~ :  ~~~~t+0.004~ \, .
\label{result:DsK1p}
\end{eqnarray}

The isospin-averaged width of the $\bar K^*$ meson is about $\Gamma_{K^*} \approx 49.1$~MeV~\cite{pdg}, which value is quite large and can not be neglected. We refer to Appendix~\ref{app:composite} for more discussions. Taking this into account, we estimate widths of $| D \bar K^*; 1^{+} \rangle$, $| D^* \bar K^*; 0^{+} \rangle$, $| D^* \bar K^*; 1^{+} \rangle$, and $| D^* \bar K^*; 2^{+} \rangle$ to be
\begin{equation}
\Gamma_{| D^{(*)} \bar K^*; J^{+} \rangle} \approx \Gamma_{\bar K^*} + \Gamma_{| D^{(*)} \bar K^*; J^{+} \rangle \to D \bar K/D^* \bar K/D \bar K^*/D^* \bar K^*/\cdots} \, ,
\label{eq:DKwidth}
\end{equation}
with the following relative branching ratios
\begin{eqnarray}
\nonumber && \mathcal{B}\Big({| D \bar K^*; 1^{+} \rangle} \to ~D \bar K^* (\to \bar K \pi)~ : ~~D^* \bar K~~ : ~D^* \bar K^* (\to \bar K \pi)~ \Big)
\\        &\approx&      ~~~~~~~~~~~~~~~~~~~~~~~~~~~ t+0.002 ~~~ : ~~0.057~~ : ~~~~~~10^{-4}~\, ,
\label{result:DKs1p}
\\[1mm] \nonumber && \mathcal{B}\Big({| D^* \bar K^*; 0^{+} \rangle} \to ~~D \bar K~~ : ~D^* \bar K^* (\to \bar K \pi)~ \Big)
\\        &\approx&      ~~~~~~~~~~~~~~~~~~~~~~~~~~ 8.5 ~~~ : ~~~~\,t+0.028~\, ,
\label{result:DsKs0p}
\\[1mm] \nonumber && \mathcal{B}\Big({| D^* \bar K^*; 1^{+} \rangle} \to ~~D \bar K^*~~ : ~~D^* \bar K~~ : ~D^* \bar K^* (\to \bar K \pi)~ : ~\bar K^* D^* (\to D \pi)~ \Big)
\\        &\approx&      ~~~~~~~~~~~~~~~~~~~~~~~~~\, 0.34 ~~~ : ~~~~1.1\,~~~ : ~~~~~~~~~~t\,~~~~~~~~~ : ~~~~~~10^{-4}t~ \, ,
\label{result:DsKs1p}
\\[1mm] \nonumber && \mathcal{B}\Big({| D^* \bar K^*; 2^{+} \rangle} \to ~~D \bar K~~ : ~D^* \bar K^* (\to \bar K \pi)~ : ~\bar K^* D^* (\to D \pi)~ \Big)
\\        &\approx&      ~~~~~~~~~~~~~~~~~~~~~~~~ 0.018~~ : ~~~~\,t+0.074\,~~~~ : ~~~~10^{-4}t~ \, .
\label{result:DsKs2p}
\end{eqnarray}

We obtain the following relative branching ratios for $D^{(*)} \bar D_s^{(*)}$ molecular states:
\begin{eqnarray}
\nonumber && \mathcal{B}\Big({| D \bar D_s; 0^{+} \rangle} \to ~\eta_c \bar K~ : ~J/\psi \bar K^* (\to \bar K \pi)~ \Big)
\\        &\approx&   ~~~~~~~~~~~~~~~~~~~~~~~~~\,  1 ~~ :   ~~~~~~~0.001~ \, ,
\label{result:DDs1}
\\[1mm] \nonumber && \mathcal{B}\Big({| D \bar D_s^*; 1^{++} \rangle} \to ~J/\psi \bar K^* (\to \bar K \pi)~ : ~~\chi_{c0} \bar K~~ : ~~D \bar D_s^*~~ : ~~\bar D_s D^*(\to D \pi)~~ \Big)
\\        &\approx&   ~~~~~~~~~~~~~~~~~~~~~~~~~~~~~~~~~~\,  1 ~~~~~~~~~~ :   ~~0.012~~ : ~~~~25~~~~ : ~~~~~~~~68~ \, ,
\label{result:DDs2}
\\[1mm] \nonumber && \mathcal{B}\Big({| D \bar D_s^*; 1^{+-} \rangle} \to ~J/\psi \bar K~ : ~~\eta_c \bar K^*~~ : ~~D \bar D_s^*~~ : ~~ \bar D_s D^*(\to D \pi)~~ \Big)
\\        &\approx&   ~~~~~~~~~~~~~~~~~~~~~~~~~~~~~\, 1 ~~~ :   ~~0.093~~ : ~~~~41~~~~ : ~~~~~~~~87~ \, ,
\label{result:DDs3}
\\[1mm] \nonumber && \mathcal{B}\Big({| D^* \bar D_s^*; 0^{+} \rangle} \to ~\eta_c \bar K~ : ~J/\psi \bar K^*~ : ~\bar D_s^* D^*(\to D \pi)~ \Big)
\\        &\approx&   ~~~~~~~~~~~~~~~~~~~~~~~~~~~  1 ~~ :   ~~\,0.088\,~~ : ~~~~~~0.001~ \, ,
\label{result:DDs4}
\\[1mm] \nonumber && \mathcal{B}\Big({| D_s \bar D_s^*; 1^{+} \rangle} \to ~J/\psi \bar K~ : ~~\eta_c \bar K^*~~ : ~~h_c \bar K~~ : ~\chi_{c1} \bar K^*(\to \bar K \pi)~ : ~~D_s^* \bar D^*(\to D \pi)~~ \Big)
\\        &\approx&   ~~~~~~~~~~~~~~~~~~~~~~~~~~~~\, 1 ~~~ :   ~~~0.36\,~~ : ~\,0.002\,~ : ~~~~~~~\,10^{-7}\,~~~~~~~ : ~~~~~~~~31~ \, ,
\label{result:DDs3}
\\[1mm] \nonumber && \mathcal{B}\Big({| D^* \bar D_s^*; 2^{+} \rangle} \to ~\eta_c \bar K~ : ~J/\psi \bar K^*~ : ~\bar D_s^* D^*(\to D \pi)~ \Big)
\\        &\approx&   ~~~~~~~~~~~~~~~~~~~~~~~~~~~  1 ~~ :   ~~~~\,27\,~~~~ : ~~~~~0.055~ \, .
\label{result:DDs6}
\end{eqnarray}
In the calculations we have adopted the possible interpretations of the $Z_{cs}(3985)$, $Z_{cs}(4000)$, and $Z_{cs}(4220)$ as the $|D \bar D_s^*; 1^{++} \rangle$, $|D \bar D_s^*; 1^{+-} \rangle$, and $|D^* \bar D_s^*; 1^{+} \rangle$ molecular states, respectively. Masses of $D^{(*)} \bar D_s^{(*)}$ molecular states are accordingly chosen, as given in Eqs.~(\ref{eq:DDsmass}).
\end{widetext}

\section{Summary and discussions}
\label{sec:summary}

\begin{table*}[hbt]
\begin{center}
\renewcommand{\arraystretch}{1.6}
\caption{Relative branching ratios of $D^{(*)} \bar D^{(*)}$, $D^{(*)} \bar K^{(*)}$, and $D^{(*)} \bar D_s^{(*)}$ hadronic molecular states and their relative production rates in $B$ and $B^{*}$ decays. In the 2nd and 3rd columns we show their relative production rates $\mathcal{R}_1 \equiv {\mathcal{B}\left(B^- \rightarrow K^- X / D^- X / \phi X \right) \over \mathcal{B}\left(B^- \rightarrow D^-|D^* \bar K^*; 0^{+} \rangle \right)/d_2}$ and $\mathcal{R}_2 \equiv {\mathcal{B}\left(B^{*-} \rightarrow K^- X / D^- X / \phi X \right) \over \mathcal{B}\left(B^{*-} \rightarrow K^-|D \bar D; 0^{++} \rangle \right)/d_4}$. In the 4th-16th/9th/13th columns we show their relative branching ratios, where we use the symbol ``$AB$'' to generally denote the two-body decay $X \to AB$, the three-body decay process $X \to AB \to A b_1 b_2$, or the four-body decay process $X \to AB \to A b_1 b_2 b_3$, depending on which channel is kinematically allowed. The number $0.63$ with $\dagger$ is fixed by the BESIII measurement ${\mathcal{B}(X(3872) \to J/\psi \omega \to J/\psi \pi \pi \pi) \over \mathcal{B}(X(3872) \rightarrow J/\psi \rho \rightarrow J/\psi \pi \pi)} = 1.6^{+0.4}_{-0.3} \pm 0.2$~\cite{Ablikim:2019zio}, which is related to the isospin breaking effect of the $X(3872)$ interpreted as the $| D \bar D^*; 1^{++} \rangle$ molecular state. The parameters $t^\prime \approx d_2 /d_1 \approx d_5 / d_4$ and $t^{\prime\prime} \approx d_3 /d_1 \approx d_6 / d_4$ (partly) measure the dynamical difference among Fig.~\ref{fig:production}(a,b,c), and the parameter $t \approx {a_i^\prime / a_i} \approx {b_i^\prime / b_i}$~($i=1\cdots6$) measures the dynamical difference between Fig.~\ref{fig:decay}(a) and Fig.~\ref{fig:decay}(b).}
\begin{tabular}{c || c | c || c | c | c | c | c | c | c | c | c | c | c | c | c || c }
\hline\hline
\multirow{2}{*}{Configuration} & \multicolumn{2}{c||}{Productions} & \multicolumn{13}{c||}{Decay Channels} & \multirow{2}{*}{Candidate}
\\ \cline{2-16}
& ~\,$\mathcal{R}_1$\,~ & ~\,$\mathcal{R}_2$\,~
& ~$\eta_c \eta$~ & $J/\psi \omega$ & $\eta_c f_0$ & $\chi_{c0} f_0$ & $\chi_{c1} f_0$
& $J/\psi \pi$ & $\chi_{c0} \pi$ & $h_c \pi$ & $\eta_c \rho$ & $J/\psi \rho$ & $\chi_{c1} \rho$
& $D \bar D^*$ & $D^* \bar D^*$
\\ \hline\hline
$|D \bar D; 0^{++} \rangle$      & --        &  $d_4$    & $1$ & $10^{-6}$ & -- & $10^{-5}$ & -- & -- & -- & -- & -- & -- & -- & -- & -- & --
\\ \hline
$|D \bar D^*; 1^{++} \rangle$    & $0.19d_1$ &  $5.7d_4$ & -- & $1$ & $0.091$ & -- & $0.086$ & -- & $0.015$ & -- & -- & $0.63^\dagger$ & -- & $7.4t$ & -- & $X(3872)$
\\ \hline
$|D \bar D^*; 1^{+-} \rangle$    & $0.12d_1$ &  $3.7d_4$ & -- & -- & -- & -- & -- & $1$ & -- & $0.011$ & $0.092$ & -- & $10^{-6}$ & $74t$ & -- & $Z_c(3900)$
\\ \hline
$|D^* \bar D^*; 0^{++} \rangle$  & $1.3d_1$  &  --       & $1$ & $0.001$ & -- & $0.001$ & -- & -- & -- & -- & -- & -- & -- & -- & $10^{-4}t$ & --
\\ \hline
$|D^* \bar D^*; 1^{+-} \rangle$  & $0.42d_1$ &  $17d_4$  & -- & -- & -- & -- & -- & $1$ & -- & $0.002$ & $0.33$ & -- & $10^{-5}$ & -- & $11t$ & $Z_c(4020)$
\\ \hline
$|D^* \bar D^*; 2^{++} \rangle$  & --        &  --       & $1$ & $73$ & --  & -- & -- & -- & -- & -- & -- & -- & -- & -- & $9.5t$ & --
\\ \hline \hline
Configuration & ~\,$\mathcal{R}_1$\,~ & ~\,$\mathcal{R}_2$\,~
& \multicolumn{2}{c|}{$\Gamma_{\bar K^*}$~[MeV]}
& $D \bar K$
& \multicolumn{2}{c|}{$D \bar K^*$}
& \multicolumn{3}{c|}{$D^* \bar K$}
& \multicolumn{3}{c|}{$D^* \bar K^*$}
& \multicolumn{2}{c||}{$D^* \bar K^*$}
& Candidate
\\ \hline\hline
$|D \bar K; 0^{+} \rangle$       & --        & $1.1d_5$ & \multicolumn{2}{c|}{--}     & -- & \multicolumn{2}{c|}{--}  & \multicolumn{3}{c|}{--}  & \multicolumn{3}{c|}{--}  & \multicolumn{2}{c||}{--} & --
\\ \hline
$|D \bar K^*; 1^{+} \rangle$     & --        & $10d_5$  & \multicolumn{2}{c|}{$49.1$~MeV} & -- & \multicolumn{2}{c|}{${t}+0.002$}  & \multicolumn{3}{c|}{$0.057$}   & \multicolumn{3}{c|}{$10^{-4}$} & \multicolumn{2}{c||}{--} & --
\\ \hline
$|D^* \bar K; 1^{+} \rangle$     & $0.67d_2$ & --       & \multicolumn{2}{c|}{--}     & -- & \multicolumn{2}{c|}{$10^{-6}$}  & \multicolumn{3}{c|}{${t}+0.004$}  & \multicolumn{3}{c|}{--}  & \multicolumn{2}{c||}{--} & --
\\ \hline
$|D^* \bar K^*; 0^{+} \rangle$   & $d_2$     & --       & \multicolumn{2}{c|}{$49.1$~MeV} & $8.5$  & \multicolumn{2}{c|}{--}  & \multicolumn{3}{c|}{--}   & \multicolumn{3}{c|}{${t}+0.028$} & \multicolumn{2}{c||}{--} & $X_0(2900)$
\\ \hline
$|D^* \bar K^*; 1^{+} \rangle$   & $0.70d_2$ & $27d_5$  & \multicolumn{2}{c|}{$49.1$~MeV} & -- & \multicolumn{2}{c|}{$0.34$}  & \multicolumn{3}{c|}{$1.1$}   & \multicolumn{3}{c|}{${t}$} & \multicolumn{2}{c||}{$10^{-4}t$} & --
\\ \hline
$|D^* \bar K^*; 2^{+} \rangle$   & --        & --       & \multicolumn{2}{c|}{$49.1$~MeV} & $0.018$ & \multicolumn{2}{c|}{--}  & \multicolumn{3}{c|}{--}   & \multicolumn{3}{c|}{${t}+0.074$} & \multicolumn{2}{c||}{$10^{-4}t$} & --
\\ \hline \hline
Configuration & ~\,$\mathcal{R}_1$\,~ & ~\,$\mathcal{R}_2$\,~
& $\eta_c \bar K$
& \multicolumn{2}{c|}{$J/\psi \bar K$}
& $\chi_{c0} \bar K$
& $h_c \bar K$
& $\eta_c \bar K^*$
& \multicolumn{2}{c|}{$J/\psi \bar K^*$}
& \multicolumn{2}{c|}{$\chi_{c1} \bar K^*$}
& $D \bar D_s^*$
& $D^* \bar D_s$
& $D^* \bar D_s^*$
& Candidate
\\ \hline\hline
$|D \bar D_s; 0^{+} \rangle$      & --        &  $11d_6$   & $1$ & \multicolumn{2}{c|}{--} & -- & -- & -- & \multicolumn{2}{c|}{$0.001$} & \multicolumn{2}{c|}{--} & -- & -- & -- & --
\\ \hline
$|D \bar D_s^*; 1^{++} \rangle$   & $0.79d_3$ &  $6.6d_6$  & -- & \multicolumn{2}{c|}{--} & $0.012$ & -- & -- & \multicolumn{2}{c|}{$1$} & \multicolumn{2}{c|}{--} & $25$ & $68$ & -- & $Z_{cs}(3985)$
\\ \hline
$|D \bar D_s^*; 1^{+-} \rangle$   & $0.75d_3$ &  $6.2d_6$  & -- & \multicolumn{2}{c|}{$1$} & -- & -- & $0.093$ & \multicolumn{2}{c|}{--} & \multicolumn{2}{c|}{--} & $41$ & $87$ & -- & $Z_{cs}(4000)$
\\ \hline
$|D^* \bar D_s^*; 0^{+} \rangle$  & --        &  --        & $1$ & \multicolumn{2}{c|}{--} & -- & -- & -- & \multicolumn{2}{c|}{$0.088$} & \multicolumn{2}{c|}{--} & -- & -- & $0.001$ & --
\\ \hline
$|D^* \bar D_s^*; 1^{+} \rangle$  & $1.8d_3$  &  --        & -- & \multicolumn{2}{c|}{$1$} & -- & $0.002$ & $0.36$ & \multicolumn{2}{c|}{--} & \multicolumn{2}{c|}{$10^{-7}$} & -- & -- & $31$ & $Z_{cs}(4220)$
\\ \hline
$|D^* \bar D_s^*; 2^{+} \rangle$  & --        &  --        & $1$ & \multicolumn{2}{c|}{--} & -- & -- & -- & \multicolumn{2}{c|}{$27$} & \multicolumn{2}{c|}{--} & -- & -- & $0.055$ & --
\\ \hline\hline
\end{tabular}
\label{tab:width}
\end{center}
\end{table*}

In this paper we systematically investigate the eighteen possibly existing $D^{(*)} \bar D^{(*)}$, $D^{(*)} \bar K^{(*)}$, and $D^{(*)} \bar D_s^{(*)}$ ($\bar D_s^{(*)} \equiv D_s^{(*)-}$) hadronic molecular states, including $|D \bar D;0^{++}\rangle$, $|D \bar D^*; 1^{++}/1^{+-}\rangle$, $|D^* \bar D^*; 0^{++}/1^{+-}/2^{++}\rangle$, $|D \bar K ;0^{+}\rangle$, $|D \bar K^*;1^{+}\rangle$, $|D^* \bar K^*; 0^{+}/1^{+}/2^{+}\rangle$, $|D \bar D_s;0^{+}\rangle$, $|D \bar D_s^* ; 1^{++}/1^{+-}\rangle$, and $|D^* \bar D_s^*; 0^{+}/1^{+}/2^{+}\rangle$. Note that the $D \bar D_s^*$ molecules are not charge-conjugated, but for convenience we use $| D \bar D_s^{*}; 1^{++} \rangle$ and $| D \bar D_s^{*}; 1^{+-} \rangle$ to denote strange partners of $| D \bar D^*; 1^{++} \rangle$ and $| D \bar D^*; 1^{+-} \rangle$, respectively. By doing this, we can differentiate the $X(3872)$, $Z_c(3900)$, and $Z_c(4020)$ as well as the $Z_{cs}(3985)$, $Z_{cs}(4000)$, and $Z_{cs}(4220)$.

We systematically construct their corresponding interpolating currents, through which we study their mass spectra as well as their production and decay properties. The isospin of these molecular states is important, and in the present study we assume: a) $|D \bar D^*; 1^{+-} \rangle$ and $|D^* \bar D^*; 1^{+-} \rangle$ have $I=1$, b) $|D \bar D; 0^{++} \rangle$, $|D^* \bar D^*; 0^{++} \rangle$, and $|D^* \bar D^*; 2^{++} \rangle$ have $I=0$, c) the isospin of the $X(3872)$ as $|D \bar D^*; 1^{++} \rangle$ is separately investigated in Sec.~\ref{sec:DD1pp}, d) all the $D^{(*)} \bar K^{(*)}$ molecular states have $I=0$, and e) all the $D^{(*)} \bar D_s^{(*)}$ molecular states have $I=1/2$.

Firstly, we use the method of QCD sum rules to calculate masses and decay constants of the $D^{(*)} \bar D^{(*)}$, $D^{(*)} \bar K^{(*)}$, and $D^{(*)} \bar D_s^{(*)}$ molecular states, and the obtained results are summarized in Table~\ref{tab:mass}. Our results support: a) the interpretations of the $X(3872)$, $Z_c(3900)$, and $Z_c(4020)$ as the $|D \bar D^*; 1^{++} \rangle$, $|D \bar D^*; 1^{+-} \rangle$, and $|D^* \bar D^*; 1^{+-} \rangle$ molecular states, respectively; b) the interpretation of the $X_0(2900)$ as the $|D^* \bar K^*; 0^{+} \rangle$ molecular state; c) the interpretations of the $Z_{cs}(3985)$, $Z_{cs}(4000)$, and $Z_{cs}(4220)$ as the $|D \bar D_s^*; 1^{++} \rangle$, $|D \bar D_s^*; 1^{+-} \rangle$, and $|D^* \bar D_s^*; 1^{+} \rangle$ molecular states, respectively. The uncertainty/accuracy is moderate but not enough to extract the binding energy as well as to differentiate the $Z_{cs}(3985)$ and $Z_{cs}(4000)$. Hence, our QCD sum rule results can only suggest but not determine: a) whether these molecular states exist or not, and b) whether they are bound states or resonance states. To better understand them, we further study their production and decay properties, and the decay constants $f_X$ extracted from QCD sum rules are important input parameters.

Secondly, we use the current algebra to study productions of $D^{(*)} \bar D^{(*)}$, $D^{(*)} \bar K^{(*)}$, and $D^{(*)} \bar D_s^{(*)}$ molecular states in $B$ and $B^{*}$ decays. We derive the relative production rates
\begin{eqnarray}
\mathcal{R}_1 &\equiv& {\mathcal{B}\left(B^- \rightarrow K^- X / D^- X / \phi X \right) \over \mathcal{B}\left(B^- \rightarrow D^-|D^* \bar K^*; 0^{+} \rangle \right)/d_2} \, ,
\\[1mm] \mathcal{R}_2 &\equiv& {\mathcal{B}\left(B^{*-} \rightarrow K^- X / D^- X / \phi X \right) \over \mathcal{B}\left(B^{*-} \rightarrow K^-|D \bar D; 0^{++} \rangle \right)/d_4} \, ,
\end{eqnarray}
and the obtained results are summarized in Table~\ref{tab:width}. In the calculations we only consider $\eta_{1\cdots6}$, $\xi_{1\cdots6}$, and $\zeta_{1\cdots6}$ defined in Eqs.~(\ref{def:eta1}--\ref{def:eta6}), Eqs.~(\ref{def:xi1}--\ref{def:xi6}), and Eqs.~(\ref{def:zeta1}--\ref{def:zeta6}), which couple to these molecular states through $S$-wave. Besides, there may exist some other currents coupling to them through $P$-wave, which are not taken into account in the present study. Consequently, some other molecular states such as $|D \bar D; 0^{++} \rangle$ may still be produced in $B^-$ decays, and omissions of these currents cause some theoretical uncertainties.

Thirdly, we use the Fierz rearrangement of the Dirac and color indices to study strong decay properties of $D^{(*)} \bar D^{(*)}$, $D^{(*)} \bar K^{(*)}$, and $D^{(*)} \bar D_s^{(*)}$ molecular states. We calculate some of their relative branching ratios, and the obtained results are summarized in Table~\ref{tab:width}. In the calculations we only consider the leading-order fall-apart decays described by color-singlet-color-singlet meson-meson currents, but neglect the $\mathcal{O}(\alpha_s)$ corrections described by color-octet-color-octet meson-meson currents. Hence, there may be some other possible decay channels, and omissions of the $\mathcal{O}(\alpha_s)$ corrections cause some theoretical uncertainties.

Generally speaking, the uncertainty of our QCD sum rule results is moderate, as given in Table~\ref{tab:mass}, while uncertainties of relative branching ratios and relative production rates are much larger. In the present study we use local currents and work under the naive factorization scheme, so our uncertainties are significantly larger than the well-developed QCD factorization scheme~\cite{Beneke:1999br,Beneke:2000ry,Beneke:2001ev}, whose uncertainty is at the 5\% level~\cite{Li:2020rcg}. On the other hand, we only calculate the ratios, which significantly reduces our uncertainties~\cite{Yu:2017zst}. Hence, we roughly estimate uncertainties of relative branching ratios to be at the $X^{+100\%}_{-~50\%}$ level, and uncertainties of relative production rates to be at the $X^{+200\%}_{-~67\%}$ level.

There are three unfixed parameters in Table~\ref{tab:width}: the parameters $t^\prime \approx d_2 /d_1 \approx d_5 / d_4$ and $t^{\prime\prime} \approx d_3 /d_1 \approx d_6 / d_4$ (partly) measure the dynamical difference among Fig.~\ref{fig:production}(a,b,c), and the parameter $t \approx {a_i^\prime / a_i} \approx {b_i^\prime / b_i}$~($i=1\cdots6$) measures the dynamical difference between Fig.~\ref{fig:decay}(a) and Fig.~\ref{fig:decay}(b). Simply assuming them to be $t \approx t^\prime \approx t^{\prime\prime} \approx 1$ (local currents), we use the results given in Table~\ref{tab:width} to draw conclusions as follows:

\subsubsection{$X_0(2900)$ as $|D^* \bar K^*; 0^{+} \rangle$}

Our QCD sum rule analyses support the interpretation of the $X_0(2900)$ as the $|D^* \bar K^*; 0^{+} \rangle$ molecular state. Given its width to be about $\Gamma_{X_0(2900)} = 57.2$~MeV~\cite{lhcb}, we use Eqs.~(\ref{eq:DKwidth}) and (\ref{result:DsKs0p}) to estimate:
\begin{equation}
\Gamma_{X_0(2900)} \approx \Gamma_{\bar K^*} + \Gamma_{X \to D \bar K} + \Gamma_{X \to D^* \bar K^* \to D^* \bar K \pi} \, ,
\end{equation}
where
\begin{eqnarray}
\nonumber \Gamma_{\bar K^*} &\approx& 49.1{~\rm MeV} \, ,
\\ \nonumber \Gamma_{X \to D^0 \bar K^0}  &\approx& 3.6{~\rm MeV} \, ,
\\ \Gamma_{X \to D^+ K^-}  &\approx& 3.6{~\rm MeV} \, ,
\\ \nonumber \Gamma_{X \to D^{*0} \bar K^{*0} \to D^{*0} (\bar K \pi)^0}  &\approx& 0.4{~\rm MeV} \, ,
\\ \nonumber \Gamma_{X \to D^{*+} K^{*-} \to D^{*+} (\bar K \pi)^-}  &\approx& 0.4{~\rm MeV} \, .
\end{eqnarray}
According to the first term $\Gamma_{\bar K^*}$, we propose to confirm the $X_0(2900)$ in its dominant $D^* \bar K \pi$ decay channel.

It is interesting to notice that in the present study the neutral $X_0(2900)^0$ is produced in the $B^- \to D^- [D^{*0} {\bar K}^{*0}]$ channel (Fig.~\ref{fig:production}(b)) and not in the suppressed $B^- \to D^- [D^{*+} {K}^{*-}]$ channel (Fig.~\ref{fig:production}(e)). However, it equally decays into the $D^0 \bar K^0$ and $D^+ K^-$ final states. Recalling that the branching ratio of the $B^- \to D^- D^{0} {\bar K}^{0}$ decay is significantly larger than that of the $B^- \to D^- D^{+} K^{-}$ decay~\cite{pdg}:
\begin{eqnarray}
\nonumber \mathcal{B} (B^- \to D^- D^{0} {\bar K}^{0}) &=& (  1.55 \pm 0.21 ) \times 10^{-3} \, ,
\\ \mathcal{B} (B^- \to D^- D^{+} K^{-}) &=& (  2.2 \pm 0.7 ) \times 10^{-4} \, ,
\end{eqnarray}
the non-suppressed signal of the neutral $X_0(2900)^0$ in the suppressed $B^- \to D^- D^{+} K^{-}$ decay channel is probably much more significant than in the $B^- \to D^- D^{0} {\bar K}^{0}$ decay channel. This behavior can be used to test its isospin breaking effect.

\subsubsection{$X_0(2900)$ versus $X(3872)$}

The $X(3872)$ and $X_0(2900)$ have both been observed in $B$ decays. Some of their experimental measurements are~\cite{pdg,Aaij:2020hon,Aaij:2020ypa}:
\begin{eqnarray}
&& \mathcal{B}(B^- \to K^- X(3872) \xrightarrow{~\,\omega\,~} K^- J/\psi \pi\pi\pi)
\label{eq:X3872br}
\\ \nonumber && ~~~~~~~~~~~~~~~~~~~~~~~~~~~\, = (6.0 \pm 2.2) \times 10^{-6} \, ,
\\[1mm] &&
\mathcal{B}(B^- \to D^- D^+ K^-) = (2.2 \pm 0.7) \times 10^{-4} \, ,
\\[1mm] && {\mathcal{B}(B^- \to D^- X_0(2900) \to D^- D^+ K^-) \over \mathcal{B}(B^- \to D^- D^+ K^-)}
\\ \nonumber && ~~~~~~~~~~~~~~~~~~~~~~~~~~~\, = (5.6 \pm 1.4 \pm 0.5) \times 10^{-2} \, ,
\end{eqnarray}
from which we can further derive
\begin{equation}
{\mathcal{B}(B^- \to D^- X_0(2900) \to D^- D^+ K^-) \over \mathcal{B}(B^- \to K^- X(3872) \xrightarrow{~\,\omega\,~} K^- J/\psi \pi\pi\pi)}
= 2.1 \pm 1.1 \, .
\label{eq:result1}
\end{equation}

From Table~\ref{tab:width} we can derive:
\begin{equation}
{\mathcal{B}(B^- \to D^- |D^* \bar K^* \rangle_{0^{+}} \to D^- D^+ K^-) \over \mathcal{B}(B^- \to K^- |D \bar D^* \rangle_{1^{++}} \xrightarrow{~\,\omega\,~} K^- J/\psi \pi\pi\pi)}
\approx 3.1 \, ,
\label{eq:result2}
\end{equation}
with the uncertainty roughly at the $X^{+200\%}_{-~66\%}$ level.

The two values given in Eq.~(\ref{eq:result1}) and Eq.~(\ref{eq:result2}) are well consistent with each other, supporting the interpretations of the $X(3872)$ and $X_0(2900)$ as the $|D \bar D^*; 1^{++} \rangle$ and $|D^* \bar K^*; 0^{+} \rangle$ molecular states, respectively.

\subsubsection{$Z_{cs}(3985)$ as $|D \bar D_s^*; 1^{++} \rangle$ and $Z_{cs}(4000)$ as $|D \bar D_s^*; 1^{+-} \rangle$}

Since the $D D_s^{*-}$ and $D^* D_s^-$ thresholds are very close to each other, in the present study we use them to combine two mixed states, as defined in Eqs.~(\ref{def:DDs2}--\ref{def:DDs3}):
\begin{eqnarray}
\nonumber \sqrt2| D \bar D_s^{*}; 1^{++} \rangle &=& | D D_s^{*-} \rangle_{J=1} + | D^* D_s^- \rangle_{J=1} \, ,
\\[1mm]
\nonumber \sqrt2| D \bar D_s^{*}; 1^{+-} \rangle &=& | D D_s^{*-} \rangle_{J=1} - | D^* D_s^- \rangle_{J=1} \, .
\end{eqnarray}
They are strange partner states of $| D \bar D^*; 1^{++} \rangle$ and $| D \bar D^*; 1^{+-} \rangle$, so we simply denote their quantum numbers as $J^{PC} = 1^{++}$ and $1^{+-}$. Our QCD sum rule analyses support the interpretations of the $Z_{cs}(3985)$ and $Z_{cs}(4000)$ as the $|D \bar D_s^*; 1^{++} \rangle$ and $|D \bar D_s^*; 1^{+-} \rangle$ molecular states, respectively, {\it i.e.}, they are strange partners of the $X(3872)$ and $Z_c(3900)$, respectively.

Our results given in Table~\ref{tab:width} suggest that the $Z_{cs}(4000)$ as $|D \bar D_s^*; 1^{+-} \rangle$ can be observed in the $B^- \to \phi K^- J/\psi$ decay, but the $Z_{cs}(3985)$ as $|D \bar D_s^*; 1^{++} \rangle$ can not. This is just consistent with the BESIII and LHCb observations~\cite{Ablikim:2020hsk,Aaij:2021ivw}. To verify the above interpretations, we propose to search for the $Z_{cs}(3985)$ in the $B^- \to  \phi J/\psi K^{*-} \to \phi J/\psi (\bar K \pi)^-$ decay process. From Table~\ref{tab:width} we can derive:
\begin{equation}
{\mathcal{B}(B^- \to \phi Z_{cs}(3985) \xrightarrow{~K^{*}~} \phi J/\psi (\bar K \pi)^-) \over \mathcal{B}(B^- \to K^- X(3872) \xrightarrow{~\,\omega\,~} K^- J/\psi \pi\pi\pi)}
\approx  0.41 \, ,
\\
\end{equation}
Together with Eq.~(\ref{eq:X3872br}) we can further derive:
\begin{equation}
\mathcal{B}(B^- \to \phi Z_{cs}(3985) \xrightarrow{~K^{*}~} \phi J/\psi (\bar K \pi)^-)
\approx  2.4 \times 10^{-6} \, .
\end{equation}

\subsubsection{$Z_{cs}(4000)$ and $Z_{cs}(4220)$ versus $X(3872)$}

The $Z_{cs}(4000)$ and $Z_{cs}(4220)$ have both been observed in the $B^- \to \phi J/\psi K^-$ decay. Some of their experimental measurements are~\cite{pdg,Aaij:2021ivw}:
\begin{eqnarray}
&& \mathcal{B}(B^- \to \phi J/\psi K^-) = (5.0 \pm 0.4) \times 10^{-5} \, ,
\\[1mm] && {\mathcal{B}(B^- \to \phi Z_{cs}(4000) \to \phi J/\psi K^-) \over \mathcal{B}(B^- \to \phi J/\psi K^-)}
\\ \nonumber && ~~~~~~~~~~~~~~~~~~~~~~~~~\, = (9.4 \pm 2.1 \pm 3.4) \times 10^{-2} \, ,
\\[1mm] && {\mathcal{B}(B^- \to \phi Z_{cs}(4220) \to \phi J/\psi K^-) \over \mathcal{B}(B^- \to \phi J/\psi K^-)}
\\ \nonumber && ~~~~~~~~~~~~~~~~~~~~~~~~~\, = (10 \pm 4 ^{+10}_{-~7}) \times 10^{-2} \, .
\end{eqnarray}
Together with Eq.~(\ref{eq:X3872br}), we can further derive
\begin{eqnarray}
\nonumber {\mathcal{B}(B^- \to \phi Z_{cs}(4000) \to \phi J/\psi K^-) \over \mathcal{B}(B^- \to K^- X(3872) \xrightarrow{\,\omega\,} K^- J/\psi \pi\pi\pi)}
&=& 0.78 \pm 0.44 \, ,
\\[1mm]
\nonumber {\mathcal{B}(B^- \to \phi Z_{cs}(4220) \to \phi J/\psi K^-) \over \mathcal{B}(B^- \to K^- X(3872) \xrightarrow{\,\omega\,} K^- J/\psi \pi\pi\pi)}
&=& 0.83 ^{+0.95}_{-0.74} \, .
\\
\label{eq:result3}
\end{eqnarray}

From Table~\ref{tab:width} we can derive:
\begin{eqnarray}
\nonumber {\mathcal{B}(B^- \to \phi |D \bar D_s^* \rangle_{1^{+-}} \to \phi J/\psi K^-) \over \mathcal{B}(B^- \to K^- |D \bar D^* \rangle_{1^{++}} \xrightarrow{~\,\omega\,~} K^- J/\psi \pi\pi\pi)}
&\approx& 0.28 \, ,
\\[1mm]
\nonumber {\mathcal{B}(B^- \to \phi |D^* \bar D_s^* \rangle_{1^{+-}} \to \phi J/\psi K^-) \over \mathcal{B}(B^- \to K^- |D \bar D^* \rangle_{1^{++}} \xrightarrow{~\,\omega\,~} K^- J/\psi \pi\pi\pi)}
&\approx& 2.7 \, ,
\\
\label{eq:result4}
\end{eqnarray}
with the uncertainty roughly at the $X^{+200\%}_{-~66\%}$ level.

The two sets of values given in Eqs.~(\ref{eq:result3}) and Eqs.~(\ref{eq:result4}) are consistent with each other, supporting the interpretations of the $Z_{cs}(4000)$ and $Z_{cs}(4220)$ as the $|D \bar D_s^*; 1^{+-} \rangle$ and $|D^* \bar D_s^*; 1^{+-} \rangle$ molecular states, respectively. To verify these interpretations, we propose to confirm the $Z_{cs}(4000)$ in the $B^- \to \phi D^0 D_s^{*-}$ and $B^- \to \phi D^{*0} D_s^-$ decays, and the $Z_{cs}(4220)$ in the $B^- \to \phi D^{*0} D_s^{*-}$ decay.

\subsubsection{$Z_c(3900)$ and $Z_c(4020)$}

Our QCD sum rule analyses support the interpretations of the $Z_c(3900)$ and $Z_c(4020)$ as the $|D \bar D^*; 1^{+-} \rangle$ and $|D^* \bar D^*; 1^{+-} \rangle$ molecular states, respectively. From Table~\ref{tab:width} we can derive:
\begin{eqnarray}
\nonumber {\mathcal{B}(B^- \to K^- Z_c(3900) \to K^- J/\psi \pi^0) \over \mathcal{B}(B^- \to K^- X(3872) \xrightarrow{~\,\omega\,~} K^- J/\psi \pi\pi\pi)}
&\approx&  0.078 \, ,
\\[1mm] \nonumber {\mathcal{B}(B^- \to K^- Z_c(4020) \to K^- J/\psi \pi^0) \over \mathcal{B}(B^- \to K^- X(3872) \xrightarrow{~\,\omega\,~} K^- J/\psi \pi\pi\pi)}
&\approx&  1.7 \, .
\\
\end{eqnarray}
Together with Eq.~(\ref{eq:X3872br}) and the experiment measurement~\cite{pdg},
\begin{equation}
\mathcal{B}(B^- \to J/\psi \bar K^0 \pi^-) = (1.14 \pm 0.11) \times 10^{-3} \, ,
\end{equation}
we can further derive:
\begin{eqnarray}
\nonumber {\mathcal{B}(B^- \to K^- Z_c(3900) \to K^- J/\psi \pi^0) \over \mathcal{B}(B^- \to K^- J/\psi \pi^0)}
&\approx&  0.1\% \, ,
\\[1mm] \nonumber {\mathcal{B}(B^- \to K^- Z_c(4020) \to K^- J/\psi \pi^0) \over \mathcal{B}(B^- \to K^- J/\psi \pi^0)}
&\approx&  1.7\% \, .
\\
\end{eqnarray}
Accordingly, we propose to search for the $Z_c(4020)$ in the $B^- \to K^- Z_c \to K^- J/\psi \pi^0$ decay process, but note that uncertainties of the above estimations are roughly at the $X^{+200\%}_{-~66\%}$ level.

\subsubsection{$B^*$ decays}

We propose to confirm the $X(3872)$, $Z_c(3900)$, $Z_c(4020)$, $Z_{cs}(3985)$, and $Z_{cs}(4000)$ in $B^{*}$ decays, although the identification of $B^*$ in its hadronic decays is still not easy. Based on the interpretations of the present study, we can derive from Table~\ref{tab:width}:
\begin{eqnarray}
\nonumber {\mathcal{B}(B^{*-} \to K^- Z_c(3900) \to K^- J/\psi \pi^0) \over \mathcal{B}(B^{*-} \to K^- X(3872) \xrightarrow{~\,\omega\,~} K^- J/\psi \pi\pi\pi)}
&\approx&  0.08 \, ,
\\[1mm] \nonumber {\mathcal{B}(B^{*-} \to K^- Z_c(4020) \to K^- J/\psi \pi^0) \over \mathcal{B}(B^{*-} \to K^- X(3872) \xrightarrow{~\,\omega\,~} K^- J/\psi \pi\pi\pi)}
&\approx&  2.2 \, ,
\\[1mm] \nonumber {\mathcal{B}(B^{*-} \to \phi Z_{cs}(3985) \xrightarrow{~K^{*}~} \phi J/\psi (\bar K \pi)^-) \over \mathcal{B}(B^{*-} \to K^- X(3872) \xrightarrow{~\,\omega\,~} K^- J/\psi \pi\pi\pi)}
&\approx& 0.11 \, ,
\\[1mm] \nonumber {\mathcal{B}(B^{*-} \to \phi Z_{cs}(4000) \to \phi J/\psi K^-) \over \mathcal{B}(B^{*-} \to K^- X(3872) \xrightarrow{~\,\omega\,~} K^- J/\psi \pi\pi\pi)}
&\approx& 0.08 \, .
\\
\end{eqnarray}

\subsubsection{Other possibly existing states}

To end this paper, we propose to search for the possibly existing $|D^* \bar D^*; 0^{++} \rangle$, $|D^* \bar K; 1^{+} \rangle$, and $|D^* \bar K^*; 1^{+} \rangle$ molecular states in $B$ decays:
\begin{itemize}

\item We propose to search for the isoscalar $|D^* \bar D^*; 0^{++} \rangle$ molecular state in the $B^- \to K^- X \to K^- \eta_c \eta$ decay, and the isovector one in the $B^- \to K^- X \to K^- \eta_c \pi$ decay.

\item We propose to search for the $|D^* \bar K; 1^{+} \rangle$ molecular state in the $B^- \to D^- X \to D^- D \bar K \pi$ decay.

\item We propose to search for the $|D^* \bar K^*; 1^{+} \rangle$ molecular state in the $B^- \to D^- X \to D^- D \bar K \pi$ and $B^- \to D^- X \to D^- D^* \bar K \pi$ decays.

\end{itemize}
Estimations on their relative production rates and relative branching ratios can be found in Table~\ref{tab:width}, although there are large uncertainties coming from their unknown (hadron) masses. Before doing this, we propose to search for the relevant decay processes first, {\it i.e.}, the $B^- \to K^- \eta_c \eta$, $B^- \to K^- \eta_c \pi$, $B^- \to D^- D \bar K \pi$, and $B^- \to D^- D^* \bar K \pi$ decays. Besides, we propose to search for the possibly existing $|D \bar D; 0^{++} \rangle$, $|D \bar K; 0^{+} \rangle$, $|D \bar K^*; 1^{+} \rangle$, $|D^* \bar K^*; 1^{+} \rangle$, and $|D \bar D_s; 0^{+} \rangle$ molecular states in $B^{*}$ decays.

\section*{Acknowledgments}

We thank Wei Chen and Li-Ming Zhang for helpful discussions.
This project is supported by the National Natural Science Foundation of China under Grants No.~11722540 and No.~12075019,
and
the Fundamental Research Funds for the Central Universities.

\appendix

\section{Width of composite particle}
\label{app:composite}

%
\begin{figure*}[hbt]
\begin{center}
\subfigure[~\mbox{$A \rightarrow B + C \rightarrow B + c_1 + c_2$.}]{\includegraphics[width=0.4\textwidth]{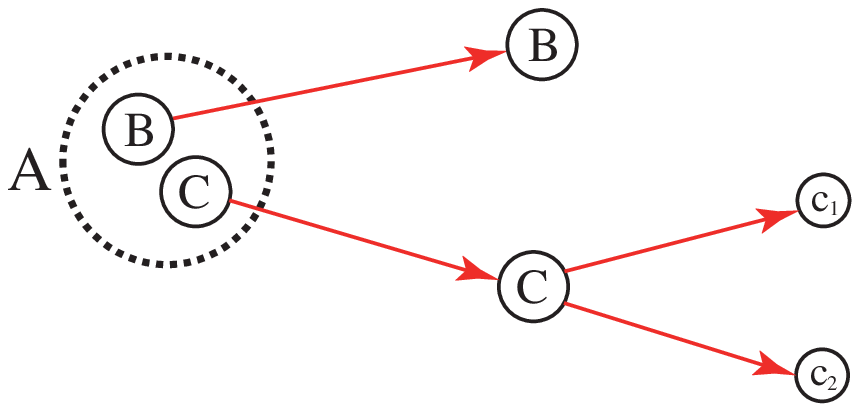}}
~~~~~~~~~~
\subfigure[~\mbox{C decays, so that $A$ also decays.}]{\includegraphics[width=0.4\textwidth]{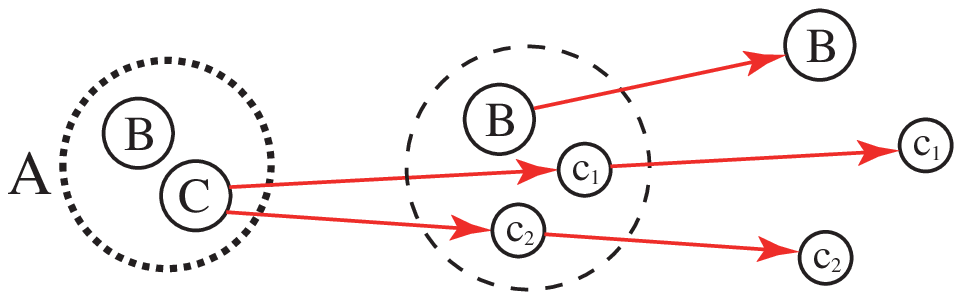}}
\caption{Two possible decay processes of the composite particle $A$ composed by two sub-particles $B$ and $C$.}
\label{fig:composite}
\end{center}
\end{figure*}
%

In this appendix we generally study the width of a composite particle $A$, which is composed by two sub-particles $B$ and $C$. We discuss several cases:
\begin{itemize}[leftmargin = 40 pt]

\item[Case A:] The two sub-particles $B$ and $C$ are both stable. If the mass of $A$ is below the $BC$ threshold, that is $M_A \leq M_B + M_C$, the composite particle $A$ is also stable. If $M_A > M_B + M_C$, the width of $A$ is $\Gamma_A = \Gamma_{A \to B + C}$.

\item[Case B:] The sub-particle $C$ can decay into $c_1$ and $c_2$, but its width is not very large. If $M_A \leq M_B + M_{c_1} + M_{c_2}$, the composite particle $A$ is stable. If $M_A > M_B + M_C$, the width of $A$ can be estimated as $\Gamma_A \approx \Gamma_{A \to B + C}$. If $M_B + M_{c_1} + M_{c_2}< M_A \leq M_B + M_C$, the width of $A$ can be estimated as the width of the decay process $\Gamma_A \approx \Gamma_{A \to B + C \to B + c_1 + c_2}$.

\item[Case C:] The sub-particle $C$ can decay into $c_1$ and $c_2$, but its width is not so small. We further assume that $M_A > M_B + M_C$, so that $A$ can decay into $B$ and $C$. The width of this process is $\Gamma_{A \to B + C}$, proportional to $g_{A \to BC}^2$, with $g_{A \to BC}$ the relevant coupling constant.

    Besides, the sub-particle $C$ can decay inside the composite particle $A$, so that $A$ also decays. The width of this process is probably smaller than $\Gamma_C$. Altogether, we arrive at $\Gamma_A \lesssim \Gamma_C + \Gamma_{A \to B + C}$.

\item[Case D:] The sub-particle $C$ can decay into $c_1$ and $c_2$, but its width is not so small. We further assume that $M_B + M_{c_1} + M_{c_2}< M_A \leq M_B + M_C$, so that $A$ can decay into $BC$ and further into $B c_1 c_2$. The width of this process is $\Gamma_{A \to B + C \to B + c_1 + c_2}$, again proportional to $g_{A \to BC}^2$.

    Besides, the sub-particle $C$ can decay inside the composite particle $A$, so that $A$ also decays. Altogether, we arrive at $\Gamma_A \lesssim \Gamma_C + \Gamma_{A \to B + C \to B + c_1 + c_2}$.

\end{itemize}
In the above discussions we just want to argue that the decay process depicted in Fig.~\ref{fig:composite}(a) and the one depicted in Fig.~\ref{fig:composite}(b) are not the same. The former depends on the coupling constant $g_{A \to BC}$, while the latter depends on widths of sub-particles but not depends on $g_{A \to BC}$. It might not be reasonable to take these two decay processes as the same one, although they might not be fully independent.

Generally assuming that the two sub-particles $B$ and $C$ are both unstable as well as $M_A > M_B + M_C$ or $M_B + M_{c_1} + M_{c_2}< M_A \leq M_B + M_C$, we arrive at
\begin{equation}
\Gamma_A \lesssim \Gamma_B + \Gamma_C + \Gamma_{A \to BC(\to c_1 c_2)} + \cdots \, ,
\end{equation}
with $\cdots$ partial widths of other possible decay channels. Consequently, we can estimate the lifetime of the composite particle $A$ through
\begin{equation}
1/t_A \lesssim 1/t_B + 1/t_C + \Gamma_{A \to BC(\to c_1 c_2)} + \cdots \, .
\end{equation}

In the present study we simply use
\begin{equation}
\Gamma_A \approx \Gamma_B + \Gamma_C + \Gamma_{A \to BC(\to c_1 c_2)} + \cdots \, ,
\end{equation}
for a weakly-coupled composite system. This formula may be useful to examine the nature of the particle $A$, {\it i.e.}, to discriminate whether it is a compact multiquark state or a composite hadronic molecular state.

\section{Uncertainties from phase angles}
\label{app:phase}

As discussed in Sec.~\ref{sec:DD0pp}, there are two different terms, $A \equiv \bar q_a \gamma_5 q_a \times \bar c_b \gamma_5 c_b$ and $B \equiv \bar q_a \gamma_\mu \gamma_5 q_a \times \bar c_b \gamma^\mu \gamma_5 c_b$, both of which can contribute to the decay of $| D \bar D; 0^{++} \rangle$ into the $\eta_c \eta$ final states; there are also two different terms, $C \equiv \bar q_a \gamma_\mu q_a \times \bar c_b \gamma^\mu c_b$ and $D \equiv \bar q_a \sigma_{\mu\nu} q_a \times \bar c_b \sigma^{\mu\nu} c_b$, both of which can contribute to the decay of $| D \bar D; 0^{++} \rangle$ into the $J/\psi \omega$ final states.

Besides, there can be more than one terms contributing to some other decay channels. Phase angles among them, such as the phase angle between the two coupling constants $f_{J/\psi}$ and $f_{J/\psi}^T$, can not be well determined in the present study. In this appendix we rotate these phase angles and redo all the calculations. Their relevant uncertainties are summarized in Table~\ref{tab:uncertainty}.

Especially, if we take into account these unknown phase angles, the BESIII measurement~\cite{Ablikim:2019ipd},
\begin{equation}
{\mathcal{B}(Z_c(3900)^\pm \rightarrow \eta_c\rho^\pm) \over \mathcal{B}(Z_c(3900)^\pm \rightarrow J/\psi\pi^\pm)} = 2.2 \pm 0.9 \, ,
\end{equation}
can be explained with the interpretation of $Z_c(3900)$ as the $| D \bar D^*; 1^{+-} \rangle$ molecular state.

\begin{table*}[hbt]
\begin{center}
\renewcommand{\arraystretch}{1.6}
\caption{Relative branching ratios of $D^{(*)} \bar D^{(*)}$, $D^{(*)} \bar K^{(*)}$, and $D^{(*)} \bar D_s^{(*)}$ hadronic molecular states and their relative production rates in $B$ and $B^{*}$ decays. See the caption of Table~\ref{tab:width} for detailed explanations. In this table we take into account those unknown phase angles among different coupling constants, so that the BESIII measurement ${\mathcal{B}(Z_c(3900)^\pm \rightarrow \eta_c\rho^\pm) \over \mathcal{B}(Z_c(3900)^\pm \rightarrow J/\psi\pi^\pm)} = 2.2 \pm 0.9$~\cite{Ablikim:2019ipd} can be explained with the interpretation of $Z_c(3900)$ as the $| D \bar D^*; 1^{+-} \rangle$ molecular state.}
\begin{tabular}{c || c | c || c | c | c | c | c | c | c | c | c | c | c | c | c}
\hline\hline
\multirow{2}{*}{Configuration} & \multicolumn{2}{c||}{Productions} & \multicolumn{13}{c}{Decay Channels}
\\ \cline{2-16}
& ~\,$\mathcal{R}_1$\,~ & ~\,$\mathcal{R}_2$\,~
& ~$\eta_c \eta$~ & $J/\psi \omega$ & $\eta_c f_0$ & $\chi_{c0} f_0$ & $\chi_{c1} f_0$
& $J/\psi \pi$ & $\chi_{c0} \pi$ & $h_c \pi$ & $\eta_c \rho$ & $J/\psi \rho$ & $\chi_{c1} \rho$
& $D \bar D^*$ & $D^* \bar D^*$
\\ \hline\hline
$|D \bar D; 0^{++} \rangle$      & --        &  $d_4$    & $0.9$--$1.0$ & $10^{-6}$ & -- & $10^{-5}$ & -- & -- & -- & -- & -- & -- & -- & -- & --
\\ \hline
$|D \bar D^*; 1^{++} \rangle$    & $0.19d_1$ &  $5.7d_4$ & -- & $1$ & $0.09$ & -- & $0.09$ & -- & $0.02$ & -- & -- & $0.63^\dagger$ & -- & $7.4t$ & --
\\ \hline
$|D \bar D^*; 1^{+-} \rangle$    & $0.12d_1$ &  $3.7d_4$ & -- & -- & -- & -- & -- & $1.0$--$5.6$ & -- & $0.01$ & $0.1$--$2.4$ & -- & $10^{-6}$ & $74t$ & --
\\ \hline
$|D^* \bar D^*; 0^{++} \rangle$  & $1.3d_1$  &  --       & $0.9$--$1.0$ & $0.001$ & -- & $0.001$ & -- & -- & -- & -- & -- & -- & -- & -- & $10^{-4}t$
\\ \hline
$|D^* \bar D^*; 1^{+-} \rangle$  & $0.42d_1$ &  $17d_4$  & -- & -- & -- & -- & -- & $1.0$--$1.9$ & -- & $0.002$ & $0.33$--$0.89$ & -- & $10^{-5}$ & -- & $11t$
\\ \hline
$|D^* \bar D^*; 2^{++} \rangle$  & --        &  --       & $1$ & $5$--$73$ & -- & -- & -- & -- & -- & -- & -- & -- & -- & -- & $9.5t$
\\ \hline \hline
Configuration & ~\,$\mathcal{R}_1$\,~ & ~\,$\mathcal{R}_2$\,~
& \multicolumn{2}{c|}{$\Gamma_{\bar K^*}$~[MeV]}
& $D \bar K$
& \multicolumn{2}{c|}{$D \bar K^*$}
& \multicolumn{3}{c|}{$D^* \bar K$}
& \multicolumn{3}{c|}{$D^* \bar K^*$}
& \multicolumn{2}{c}{$D^* \bar K^*$}
\\ \hline\hline
$|D \bar K; 0^{+} \rangle$       & --        & $1.1d_5$ & \multicolumn{2}{c|}{--}     & -- & \multicolumn{2}{c|}{--}  & \multicolumn{3}{c|}{--}  & \multicolumn{3}{c|}{--}  & \multicolumn{2}{c}{--}
\\ \hline
$|D \bar K^*; 1^{+} \rangle$     & --        & $10d_5$  & \multicolumn{2}{c|}{$49.1$~MeV} & -- & \multicolumn{2}{c|}{${t}$}  & \multicolumn{3}{c|}{$0.06$--$0.25$}   & \multicolumn{3}{c|}{$10^{-4}$} & \multicolumn{2}{c}{--}
\\ \hline
$|D^* \bar K; 1^{+} \rangle$     & $0.67d_2$ & --       & \multicolumn{2}{c|}{--}     & -- & \multicolumn{2}{c|}{$10^{-6}$}  & \multicolumn{3}{c|}{${t}$}  & \multicolumn{3}{c|}{--}  & \multicolumn{2}{c}{--}
\\ \hline
$|D^* \bar K^*; 0^{+} \rangle$   & $d_2$     & --       & \multicolumn{2}{c|}{$49.1$~MeV} & $3.5$--$8.5$  & \multicolumn{2}{c|}{--}  & \multicolumn{3}{c|}{--}   & \multicolumn{3}{c|}{${t}$} & \multicolumn{2}{c}{--}
\\ \hline
$|D^* \bar K^*; 1^{+} \rangle$   & $0.70d_2$ & $27d_5$  & \multicolumn{2}{c|}{$49.1$~MeV} & -- & \multicolumn{2}{c|}{$0.34$--$0.75$}  & \multicolumn{3}{c|}{$1.1$--$2.0$}   & \multicolumn{3}{c|}{${t}$} & \multicolumn{2}{c}{$10^{-4}t$}
\\ \hline
$|D^* \bar K^*; 2^{+} \rangle$   & --        & --       & \multicolumn{2}{c|}{$49.1$~MeV} & $0.02$ & \multicolumn{2}{c|}{--}  & \multicolumn{3}{c|}{--}   & \multicolumn{3}{c|}{${t}$} & \multicolumn{2}{c}{$10^{-4}t$}
\\ \hline \hline
Configuration & ~\,$\mathcal{R}_1$\,~ & ~\,$\mathcal{R}_2$\,~
& $\eta_c \bar K$
& \multicolumn{2}{c|}{$J/\psi \bar K$}
& $\chi_{c0} \bar K$
& $h_c \bar K$
& $\eta_c \bar K^*$
& \multicolumn{2}{c|}{$J/\psi \bar K^*$}
& \multicolumn{2}{c|}{$\chi_{c1} \bar K^*$}
& $D \bar D_s^*$
& $D^* \bar D_s$
& $D^* \bar D_s^*$
\\ \hline\hline
$|D \bar D_s; 0^{+} \rangle$      & --        &  $11d_6$   & $0.2$--$1.0$ & \multicolumn{2}{c|}{--} & -- & -- & -- & \multicolumn{2}{c|}{$0.001$} & \multicolumn{2}{c|}{--} & -- & -- & --
\\ \hline
$|D \bar D_s^*; 1^{++} \rangle$   & $0.79d_3$ &  $6.6d_6$  & -- & \multicolumn{2}{c|}{--} & $0.01$ & -- & -- & \multicolumn{2}{c|}{$1$} & \multicolumn{2}{c|}{--} & $25$ & $68$ & --
\\ \hline
$|D \bar D_s^*; 1^{+-} \rangle$   & $0.75d_3$ &  $6.2d_6$  & -- & \multicolumn{2}{c|}{$1.0$--$8.3$} & -- & -- & $0.1$--$3.1$ & \multicolumn{2}{c|}{--} & \multicolumn{2}{c|}{--} & $41$ & $87$ & --
\\ \hline
$|D^* \bar D_s^*; 0^{+} \rangle$  & --        &  --        & $0.3$--$1.0$ & \multicolumn{2}{c|}{--} & -- & -- & -- & \multicolumn{2}{c|}{$0.09$} & \multicolumn{2}{c|}{--} & -- & -- & $0.001$
\\ \hline
$|D^* \bar D_s^*; 1^{+} \rangle$  & $1.8d_3$  &  --        & -- & \multicolumn{2}{c|}{$1.0$--$2.3$} & -- & $0.002$ & $0.4$--$1.0$ & \multicolumn{2}{c|}{--} & \multicolumn{2}{c|}{$10^{-7}$} & -- & -- & $31$
\\ \hline
$|D^* \bar D_s^*; 2^{+} \rangle$  & --        &  --        & $1$ & \multicolumn{2}{c|}{--} & -- & -- & -- & \multicolumn{2}{c|}{$2$--$27$} & \multicolumn{2}{c|}{--} & -- & -- & $0.06$
\\ \hline\hline
\end{tabular}
\label{tab:uncertainty}
\end{center}
\end{table*}

\end{document}